# IMPROVED LABORATORY TRANSITION PROBABILITIES FOR Ce II, APPLICATION TO THE CERIUM ABUNDANCES OF THE SUN AND FIVE r-PROCESS RICH, METAL-POOR STARS, AND RARE EARTH LAB DATA SUMMARY

(short title: Ce II Transition Probabilities and Abundances)

J. E. Lawler[1], C. Sneden[2], J. J. Cowan[3], I. I. Ivans[4], and E. A. Den Hartog[1]

[1]Department of Physics, University of Wisconsin, Madison, WI 53706; jelawler@wisc.edu, eadenhar@wisc.edu

[2]Department of Astronomy and McDonald Observatory, University of Texas, Austin, TX 78712 & INAF, Osservatorio Astronomico di Padova,

VicoloOsservatorio 5, I-35122 Padova, Italy; chris@verdi.as.utexas.edu,

[3]Homer L. Dodge Department of Physics and Astronomy, University of Oklahoma, Norman, OK 73019; cowan@nhn.ou.edu

[4]The Observatories of the Carnegie Institution of Washington, 813 Santa Barbara St., Pasadena, CA 91101 & Princeton University Observatory, Peyton Hall, Princeton, NJ 08544; iii@ociw.edu

ABSTRACT

Recent radiative lifetime measurements accurate to ±5% using laser-induced fluorescence (LIF) on 43 even-parity and 15 odd-parity levels of Ce II have been combined with new branching fractions measured using a Fourier transform spectrometer (FTS) to determine transition probabilities for 921 lines of Ce II. This improved laboratory data set has been used to determine a new solar photospheric Ce abundance, log $\varepsilon$ = 1.61 ± 0.01 ($\sigma$ = 0.06 from 45 lines), a value in excellent agreement with the recommended meteoritic abundance, log $\varepsilon$ = 1.61 ± 0.02. Revised Ce abundances have also been derived for the *r*-process-rich metal-poor giant stars BD+17º3248, CS 22892-052, CS 31082-001, HD 115444, and HD 221170. Between 26 and 40 lines were used for determining the Ce abundance in these five stars, yielding a small statistical uncertainty of ± 0.01 dex similar to the Solar result. The relative abundances in the metal-poor stars of Ce and Eu, a nearly pure *r*-process element in the Sun, matches *r*-process only model predictions for Solar System material. This consistent match with small scatter over a wide range of stellar metallicities lends support to these predictions of elemental fractions. A companion paper includes an interpretation of these new precision abundance results for Ce as well as new abundance results and interpretation for Pr, Dy, and Tm.

Subject headings: atomic data — stars: abundances stars: Population II — Sun: abundances — Galaxy: evolution—nuclear reactions, nucleosynthesis, abundances—stars: individual (BD +17º3248, CS 22892-052, CS 31082-001, HD 115444, HD 221170)

1. INTRODUCTION

The study of elemental abundances in stellar photospheres continues to be a rich area of investigation. Halo stars are among the oldest objects in the Galaxy and these stars provide a "fossil" record of the Galactic chemical evolution. The discovery and detailed study of a class of metal-poor Galactic halo stars with variable *n*(eutron)-capture elemental abundances (e.g. Sneden et al. 1995, Smith et al. 1995, Cowan et al. 1996, Woolf et al. 1995, Sneden et al. 1996, Burris et al. 2000, Hill et al. 2002, Sneden et al. 2003a) is particularly significant.

Rare-Earth (RE) elements are among the most spectroscopically accessible of the *n*(uetron)-capture elements. The open f-shell of the RE neutral atoms and ions yields many strong lines in the visible and near-IR where spectral line blending is less of a problem than in the UV, and where ground-based observations are possible. Observations with a high Signal-to-Noise ratio (S/N) and a high spectral resolving power on an increasing number of stars are now available from large ground-based telescopes. The quality of these new astronomical data can be fully exploited only if similar quality basic spectroscopic data, especially data on transition probabilities, are available. The combination of improved astronomical data and improved laboratory data has reduced line-to-line and star-to-star scatter in abundance values for many RE elements.

Cerium is the last of the RE elements in need of additional work and it has the richest line spectra of the RE elements. Experimental work since the mid-1990s has emphasized LIF lifetime measurements (e.g. Langhans et al. 1995, Li et al. 2000, Zhang et al. 2001, Xu et al. 2003, Den Hartog & Lawler 2008). These LIF lifetimes have typically been combined with theoretical branching fractions to determine transition probabilities for individual Ce II lines (e.g. Zhang et al. 2001, Palmeri et al. 2000, Biémont & Quinet 2005).

In spite of the somewhat daunting line density, a large set of Ce II branching fractions measurements based on FTS data is now complete. The branching fraction measurements are combined with the most recent and extensive LIF radiative lifetimes from Den Hartog & Lawler (2008) to determine transition probabilities for 921 lines of Ce II reported herein.

Although progress has been made in the theoretical determination of atomic transition probabilities, a recent comparison strongly suggests that modern experimental methods yield more reliable atomic transition probabilities (Lawler et al. 2008a). The method used in this study on Ce is to combine radiative lifetimes from time-resolved laser-induced-fluorescence (TR-LIF) with emission branching fraction measurements from high resolution data recorded using a Fourier transform spectrometer (FTS). This method for measuring transition probabilities is accurate, flexible, and efficient. The systematic determination of experimental transition probabilities by combining radiative lifetimes from TR-LIF with branching fractions from emission data recorded with a FTS has played a central role in providing the basic atomic data needed for RE abundance determinations. This method yields absolute transition probabilities which are accurate to ±5% (~0.02 dex) for strong lines.

These laboratory data are applied to re-determine the Solar abundance of Ce and to refine the Ce abundance in five *r*-process rich, metal poor Galactic halo stars. The Appendix to this paper includes a summary to all of our RE lab data in machine readable form. This work on Ce II completes a multi-year effort to improve laboratory spectroscopic data for Rare Earth (RE) ions and to apply these data in abundance studies (Lawler et al. 2008b and references therein).

The emergence of a tightly defined *r*-process only abundance pattern in many very metal-poor Galactic halo stars, at least for the RE elements, has been an exciting development (e.g. Sneden et al. 2003b, Ivans et al. 2006, Lawler et al. 2006, Den Hartog et al. 2006). As this abundance pattern becomes even more tightly defined, it will: i) provide a powerful constraint on future modeling of the *r*-process nucleosynthesis; ii) help determine a definitive *r*-process site; and iii) unlock other details of the *r*-process and of the Galactic chemical evolution.

## 2. Ce II BRANCHING FRACTIONS AND ATOMIC TRANSITION PROBABILITIES

As discussed above, radiative lifetimes from Den Hartog & Lawler (2008) provide a foundation for this study of branching fractions and the transition probabilities of Ce II. Branching fraction measurements were attempted on lines from all 74 levels of the lifetime experiment, and were completed for lines from 43 even-parity and 15 odd-parity upper levels. As in earlier work on other RE spectra, we used the 1.0 meter FTS at the National Solar Observatory (NSO) on Kitt Peak for Ce II branching fraction measurements. The Kitt Peak FTS has the large etendue of all interferometric spectrometers, a limit of resolution as small as 0.01 cm$^{-1}$, wavenumber accuracy to 1 part in $10^8$, broad spectral coverage from the UV to IR, and the capability of recording a million point spectrum in 10 minutes (Brault 1976). The line density in Ce II is so high that a very high performance FTS is essential. Grating spectrographs are not adequate.

### 2.1 Energy Levels of Ce II

Figure 1 shows a partial Grotrian diagram constructed from the energy level compilation of Martin et al. (1978) for singly ionized Ce. A total of 288 even-parity and 192 odd-parity levels are included in the compilation. Most of the low-lying (< 20,000 cm$^{-1}$) energy levels are known and have assignments. Although Ce II is the most complex of the singly ionized RE spectra, classical analyses of the spectrum have yielded extensive results (e.g. Martin et al. and reference therein). The absence of hfs results in narrow spectral lines and a tolerable amount of line blending for such a rich spectrum.

There are three low odd-parity configurations including the 4f($^2$F)5d$^2$, 4f($^2$F)6s$^2$, and 4f($^2$F)5d6s which contribute over 100 levels below 20,000 cm$^{-1}$ including the ground level. This near degeneracy of the 4f, 5d, and 6s orbitals is typical of REs and it leads to substantial configuration mixing. Of the 118 low-lying odd-parity levels listed in Martin et al. (1978), 74 are assigned to the 4f($^2$F)5d$^2$ configuration, 2 are assigned to the 4f($^2$F)6s$^2$ configuration, and 42 are assigned to the 4f($^2$F)5d6s configuration. Not all of these 118 levels have term assignments, but a counting of angular momentum projections indicates that very few (perhaps only 2) levels from these three configurations are unobserved. There is a gap of about 5,700 cm$^{-1}$ before the 4f$^2$6p configuration starts at 25,766 cm$^{-1}$. The nearly complete knowledge of low odd-parity levels simplifies the branching fraction study, since there is little chance of missing a strong branch to an unobserved lower level. Even-parity upper levels of this study are between 24,000 and 33,000 cm$^{-1}$ and all have short, < 9 nsec, radiative lifetimes. Due to the frequency cubed scaling of transition probabilities, there is no real possibility of near-IR branches to higher unobserved levels contributing more than a small fraction of 1% to the total decay of such short lived levels.

Two configurations, the 4f$^2$6s and 4f$^2$5d, contribute low-lying even-parity levels. Although the 4f$^2$6s configuration extends from 3,800 cm$^{-1}$ to just over 21,000 cm$^{-1}$, the 4f$^2$5d extends up to nearly 40,000 cm$^{-1}$. There are 99 known even-parity levels below 25,000 cm$^{-1}$. A counting of angular momentum projections including levels of the 4f$^2$5d configuration up to 40,000 cm$^{-1}$ indicates that few (perhaps only 4) levels of the two low even-parity configurations are unobserved. These unobserved even-parity levels are low J levels (3/2 or 1/2) and are likely around 20,000 cm$^{-1}$ or higher. Two of the missing J = 1/2 levels were found by Palmeri et al. (2000). This situation is similar to the other parity; there is a nearly complete knowledge of low even-parity levels. Odd-parity upper levels of this study are between 25,000 and 33,000 cm$^{-1}$ and all have short, < 9 nsec, radiative lifetimes. Due to the frequency cubed scaling of transition probabilities, there is again no possibility of near-IR branches to higher unobserved levels contributing more than a small fraction of 1% to the total decay of such short lived levels. This nearly complete knowledge, including assignments, of the low-lying (< 20,000 cm$^{-1}$) energy levels of both parities in Ce II greatly simplifies the branching fraction analysis.

2.2 Ce II Branching Fraction Analysis and Relative Radiometric Calibration

As in earlier studies our experimental branching fractions are based on a large set of FTS data including: spectra of lamps at high currents to reveal very weak branches to known levels, good IR spectra to reveal weak IR branches to known levels, and low current spectra in which dominant branches are optically thin covering the UV to near-IR. Table 1 is a list of the 14 FTS spectra used in our branching fraction study. All were recorded using the National Solar Observatory 1.0 meter FTS on Kitt Peak. Most of these spectra, #1-9, were recorded during a February 2002 observing run using Hollow Cathode Discharge (HCD) lamps. Visible spectra of an Electrodeless Discharge Lamp (EDL), #10-11, have quite good S/N on weaker lines. Lines of Ce II in comparison to lines of Ce I are stronger in spectra #10-11 than in spectra #1-9 and this effect was very valuable in assessing possible blends of Ce II and Ce I lines. Spectra of Ce/Ne HCD lamps were used only to assess and, when necessary, correct for blends between Ce II and Ar lines. Since no Ar II or Ar I lines appeared in the EDL spectra, these could also be used to assess and correct for blends between Ce II and Ar lines. This redundancy is important since many cases of possible blends involving Ce II, Ce I, and Ar lines occur. No significant optical depth effects on Ce II lines appear in any of the spectra, probably due to the rich level structure and large partition function of this ion. Spectra #12-14 yielded measurements on a few near-IR lines, but generally the near IR lines from the short lived levels of this study have negligible branching fractions. All 14 raw spectra are available from the electronic archives of the National Solar Observatory[1].

The establishment of an accurate relative radiometric calibration or efficiency is critical to a branching fraction experiment. As indicated in Table 1, we made use of both standard lamp calibrations and Ar I and Ar II line calibrations in this Ce II study. Tungsten (W) filament standard lamps are particularly useful near the Si detector cutoff in the 10,000 to 9,000 $cm^{-1}$ range where the FTS sensitivity is changing rapidly as a function of wave number, and near the dip in sensitivity at 12,500 $cm^{-1}$ from the aluminum coated optics. Tungsten lamps are not bright enough to be useful for FTS calibrations in the UV region, and blue or UV branches typically dominate the decay of levels studied using our lifetime experiment. In general one must be careful when using continuum lamps to calibrate the FTS over wide spectral ranges, because the "ghost" of a continuum is a continuum. The Ar I and Ar II line technique, which is internal to the HCD Ce/Ar lamp spectra, is still our preferred calibration technique. It captures the wavelength-dependent response of detectors, spectrometer optics, lamp windows, and any other components in the light path or any reflections which contribute to the detected signal (such as due to light reflecting off the back of the hollow cathode). This calibration technique is based on a comparison of well-known branching ratios for sets of Ar I and Ar II lines widely separated in wavelength, to the intensities measured for the same lines. Sets of Ar I and Ar II lines have been established for this purpose in the range of 4300 to 35000 cm$^{-1}$ by Adams & Whaling (1981), Danzmann & Kock (1982), Hashiguchi & Hasikuni (1985), and Whaling et al. (1993). The Ce/Ne spectra from 2002 and the EDL spectra from 1985 could only be calibrated using W standard lamps. The older W lamp is a strip lamp calibrated as a spectral radiance (W/(m$^2$ sr nm)) standard, and the newer is a tungsten-quartz-halogen lamp calibrated as a spectral irradiance (W/(m$^2$ nm) at a specified distance) standard. Neither of these W filament lamps is hot or bright enough to yield a reliable UV calibration, but they are useful in the visible and near IR for interpolation and as a redundant calibration.

---

[1]Available at http://nsokp.nso.edu/

All possible transition wave numbers between known energy levels of Ce II satisfying both the parity change and $\Delta J$ = -1, 0, or 1 selection rules were computed and used during analysis of FTS data. Energy levels from Martin et al. (1978) were used to determine possible transition wave numbers. Levels from Martin et al. (1978) are available in electronic form from Martin et al. (2000)[2]. Systematic errors from missing branches to known lower levels are negligible in our work, because the level structure of Ce II is so well known and because we were able to make measurements on branching fractions of 0.01 or smaller. All naturally occurring Ce isotopes are even (nuclear spin I = 0) isotopes: $^{136}$Ce (abundance 0.185 %), $^{138}$Ce (abundance 0.251%), $^{140}$Ce (abundance 88.450 %) and $^{142}$Ce (abundance 11.114%) (Bˆhlke et al. 2005). Lines of Ce II are generally quite narrow because they have no hfs and small isotope shifts.

Branching fraction measurements were completed for lines from 43 even-parity and 15 odd-parity upper levels of the 74 levels studied in the lifetime experiment by Den Hartog & Lawler (2008). The levels for which branching fractions could not be completed had a strong branch beyond the UV limit of our spectra, or had a strong branch which was severely blended. Typically an even-parity upper level, depending on its J value, has about 55 possible transitions to known lower levels, and an odd-parity upper level has about 40 possible transitions to known lower levels. More than 30,000 possible spectral line observations were studied during the analysis of 14 different Ce/Ar and Ce/Ne spectra. We set integration limits and occasionally nonzero baselines "interactively" during analysis of the FTS spectra. An occasional nonzero baseline is needed when a weak line is located on a line wing of a much stronger line. Weak red and near-IR lines of Ce II in the EDL spectra also require a nonzero baseline due to some continuum emission from this lamp. The same numerical integration routine was used to determine the un-calibrated intensities of Ce II lines and selected Ar I and Ar II lines used to establish a relative radiometric calibration of the spectra. A simple numerical integration technique was used in this and most of our other RE studies because of weakly resolved or unresolved hyperfine and isotopic structure. More sophisticated profile fitting is used only when the line sub-component structure is either fully resolved in the FTS data or known from independent measurements.

---

[2]Available at http:// physics.nist.gov/PhysRefData/ASD/index.html

2.3 Branching Fraction Uncertainties

The procedure for determining branching fraction uncertainties was described in detail by Wickliffe et al. (2000). Branching fractions from a given upper level are defined to sum to unity, thus a dominant line from an upper level has small branching fraction uncertainty almost by definition. Branching fractions for weaker lines near the dominant line(s) tend to have uncertainties limited by their S/N. Systematic uncertainties in the radiometric calibration are typically the most serious source of uncertainty for widely separated lines from a common upper level. We used a formula for estimating this systematic uncertainty that was presented and tested extensively by Wickliffe et al. (2000). The EDL spectra enabled us to include measurements of quite weak branches in the visible and near-IR. Uncertainties on branching fractions of the weak visible and near-IR lines are larger than uncertainties on the dominant branches. In the final analysis, the branching fraction uncertainties are primarily systematic. Redundant measurements with independent radiometric calibrations help in the assessment of systematic uncertainties. Redundant measurements from spectra with different discharge conditions also make it easier to spot blended lines. Typically some fraction of a weak blended feature with a 1% or less branching fraction was included in the branching fraction normalization, but the blended line was omitted from the final table of transition probabilities.

2.4 Ce II Atomic Transition Probabilities

Branching fractions from the FTS spectra were combined with the radiative lifetime measurements (Den Hartog & Lawler 2008) to determine absolute transition probabilities for 921 lines of Ce II in Table 2. Air wavelengths in Table 2 were computed from energy levels (Martin et al. 1978) using the standard index of air (Edlén 1953). Parities are included in Table 2 using "ev" and "od" notation and decimal notation for J values which are compatible with our machine readable table of transition probabilities.

Transition probabilities for the very weakest lines (branching fractions ~ 0.001 or weaker) which were observed with poor S/N and for weak blended lines (branching fractions ≤ 0.01) are not included in Table 2, however these lines are included in the branching fraction normalization. The effect of the problem lines becomes apparent if one sums all transition probabilities in Table 2 from a chosen upper level, and compares the sum to the inverse of the upper level lifetime from Den Hartog & Lawler (2008). Typically the sum of the Table 2 transition probabilities is between 90% and 100 % of the inverse lifetime. Although there is significant fractional uncertainty in the branching fractions for these problem lines, this does not have much effect on the uncertainty of the stronger lines that were kept in Table 2. Branching fraction uncertainties are combined in quadrature with lifetime uncertainties to determine the transition probability uncertainties in Table 2.

2.5 Comparisons to Other Data Sets

Three large-scale theoretical or semi-empirical sets of Ce II transition probabilities are available for comparison to our new experimental results. Figure 2 shows comparisons of our new experimental transition probabilities to theoretical values from Fawcett (1990). A total of 686 lines in common are included in the Fawcett comparisons. Fawcett's results are based on ab-initio calculations of Slater parameters which were least-square adjusted to experimental energies. He also included configuration mixing. Figure 3 shows comparisons of our results to theoretical transition probabilities recently downloaded from the D.R.E.A.M. database (Biémont & Quinet 2005)[3]. Earlier work by some of the same team members (Palmeri et al. 2000, Zhang et al. 2001) is included in the more extensive online data set. A total of 890 lines in common are included in the Biémont comparisons. The D.R.E.A.M. results may not be purely theoretical due to the use of experimental LIF lifetimes to re-scale theoretical transition probabilities in some cases (e.g. Zhang et al. 2001). Figure 4 shows comparisons of our results to recently downloaded semi-empirical results from Kurucz (1998)[4]. A total of 529 lines in common are included in the Kurucz comparisons. It is interesting that the comparisons to semi-empirical results from Kurucz show the best agreement. It is not possible to draw conclusions about a particular theoretical method because the Kurucz database includes data from a number of sources.

---

[3] Available at http://w3.umh.ac.be/~astro/dream.shtml
[4] Available at http://kurucz.harvard.edu/

A more substantive question arises about the accuracy of the new experimental results. The agreement between our new experimental results and the various large scale theoretical and semi-empirical data sets is not impressive in any of the above comparisons. One might ask if our experimental error bars are reasonable. This question was recently addressed in a detailed comparison (Lawler et al. 2008a) between Sm II transition probabilities measured as part of this program (Lawler et al. 2006), and independent transition probability measurements from Rehse et al. (2006). Figure 5 shows comparisons of these two experimental data sets for Sm II that were published almost simultaneously. A total of 347 lines in common are included. Lines in common from 2 levels reported by Rehse et al. and identified as problematic (S. D. Rosner, Private Communication 2007) are omitted. The two experiments that produced these data sets on Sm II are quite different. A detailed discussion of the relative strengths, weaknesses, and possible systematic errors of both experiments is available (Lawler et al. 2008a). The agreement between these two experimental data sets is much better than the agreement of either with theoretical results.

3. SOLAR AND STELLAR CERIUM ABUNDANCES

We used the Ce II transition probabilities to determine new Ce abundances in the solar photosphere and five very metal-poor ([Fe/H] < -2) , "*r*-process" *n*-capture-rich giant stars. Our procedures followed those used in previous papers of this series; see Den Hartog et al. (2006), Lawler et al. (2006, 2007, 2008b), and references therein.

3.1 Line Selection

Cool stars exhibit a rich Ce II spectrum. For example, the solar photospheric line compendium of Moore, Minnaert, & Houtgast (1966) lists nearly 200 lines of this species. Therefore identification of transitions suitable for abundance analysis was a straightforward task. We began by computing relative line strength factors for all 921 Ce II lines of Table 2. As discussed in Lawler et al. (2008b, and references therein), the relative absorption strengths of weak-to-moderate lines within a given species are proportional to their transition probabilities modified by their Boltzmann excitation factors. For a line on the linear part of the curve-of-growth this relationship is:

$$\log(RW) = \log(EW/\lambda) = \text{constant} + \log(gf) - \theta\chi$$

where RW is the reduced width, EW is the equivalent width (mÅ), $gf$ is the oscillator strength, $\chi$ is the excitation energy, and $\theta = 5040/T$ is the inverse temperature. The relative strengths of lines of different species also depend on their elemental abundances and Saha ionization equilibrium factors. However Ce, like all REs, has a low first ionization potential: 5.539 eV (Grigoriev & Melikhov 1997). They all are completely ionized in the stellar photospheres of the Sun and stars of our sample; corrections for their other ionization state populations are negligible. Thus their relative strength factors STR can be written as:

$$STR \equiv \log(\varepsilon gf) - \theta\chi,$$

where $\varepsilon$ is the elemental abundance. In Figure 6 we plot these relative strength factors as a function of wavelength for Gd II lines (Den Hartog et al. 2006) and Ce II lines (this paper). For these computations we used solar abundances: $\log \varepsilon(Gd) = +1.11$ (Den Hartog et al. 2006) and $\log \varepsilon(Ce) = +1.55$, which is close to the recommended meteoritic Ce abundance of $\log \varepsilon(Ce) = +1.58$ (Lodders (2003). We also adopted $\theta = 1.0$ for this exercise, roughly the average value of the Sun and the metal-poor giants considered here (since most detectable Gd II and Ce II transitions in these stars have $\chi < 1$ eV, the exact value of $\theta\theta$ is not important in this exercise).

As in our previous studies we have drawn horizontal lines in Figure 6 to indicate approximate STR values for "strong" and "barely detectable" lines. These were defined by Lawler et al. (2006) for Sm II, who used the Delbouille, Roland, & Neven (1973) solar photospheric spectrum to estimate the EWs of the weakest lines that could routinely be employed in abundance analyses. The EW limit is $\approx 1.5$ mÅ in the spectral region $\lambda \sim 4500$ Å, or $\log(RW) \approx -6.5$. For such Ce II lines, STR $\approx -0.6$. That EW and thus STR limit applies also to Gd II and Ce II lines, as shown in both panels of Figure 6 with horizontal dotted lines. A minimum STR value for relatively strong lines was estimated by Lawler et al. (2006), defining it to be a factor of 20 larger than the line detection limit value, or STR $= -0.6 + 1.3 = +0.7$. We have indicated this "strong-line" limit in Figure 6 with dashed horizontal lines.

Figure 6 displays STR values only in the wavelength range $3000$ Å $\leq \lambda \leq 7000$ Å even though our transition probability data extend to nearly 11000 Å (1.1 $\mu$). Nearly all Gd II and Ce II lines beyond 7000 Å are undetectably weak in the stars of interest here, having solar relative strengths $\log(STR) < -1.0$. Clearly the "strong" lines of Gd II, Ce II, and nearly all rare earth ions occur in the blue and near-UV spectral regions ($\lambda < 4500$ Å), which we emphasize with vertical lines drawn at 4000 Å in the figure. Fortunately, about 70 strong Ce II lines (STR $> +0.7$) lie longward of 4000 Å, and thus are easily accessible to ground-based spectroscopy in a (relatively) un-congested spectral domain. No strong Gd II lines are so fortuitously located.

The Ce II STR values were used to trim the original set of 921 transitions (Table 2) to about 620, by discarding those lines weaker than STR $< -0.6$. We then searched through this list to identify potentially useful Ce II lines for abundance analyses. We inspected these lines in solar photospheric center-of-disk spectrum of Delbouille et al. (1973) and in our spectrum of the *r*-process-rich metal-poor giant star CS 31082-001 ([Fe/H] = -2.9, [Eu/Fe] = +1.7, Hill et al. 2002). These spectra, aided by the Moore et al. (1966) solar line identifications and the comprehensive Kurucz (1998) atomic and molecular line lists, showed that the vast majority of candidate Ce II transitions are too weak and/or too blended to be of use in the Sun and stars of interest here. This process produced a final set of about 60 lines for more detailed study.

3.2 The Solar Photospheric Cerium Abundance

We computed synthetic spectra for the Ce II lines as in previous papers of this series. Briefly, we started with atomic and molecular line lists in small (4-6 Å) wavelength regions, from Kurucz's (1998) line database, supplemented in a few cases by Moore et al.'s (1966) solar identifications. We adopted *gf*-values from laboratory studies for these ionized n-capture element species: Y, Hannaford et al. (1982); Zr, Malcheva et al. (2006); La, Lawler et al. (2001a); Ce, this study; Nd, Den Hartog et al. (2003); Sm, Lawler et al. (2006); Eu, Lawler et al. (2001b); Gd, Den Hartog et al. (2006); Tb, Lawler et al. (2001c); Dy, Wickliffe et al. (2000); Ho, Lawler et al. (2004); Er, Lawler et al. (2008b); and Hf, Lawler et al. (2007). The Ce II transitions were treated as single features, since their hyperfine and isotopic substructures are very small.

We adopted the Holweger & Müller (1974) empirical model photosphere for most of the solar computations. Standard solar abundances were taken from recent reviews (e.g. Grevesse & Sauval 2002; Lodders 2003, Grevesse, Asplund, & Sauval 2007), modified for some elements to include recent updates for the *n*-capture elements (given in the papers cited above). We input the solar model, abundance set, and line lists into the current version of the LTE line analysis code MOOG (Sneden 1973) to generate synthetic spectra. These were matched to the observed solar spectrum (Delbouille et al. 1973) after application of empirical Gaussian broadening functions to account for solar macroturbulence and spectrograph instrumental profile.

We computed multiple trial synthetic spectra for each line region. The oscillator strengths for atomic lines other than the *n*-capture species listed above were adjusted to fit the observed solar spectrum. The C, N, and O abundances were varied to match observed CH, CN, NH, and OH line strengths. For absorption features without plausible identifications, we arbitrarily attributed them to Fe I with excitation potentials $\chi = 3.5$ eV and *gf*-values adjusted to fit the solar spectrum. In cases where line contamination was a significant part of the overall absorption at the Ce II wavelength, the line was discarded for the solar analysis but kept for possible use with the metal-poor giants. After iterating the line list data, final syntheses of 45 Ce II lines were retained for the solar analysis. Individual line abundances are listed in Table 3, along with their excitation energies and *gf*-values. A straight mean abundance is $\log \varepsilon(Ce) = 1.61 \pm 0.01$ ($\sigma = 0.06$).

For internal (line-to-line) errors, we estimate line profile fitting uncertainties to be ±0.02 dex, contamination by other (non-Ce) features to be ±0.02 dex, and log(*gf*) uncertainties for the Ce II transitions (see Table 2) are ±0.03. Adding these uncertainties in quadrature yields an estimated total internal uncertainty per transition of ±0.05 dex, which is close to the observed $\sigma = 0.06$. With solar abundance information from 45 Ce II lines, the mean standard deviation of 0.01 suggests that internal uncertainties are very small contributors to the overall error budget. Scale (external) errors can be due to atomic data uncertainties beyond *gf* errors, model atmosphere choices, and analysis technique. Since Ce exists almost exclusively as Ce II in these stars, only the Ce II partition function (which enters directly into the Boltzmann equations for our transitions) could contribute to abundance scale uncertainties. Our spectral synthesis code uses Irwin's (1981) partition function polynomial fits to the atomic energy level data available at that time. We re-calculated Ce II partition functions with current energy level information (Martin et al. 2000), but found negligible changes from the Irwin computations in the temperature domain of interest for this study. Finally, in low metallicity stars, Rayleigh scattering becomes an important continuous opacity source in the blue-uv spectral regions, and radiative transfer source functions can depart somewhat from the Planck function assumed in our calculations. In Sneden et al. (2009) we discuss experiments with more proper accounting for continuum scattering, showing that it produces only small effects for the stars of our sample, and affects most RE elements in a nearly equal fashion, leaving abundance ratios virtually unchanged.

As in Lawler et al. (2007), we repeated some of the abundance computations using the Kurucz (1998) and Grevesse & Sauval (1999) models, finding average abundance shifts of -0.02 dex compared to those done with the Holweger & Müller (1974) model. Finally, we emphasize that our analysis is a standard LTE one with a pure Planck source function. Departures from LTE in the ionization equilibrium cannot affect our Ce II analysis because Ce is already completely ionized. The existence of many low excitation energy levels with this ion increases the likelihood of colllisional dominance in their populations. Comments are included in the companion paper (Sneden et al. 2009) on the effect of inclusion of scattering in the continuum source function, but the effect is extremely small for the Sun (and weak for the r-process rich stars). Combining line-to-line scatter uncertainties (±0.01 from the standard deviation of the mean, Table 3) with scale uncertainties, we recommend log ε(Ce)$_{Sun}$ = +1.61 ± 0.03. This value is in excellent agreement with the recommended meteoritic abundance of Lodders (2003): log ε(Ce)$_{met}$ = +1.61 ± 0.02.

Palmeri et al. (2000) published the most recent Ce abundance study of the solar photosphere. Analyzing the EWs of 26 lines, they derived log ε(Ce)$_{Sun}$ = +1.63 ± 0.04 from one set of transition probability calculations, and +1.70 ± 0.04 from a different set. We repeated these abundance computations with their EWs, the final recommended log(*gf*) values from their web site, and our line analysis code MOOG (Sneden 1973), obtaining log ε(Ce)$_{Sun}$ = +1.67 ± 0.04 (σ = 0.14). Finally, we substituted their log(*gf*) values for the ones of the present study for our preferred transitions (Table 3), finding log ε(Ce)$_{Sun}$ = +1.59 ± 0.02 (σ = 0.10). All of these computations are consistent with the overall scale agreement between the Palmeri et al. transition probabilities and those of the present study that was discussed in Section 2.5.

3.3 Cerium Abundances in Five *r*-Process-Rich Low Metallicity Stars

The spectra of several very metal-poor, *r*-process-rich giant stars were also analyzed to determine their Ce abundances: BD+17º3248 ([Fe/H] = -2.1, [Eu/Fe] = +0.9, Cowan et al. 2002); CS 22892-052 ([Fe/H] = -3.1, [Eu/Fe] = +1.5, Sneden et al. 2003a); CS 31082-001 ([Fe/H] = -2.9, [Eu/Fe] = +1.7, Hill et al. 2002); HD 115444 ([Fe/H] = -2.9, [Eu/Fe] = +0.8, Westin et al. 2000); and HD 221170 ([Fe/H] = -2.2, [Eu/Fe] = +0.8, Ivans et al. 2006). We employed the same analytical methods for these stars as was done for the solar photosphere. The abundances from individual lines are listed in Table 3 and displayed in Figure 7. The average abundances, mean and sample standard deviations, and number of lines are also recorded at the bottom of Table 3 and in Figure 7. The line-to-line scatters are all small, σ = 0.03 – 0.07, with no obvious trends of abundances as functions of wavelength, log(*gf*), or excitation potential. The line-to-line scatters for individual stars approximately anti-correlate with their Ce II line strengths. We derived the smallest scatter (σ = 0.03) for CS 31082-001, which has the strongest lines, and the largest (σ = 0.07) for HD 115444, with mostly very weak lines.

4. DISCUSSION AND CONCLUSIONS

We compare the newly derived Ce abundances among the five *r*-rich stars in Figure 8. The differences between the observed Ce abundances for each star - BD+17º3248, CS 22892-052, CS 31082-001, HD 115444, and HD 221170 – and the solar system *r*-process only value (Arlandini et al. 1999) are illustrated in the Figure. In all cases the abundances have been normalized with respect to the Eu abundances. (These abundances have also been newly redetermined for those five stars and are reported in the companion paper Sneden et al. 2009.) The corresponding error bars are those given for log ε(Ce) in Table 3. As is clear from the figure, the Ce abundances in all five *r*-rich stars are in general agreement with a solar system *r*-only value and there is little scatter among the stars – with the largest deviations from the solar *r*-process value (indicated by a dashed line) of approximately 0.1 dex. (A value of zero would thus indicate a perfect agreement with that prediction of the solar system *r*-only values.) The two stars deviating the most from the solar system r-process values are HD 115444 and BD 17º3248. The increased precision of our new Ce abundance determinations suggests that the error limits are typically less than 0.1 dex. Thus, it is possible that there might be some small *s*-process contamination to the Ce abundances in those two stars. (Ce is a predominantly s-process element, 81%, in solar system material.) If, however, this slight *s*-process contribution is due to Galactic *s*-process nucleosynthesis, we would expect to see an increasing trend in Ce abundance as a function of metallicity. Previous studies (see Burris et al. 2000 and Simmerer et al. 2004) found evidence of Galactic *s*-process nucleosynthesis already at metallicities less than -2.0.

We see in Figure 8 that (somewhat surprisingly) there is no metallicity effect on the agreement between the Ce stellar and solar system r-process only ratios. There is an almost perfect agreement with the *r*-process only value for Ce in CS 22892-052 at [Fe/H] = -3.1 and in HD 221170 at [Fe/H] = -2.1. Since Ce is produced predominantly in the *s*-process in solar system material (see e.g., Simmerer et al. 2004), it might be expected that there would be some rise in the Ce abundance at metallicities near -2. While BD+17º3248 does show a rise at that point, there is no clear overall trend, at least for this sample of only five *r*-rich stars. Thus, there may be something specific to these two stars that would indicate Ce abundances slightly larger than, but still roughly in agreement with, the predicted *r*-process only ratios. Studies of possible small *s*-process contributions to other *s*-process dominant RE elements, such as Ba, in these five stars studied here is included in Sneden et al. (2009).

This consistency between stellar and solar-system *r*-process-only values for rare earth elements is not surprising and has been reported previously for each of these stars: BD+17º3248 (Cowan et al. 2002), CS 22892-052 (Sneden et al. 2004), CS 31082-001 (Hill et al. 2002), HD 115444 (Westin et al. 2000) and HD 221170 (Ivans et al. 2006). However, in the past there has been considerably more star-to-star scatter in individual elements, including Ce (see figure 10 in Lawler et al. 2008b). This scatter has now been dramatically reduced for Ce with much smaller error bars than previously possible, as a result of the new atomic physics data presented in this paper. Further discussion and analyses of other rare earth elements, based upon new atomic data, can be found in the companion paper Sneden et al. (2009).

ACKNOWLEDGEMENTS

This work has been supported by the National Science Foundations through grants AST-0506324 to JEL & EDH, AST- 0607708 to CS, and AST-0707447 to JJC.

APPENDIX

This work on Ce II, as indicated in the Introduction, completes a multi-year effort to improve laboratory spectroscopic data for Rare Earth (RE) ions and to apply these data in abundance studies. This appendix is a summary of lab data from the project including transition probabilities, isotopic shifts, hyperfine structure (hfs) constants, and complete isotopic and hfs line component patterns. References are included to published data tables as well as some new machine-readable (MR) tables. A few of our early papers did not include MR tables of transition probability data, and we have since had numerous requests for such MR tables. In other cases we published only isotope shifts and/or hfs constants without realizing that many astronomers would prefer complete isotopic and/or hyperfine line component patterns. We are here providing such tables.

We also wish to correct a typographical error in a hfs energy formula which appeared in several of our papers (Lawler et al. 2001b, Lawler et al. 2001d, Lawler et al. 2004). The typographical error did not affect any of our tables of hfs constants or tables of complete line component patterns. We are using the Casimir formula as presented in the elementary text by Woodgate (1980). This correct formula is

$$\Delta E = \frac{AK}{2} + B\frac{3K(K+1) - 4I(I+1)J(J+1)}{8I(2I-1)J(2J-1)},$$

where $\Delta E$ is the shift in wave numbers of a hfs sub-level (F, J) from the center of gravity of the fine structure level (J),

$$K = F(F+1) - J(J+1) - I(I+1),$$

$F$ is the total atomic angular momentum, $J$ is the total electronic angular momentum, and $I$ is the nuclear spin. The more advanced text by Cowan (1981) has an excellent discussion of hfs, but uses a slightly different definition of the hyperfine B. Relative intensities of hyperfine components are expressed in terms of Wigner 6-j symbols by Cowan (1981).

Some readers may find it more convenient to use the simple Russell-Saunders line strength formulae given in the classic text by Condon & Shortley (1935). The use of the Russell-Saunders formulae only requires substitution of F for J, J for L, and I for S.

Each RE element has its own nuclear configuration, and each first ion of these elements has a unique electronic energy structure. However, some general properties that connect many of the Z = 57-71 ions are worth noting. Among the seven stable odd-Z elements of this group, five of them have only one naturally-occurring isotope. Lu (Z = 71) has stable isotopes $^{175}$Lu and $^{176}$Lu, but $^{175}$Lu comprises 97.4% of the solar system Lu abundance. This essentially leaves only Eu as the odd-Z RE element with multiple abundant isotopes: $^{151}$Eu (47.8% of the Eu elemental abundance in the solar system), and $^{153}$Eu (52.2%). Additionally, the Eu isotopic fraction appears to be relatively insensitive to the *n*-capture synthesis method, being nearly 50/50 in both the *r*-process and *s*-process (e.g. Sneden et al. 2002, Aoki et al. 2003). Therefore once the hyperfine patterns for odd-Z rare-earth ionized transitions are known, the spectra can be synthesized with confidence.

In contrast, all but one of the even-Z elements in this element domain have at least five naturally-occurring non-negligible isotopes ( > 1% fraction in the solar system). The single exception is Ce, with just $^{140}$Ce (88.5%) and $^{142}$Ce (11.1%). The odd-A isotopes of these elements have non-zero nuclear magnetic moments, and thus their ionized-species transitions have usually small hfs. However, the isotopic wavelength shifts are usually undetectably small. Magain (1995) and Lambert & Allende Prieto (2002) studied the isotopic mix of Ba, a near RE element, from the Ba II 4554Å line in one very metal-poor star, but their results had large error bars because the isotopic wavelength shifts are only ~0.05Å. Roederer et al. (2008) derived Sm isotopic abundance fractions from several lines of Sm II in metal-poor stars, but the isotopic splits were always <~0.08Å. In both of these examples the isotopic wavelength spreads are comparable to the stellar thermal and microturbulent broadening of the lines. No other RE even-Z element appears to exhibit significant isotopic broadening in spectroscopically accessible transitions (see attempts to analyze Nd II by Roederer et al.), so for most purposes their lines can be treated as single features.

Table 4 is a summary of laboratory data measured or compiled as part of this program on RE ions with locations of MR tables of **log**$(gf)$ values and of Complete Line Component Patterns (**CLCP**). The general discussion above is supplemented with more detailed discussions of the spectra of each RE ion below. As the quality of stellar spectra from large and very large ( > 10 m) telescopes continues to improve, there will be increased opportunities for elemental abundance studies and isotopic fraction measurements on many more stars. Complete and very accurate tables of solar system isotopic fractions and Nuclear Moments are available, but are not included here (Bˆhlke et al. 2005, Stone 2005).

Lanthanum is an on odd Z element with essentially one stable isotope, $^{139}$La, with hfs from a nuclear spin I = 7/2 which is significant in stellar abundance work. The very low isotopic fraction of $^{138}$La of 0.09% in solar system material is negligible. When our work on La II was published (Lawler et al. 2001a) we were not aware of the ApJ option to include MR tables. Table 5 is a MR version of the La II transition probability table from the 2001 paper. A set of hfs constants for $^{139}$La was measured and/or compiled by Lawler et al. (2001a) and updated by Ivans et al. (2006) with a MR table of CLCP.

Cerium is covered in this paper. It is an even Z element with only nuclear spin I = 0 isotopes. Two isotopes have solar system fractions above 10%, but the isotopes shifts are generally small. There are also two rare I = 0 isotopes with solar system fractions < 1%.

Lines of Ce II are sufficiently narrow in our FTS data that they may be treated as single component lines in stellar abundance work.

Praseodymium is discussed in the companion paper (Sneden et al. 2009). It is an odd Z element with a single stable isotope, $^{141}$Pr, with hfs from a nuclear spin I = 5/2 which is significant in stellar abundance work. The companion paper has a recommended set of log(gf) values and a MR table of CLCP.

Neodymium is an even Z element with five I = 0 isotopes and two I = 7/2 isotopes, all with fractions of 5% or more in solar system material. Some lines of Nd II have structure in our FTS data as noted in Table 3 of Den Hartog et al. (2006) but these lines tended to be weak in stellar spectra. Most lines of Nd II are sufficiently narrow in our FTS data to be treated as single component lines in stellar abundance work. Roederer et al. (2008) compiled Nd II isotope shifts and hfs constants, and published a MR tables of CLCP. Isotopic fraction measurements on Nd in stellar abundance work have been attempted, but with only limited success because the isotope shifts are so small. Since there is no resolved structure in Nd II lines in stellar spectra, tiny center-of-gravity shifts of Nd II lines must be measured in stellar spectra to determine isotopic fractions.

Prometheum has no stable isotopes. Although there have been tentative observations of Pm II lines in stellar spectra (e.g. Cowley et al. 2004), none have been confirmed.

Samarium is an even Z element with five I = 0 isotopes and two I = 7/2 isotopes, all with fractions of 3% or more in solar system material. Most lines of Sm II are sufficiently narrow in our FTS data to be treated as single component lines in stellar abundance work. Roederer et al. (2008) compiled Sm II isotope shifts and hfs constants, and published a MR table of CLCP. Isotopic fraction measurements on Sm in stellar abundance work are somewhat easier than on Nd, and such measurements have been performed (Roederer et al. 2008). Small center-of-gravity shifts of Sm II lines were measured in stellar spectra to determine isotopic fractions.

Europium is an odd Z element with two I = 5/2 isotopes, both with large, 47.8% and 52.2%, isotopic fractions in solar system material. When our work on Eu II was published (Lawler et al. 2001b) we did not include MR tables. Table 6 is a MR version of the Eu II transition probability table from the 2001 paper. A set of hfs constants for $^{151}$Eu and $^{153}$Eu was compiled by Lawler et al. (2001b). The energy levels and center-of gravity wavelengths of Eu II transition were improved by Ivans et al. (2006) and CLCP were provided in MR Table 6 of that paper. Astrophysical data on halo stars from modern large telescopes have now reached the level of quality that isotopic abundances of Eu can be determined with some accuracy and precision (e.g., Sneden et al. 2002, Aoki et al. 2003). Evidence to date supports a uniform isotopic mix from all *r*-process events. The lines of Eu II connected to ground and lowest metastable levels are ideal for such studies. The unpaired s-electron of these two levels yields both large isotope shifts from the finite nuclear size (field shifts) and wide hfs. Isotope shifts and hfs yield resolvable structure in Eu II lines in stellar spectra, which is much easier to measure than the small wavelength shifts of Sm II or Nd II lines due to varying isotopic fractions.

Gadolinium is an even Z element with five I = 0 isotopes and two I = 3/2 isotopes. The two lightest I = 0 isotopes have rather low fractions < 2.5% in solar system material. Most lines of Gd II are sufficiently narrow in our FTS data to be treated as single component lines in stellar abundance work. The widths of line profiles in our highest resolution FTS data vary, and in a few cases the profiles have partially resolved structure.

Although it is not possible today, it may at some point in the future be possible to observe the isotopic mixture of Gd in a metal-poor halo star.

Terbium is an on odd Z element with one stable isotope, $^{159}$Tb, with hfs from a nuclear spin I = 3/2 which is significant in stellar abundance work. Lawler et al. (2001d) published a large set of hfs constants. Complete line component patterns for lines of Tb II are in MR Table 7 below.

Dysprosium is an even Z element with five I = 0 isotopes and two I = 5/2 isotopes. The three lightest I = 0 isotopes have rather low fractions < 2.5% in solar system material. Although some lines have detectable structure, most lines of Dy II are sufficiently narrow in our FTS data to be treated as single component lines in stellar abundance work. Wickliffe et al. (2000) measured a large set of Dy I and Dy II transition probabilities in this effort on RE species. This work on Dy was done jointly with NIST and comparisons of independent branching fraction measurements were performed. Both the Univ. of Wisconsin (UW) and NIST transition probabilities are based on the same radiative lifetimes measured at UW using time-resolved LIF. Wickliffe et al. published a merged table of Dy I and Dy II transition probabilities with both UW and NIST measurements. This table has been reconstructed here as MR Table 8.

Holmium is an odd Z element with a single stable isotope, $^{165}$Ho, and hfs from a nuclear spin I = 7/2 which is significant in stellar abundance work. The high line density of Ho I and Ho II in combination with wide hfs has inhibited analysis of Ho II. Lawler et al. (2004) focused their efforts on the low-lying levels which are connected by strong "resonance like" transitions. They reported new transition probability measurements based on radiative lifetimes from LIF in combination with branching fractions from FTS data, a small set of improved energy levels from FTS data for better center-of-gravity wavelengths, and hfs constants from FTS data for the levels of interest. Their table of transition probabilities is reproduced as MR Table 9 below, and MR Table 10 includes CLCP for the strong "resonance like" lines of Ho II.

Erbium is an even Z element with five I = 0 isotopes and one I = 7/2 isotope. The two lightest I = 0 isotopes have rather low fractions < 2% in solar system material. Although some lines have detectable structure, most lines of Er II are sufficiently narrow in our FTS data to be treated as single component lines in stellar abundance work. Some of the IR lines studied by Lawler et al. (2008b) have detectable isotopic structures, but these lines are so weak with such high excitation potentials that there is no hope of astrophysical detections in the foreseeable future.

Thulium is an odd Z element with one stable isotope, $^{169}$Tm. This isotope has a nuclear spin of I = 1/2. Although some lines have detectable structure, most lines of Tm II are sufficiently narrow in our FTS data to be treated as single component lines in stellar abundance work. Wickliffe and Lawler (1997) measured a large set of Tm I and Tm II transition probabilities in this effort on RE species. Their data on Tm II is included below as MR Table 11.

Ytterbium is discussed in the companion paper (Sneden et al. 2009). It is an even Z element with five I = 0 isotopes, one I = 1/2 isotope, and one I = 5/2 isotope. The lightest two I = 0 isotopes have rather low fractions, < 3.1%, in solar system material. Only two Yb II resonance lines from the same multiplet are useful in abundance studies and thus a MR table of transition probabilities is unnecessary. The unpaired s-electron of the ground level yields large isotope shifts and hfs. Isotopic and hfs are not negligible in stellar abundance work on Yb II. The companion paper has recommended log(gf) values and a MR table of CLCP.

Lutetium is an odd Z element with one dominant stable isotope, $^{175}$Lu, with hfs from a nuclear spin I = 7/2 which is significant in stellar abundance work. The low isotopic fraction of $^{176}$Lu of 2.59% in solar system material is negligible given the generally low abundance of Lu and quality of stellar spectra available today. During a search for Lu in metal-poor stars, it became apparent that the NIST energy levels (Martin et al. 1978) are not as accurate as the NIST energy levels for many other species. Energy levels from older measurements using photographic techniques on species with wide hfs often have this problem. Table 12 presents new energy level measurements for some of the low-lying levels of Lu II of both parities. Lutetium was studied early in this project on RE ions, shortly after Bord, Cowley, & Mirijanian (1998) found a clean Lu II line in the solar spectrum at 6221 Å. Den Hartog et al. (1998) measured the upper level lifetime of this line using LIF and the branching fractions of lines from this level using FTS data to determine accurate transition probabilities. Shortly thereafter additional branching fractions from FTS data were reported by Quinet et al. (1999) and additional radiative lifetimes from time-resolved LIF were reported by Fedchak et al. (2000). In this early phase of the RE project we were collaborating more closely with theorists to determine how well ab-initio quantum mechanical methods could be used to calculate radiative lifetimes and branching fractions. With only two valence electrons, Lu II is particularly simple RE spectrum with minimal configuration mixing. However, both core polarization and relativistic effects are important in Lu II like other RE spectra. At this time we are of opinion that careful branching fraction measurements from FTS data in combination with radiative lifetimes from LIF generally yield better RE transition probabilities than can be computed using ab-initio quantum methods. We therefore recommend the Lu II transition probabilities of Table 13 that were determined by combining branching fractions from FTS data with radiative lifetimes from our time-resolved LIF experiment. Sneden et al. (2003a) measured a set of hfs constants from FTS data for levels of $^{175}$Lu. For the user's convenience MR Table 14 includes CLCP for lines of $^{175}$Lu II. The reader should note that the components from the rare (2.59%) isotope $^{176}$Lu are not included in Table 14. New energy levels from Table 12 are included in Table 13 and 14 if available for both the upper and lower levels of the transition, otherwise energy levels from Martin et al. (1978) are used. Air wavelengths are computed from those energy levels using the standard index of air (Edlén 1953).

FIGURE CAPTIONS

Figure 1. Partial Grotrian diagram for Ce II. Upper and lower levels of both parities included in this study are shown.

Figure 2. Comparison of theoretical Ce II transition probabilities from Fawcett (1990) to our transition probabilities as function of our transition probability or log(*gf*), wavelength, and upper level energy.

Figure 3. Comparison of Ce II transition probabilities recently downloaded from the D.R.E.A.M. database (Biémont & Quinet 2005) to our transition probabilities as function of our transition probability or log(*gf*), wavelength, and upper level energy.

Figure 4. Comparison of semi-empirical Ce II transition probabilities recently downloaded from the Kurucz. database (Kurucz 1998) to our transition probabilities as function of our transition probability or log(*gf*), wavelength, and upper level energy.

Figure 5. Comparison of experimental Sm II transition probabilities from Rehse et al. (2006) to our transition probabilities as function of our transition probability or log(*gf*), wavelength, and upper level energy. This plot illustrates the level of agreement now achieved between modern, but rather different, experimental techniques.

Figure 6. Relative transition strength factors, $STR \equiv \log(\varepsilon gf) - \theta\chi$, for Gd II (Den Hartog et al. 2006) and Ce II (this study). Definitions of "detection limit" and "strong lines" are given in the text. For display purposes the long-wavelength limit has been set to 7000 Å. Lines longward of this limit are too weak to be detected in any of the stars of this study. The short-wavelength limit of 3000 Å is defined by the Earth atmospheric transmission cutoff. All Ce II lines of this study are longward of this limit.

Figure 7. Abundances of individual Ce II lines for all program stars, plotted as a function of wavelength. For each star, a dotted line is drawn at the mean abundance. The mean abundances, sample standard deviation σ, and the number of lines used in the analyses are given in the legends of each figure panel.

Figure 8. Comparison of the newly determined Ce abundances in five *r*-process rich stars to the solar system *r*-process only value (Arlandini et al. 1999). For each star the abundances have been normalized at Eu. The dashed line indicates a perfect agreement between the stellar and solar system *r*-only values for Ce. The error bars are the sigma values listed for each star in Table 3.

REFERENCES

Adams, D. L., & Whaling, W. 1981, J. Opt. Soc. Am., 71, 1036

Aoki, W., Honda, S., Beers, T. C., & Sneden C. 2003, ApJ, 586, 506

Arlandini, C., K‰ppeler, F., Wisshak, K., Gallino, R., Lugaro, M., Busso, M., & Straniero, O. 1999, ApJ, 525, 886

Biémont, E. & Quinet, P. 2005, J. Electron Spectrosc. and Related Phenomena, 144, 23

Bˆhlke, J. K., de Laeter, J. R., De BiËvre, P., Hidaka, H., Peiser, H. S., Rosman, K. J. R., & Taylor, P. D. P. 2005, J. Phys. Chem. Ref. Data., 34, 57

Bord, D. J., Cowley, C. R., & Mirijanian, D 1998, Solar Physics 178, 221

Brault, J. W. 1976, J. Opt. Soc. Am., 66, 1081

Burris, D. L., Pilachowski, C.A., Armandroff, T. E., Sneden, C., Cowan, J. J., & Roe, H. 2000, ApJ, 544, 302

Condon, E. U. & Shortley, G. H. 1935, The Theory of Atomic Spectra (Cambridge University Press, Cambridge) p. 238

Cowan, R. D. 1981, The Theory of Atomic Structure and Spectra (Univ. of California Press, Berkeley) p. 508

Cowan, J. J., Sneden, C., Truran, J. W., & Burris, D. L. 1996, ApJ, 460, L115

Cowan, J. J., Sneden, C., Burles, S., Ivans, I. I., Beers, T. C., Truran, J. W., Lawler, J. E., Primas, F., Fuller, G. M., Pfeiffer, B., & Kratz, K.-L. 2002, ApJ, 572, 861

Cowley, C. R., Bidelman, W. P., Hubrig, S., Mathys, G., & Bord, D. J. 2004, A&A 419, 1087

Danzmann, K. & Kock M. 1982, J. Opt. Soc. Am., 72, 1556

Delbouille, L, Roland, G., & Neven, L. 1973, Photometric Atlas of the Solar Spectrum from lambda 3000 to lambda 10000, (LiËge, Inst. d'Ap., Univ. de LiËge)

Den Hartog, E. A., Curry, J. J., Wickliffe, M. E., & Lawler J. E. 1998, Solar Physics 178, 239

Den Hartog, E. A. & Lawler, J. E. 2008, J. Phys. B: At. Mol. Opt. Phys., 41, 045701

Den Hartog, E. A., Lawler, J. E., Sneden, C., & Cowan, J. J. 2003, ApJS, 148, 543

Den Hartog, E. A., Lawler, J. E., Sneden, C., & Cowan, J. J. 2006, ApJS, 167, 292

Edlén, B. 1953, J. Opt. Soc. Am., 43, 339

Fawcett, B. C. 1990, Atomic and Nuclear Data Tables, 46, 217

Fedchak, J. A., Den Hartog, E. A., Lawler, J. E., Palmeri, P., Quinet, P., & Biémont, E. 2000, ApJ, 542, 1109

Grevesse, N., & Sauval, A. J. 1999, A&A, 347, 348

Grevesse, N., & Sauval, A. J. 2002, Adv. Space. Res., 30, 3

Grevesse, N., Asplund, M., & Sauval, A. J. 2007, Space. Sci. Rev., 130, 105

Grigoriev, I. S., & Melikhov, E. Z. 1997, Handbook of Physical Quantities, (Boca Raton, CRC Press) p. 516

Hannaford, P., Lowe, R. M., Grevesse, N., Biémont, E., & Whaling, W. 1982, ApJ, 261, 736

Hashiguchi, S. & Hasikuni, M. 1985, J. Phys. Soc. Japan 54, 1290

Hill, V., et al. 2002, A&A, 387, 560

Holweger, H., & Müller, E. A. 1974, Sol. Phys., 39, 19

Irwin, A. W. 1981, ApJS, 45, 621

Ivans, I. I., Simmerer, J., Sneden, C., Lawler, J. E., Cowan, J. J., Gallino, R., & Bisterzo, S. 2006, ApJ, 645, 613

Kurucz, R. L. 1998, in Fundamental Stellar Properties: The Interaction between Observation and Theory, IAU Symp. 189, ed T. R. Bedding, A. J. Booth & J. Davis (Dordrecht: Kluwer), p. 217

Lambert, D. L., & Allende Prieto C. 2002, MNRAS, 335, 325

Langhans G., Schade, W., & Helbig, V. 1995, Z. Phys. D, 34, 155

Lawler, J. E., Bonvallet, G., & Sneden, C. 2001a, ApJ, 556, 452

Lawler, J. E., Den Hartog, E. A., Labby, Z. E., Sneden C., Cowan J. J., & Ivans I. I. 2007, ApJS, 169, 120

Lawler, J. E., Den Hartog, E. A., Sneden, C., & Cowan, J. J. 2006, ApJS, 162, 227

Lawler, J. E., Sneden, C., & Cowan, J. J. 2004, ApJ, 604, 850

Lawler, J. E., Den Hartog, E. A., Sneden, C., & Cowan, J. J. 2008a, Can. J. Phys., 86, 1033

Lawler, J. E., Sneden, C., Cowan, J. J., Wyart, J.-F. , Ivans, I. I., Sobeck, J. S., Stockett, M. H., & Den Hartog, E. A. 2008b, ApJS, 178, 71

Lawler, J. E., Wickliffe, M. E., Cowley, C. R., & Sneden, C. 2001c, ApJS, 137, 341

Lawler, J. E., Wickliffe, M. E., Den Hartog, E. A., & Sneden, C. 2001b, ApJ, 563, 1075

Lawler, J. E., Wyart, J.-F., & Blaise, J. 2001d, ApJS, 137, 351

Li, Z. S., Lundberg, H., Wahlgren, G. M., Sikstrˆm, C. M. 2000, Phys. Rev A, 62, 032505

Lodders, K. 2003, ApJ, 591, 1220

Malcheva, G., Blagoev, K., Mayo, R., Ortiz, M., Xu, H. L., Svanberg, S., Quinet, P., & Biémont, E. 2006, MNRAS, 367, 754

Magain P. 1995, A&Ap, 297, 686

Martin, W. C., Sugar, J., & Musgrove, A. 2000, NIST Atomic Spectra Database, (http://physics.nist.gov/PhysRefData/ASD/index.html)

Martin, W.C., Zalubas, R., & Hagan, L. 1978, Atomic Energy Levels The Rare Earth Elements, NSRDS NBS 60 (Washington: U. S. G. P. O.) p. 70

Moore, C. E., Minnaert, M. G. J., & Houtgast, J. 1966, The Solar Spectrum 2934 Å to 8770 Å, NBS Monograph 61 (Washington: U.S. G. P. O.)

Palmeri, P., Quinet, P., Wyart, J.-F., & Biémont, E. 2000, Physica Scripta, 61, 323

Quinet, P. Palmeri, P., Biémont, E., McCurdy, M. M., Rieger, G., Pinnington, E. H., Wickliffe, M. E., & Lawler J. E. 1999, MNRAS 307, 934

Rehse, S. J., Li, R., Scholl, T. J., Sharikova, A., Chatelain, R., Holt, R. A., & Rosner, S. D. 2006, Can. J. Phys., 84 723

Roederer, I. U., Lawler, J. E., Sneden, C., Cowan, J. J., Sobeck, J. S., & Pilachowski, C. A. 2008, ApJ, 675, 723

Simmerer, J., Sneden, C., Cowan, J. J., Collier, J., Woolf, V. M., & Lawler, J. E. 2004, ApJ, 617, 1091

Smith, V. V., Cunha, K., & Lambert, D. L. 1995, AJ, 110, 2827

Sneden, C. 1973, ApJ, 184, 839

Sneden, C., Basri, G., Boesgarrd, A. M., Brown, J. A., Carney, B. W., Kraft, R. P., Smith, V., & Suntzeff, N. B. 1995, Publ. of the Astron. Soc. of the Pacific 107, 997

Sneden, C., Cowan, J. J., Lawler, J. E., Burles, S., Beers, T. C., & Fuller, G. M. 2002, ApJ, 566, L25

Sneden, C., McWilliam, A., Preston, G. W., Cowan, J. J., Burris, D. L., & Armosky, B. J. 1996, ApJ, 467, 819

Sneden, C., Cowan, J. J., Lawler, J. E., Ivans, I. I., Burles, S., Beers, T. C., Primas, F., Hill, V., Truran, J. W., Fuller, G. M., Pfeiffer, B., & Kratz, K.-L. 2003a, ApJ, 591, 936

Sneden, C., Cowan, J. J., & Lawler, J. E. 2003b, Nuclear Phys. A718, 29c

Sneden, C., Lawler, J. E., Cowan, J. J., Ivans, I. I., & Den Hartog, E. A. 2009, ApJS, companion paper, submitted

Stone, N. J. 2005, Atomic Data and Nuclear Data Tables, 90, 75

Westin, J., Sneden, C., Gustafsson, B., & Cowan, J.J. 2000, ApJ, 530, 783

Whaling, W., Carle, M. T., & Pitt, M. L. 1993, J. Quant. Spectrosc. Radiat. Transfer 50, 7

Wickliffe, M. E. & Lawler, J. E. 1997, J. Opt. Soc. Am. B 14, 737

Wickliffe, M. E., Lawler, J. E., & Nave, G. 2000, J. Quant. Spectrosc. Radiat. Transfer, 66, 363

Woolf, V. M., Tomkin, J., & Lambert, D. L. 1995, ApJ, 453, 660

1Woodgate, G. K. 1980, *Elementary Atomic Structure* 2nd Ed. (Clarendon Press, Oxford) p. 184

Xu, H. L., Persson, A., & Svanberg, S. 2003, Eur. Phys. J. D, 23, 233

Zhang, Z. G., Svanberg, S., Jiang, Z., Palmeri, P., Quinet, P., & Biémont, E. 2001, Physica Scripta 63, 122

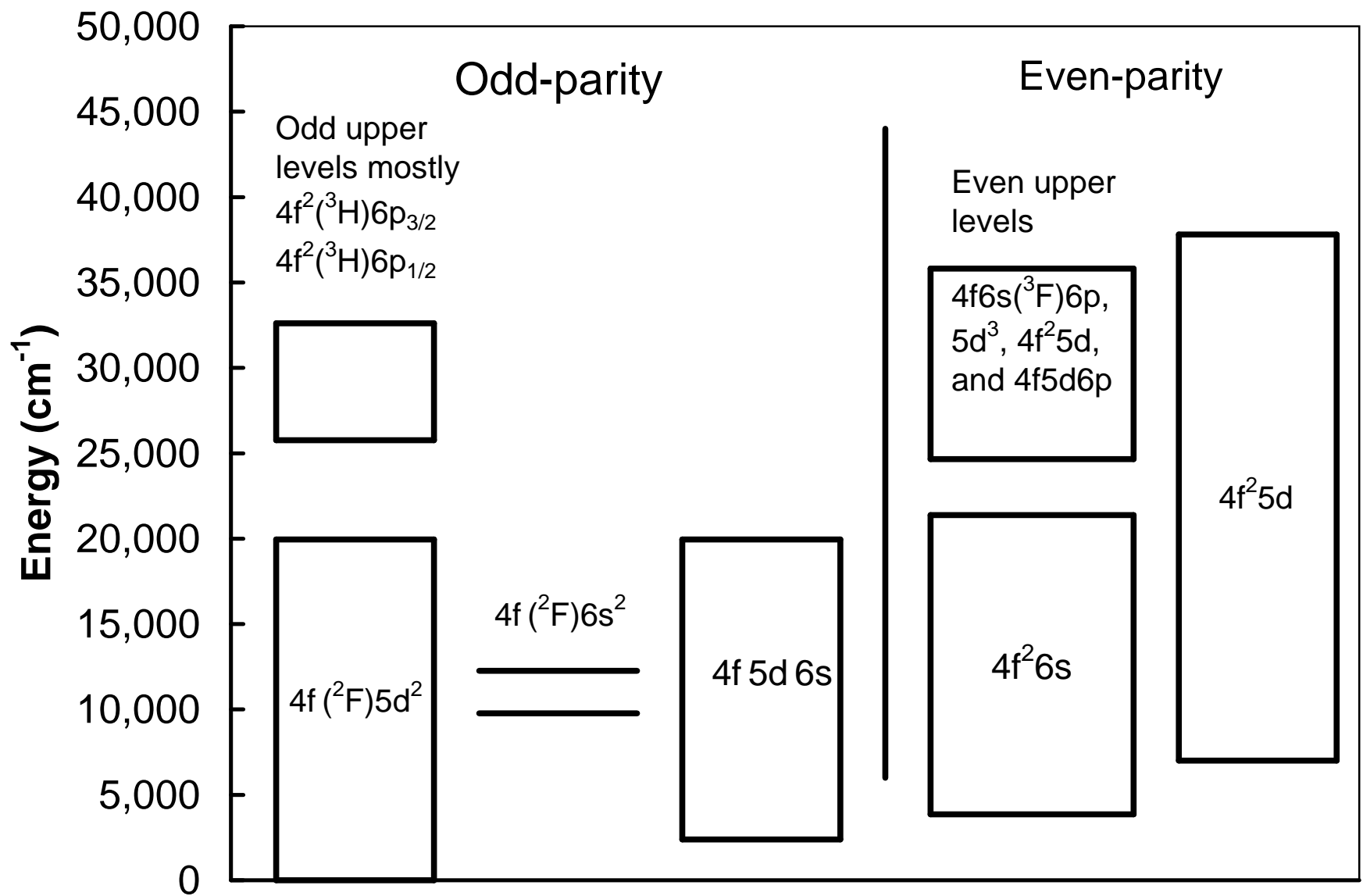

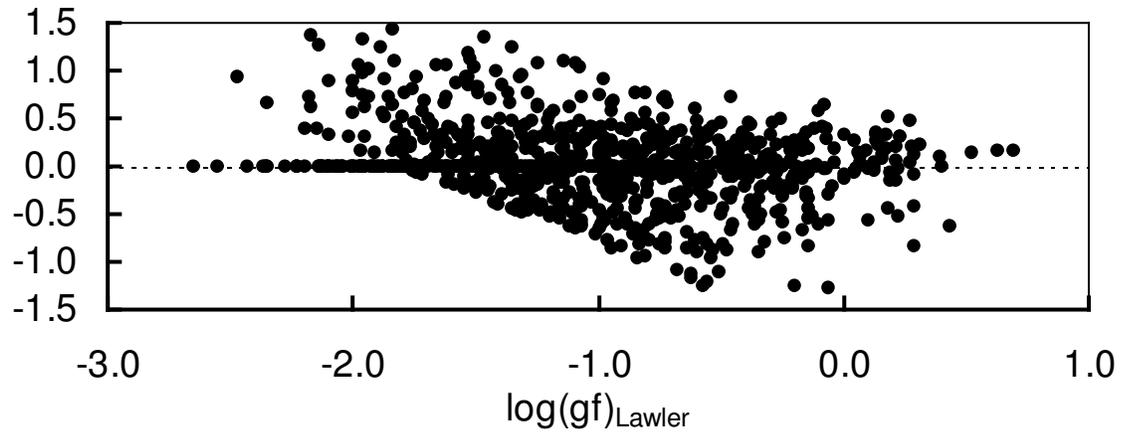
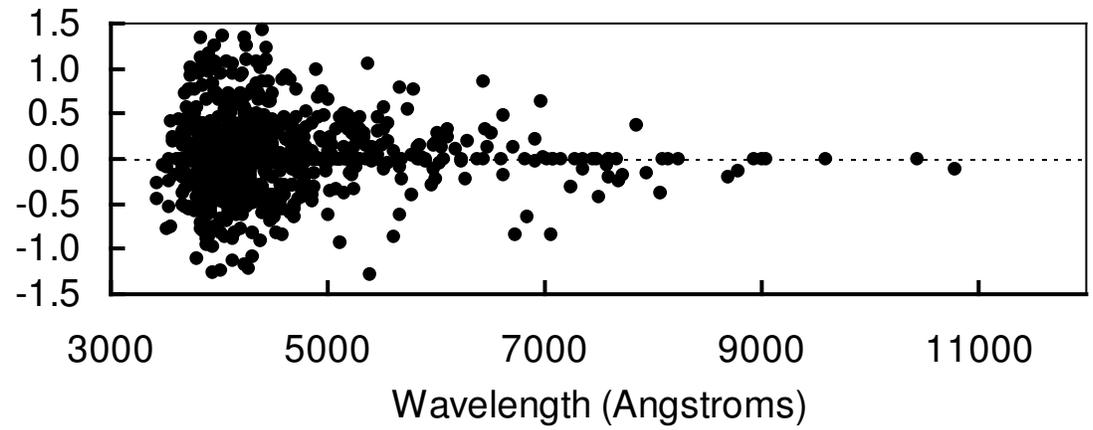
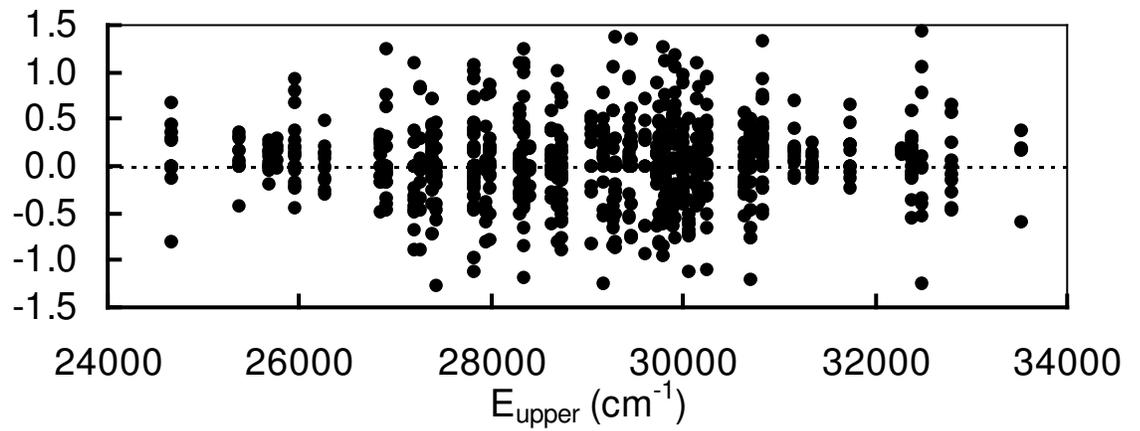

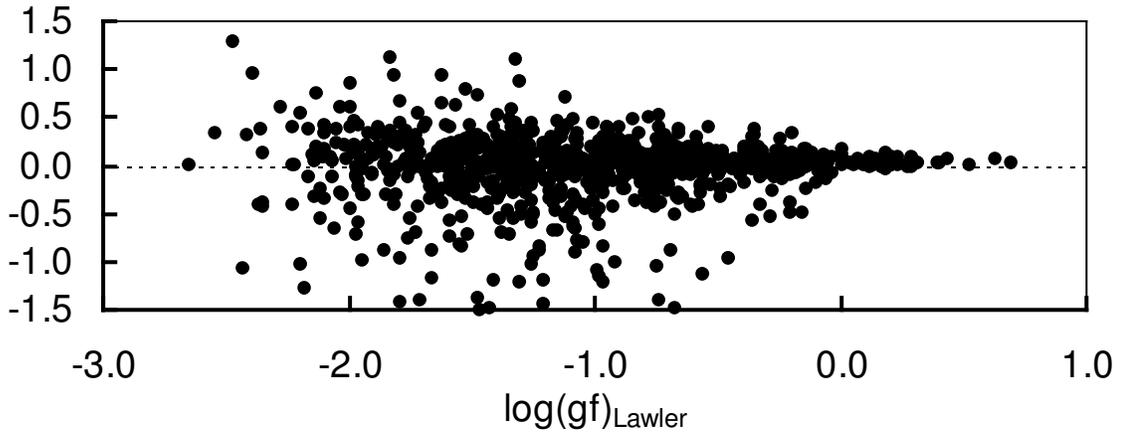
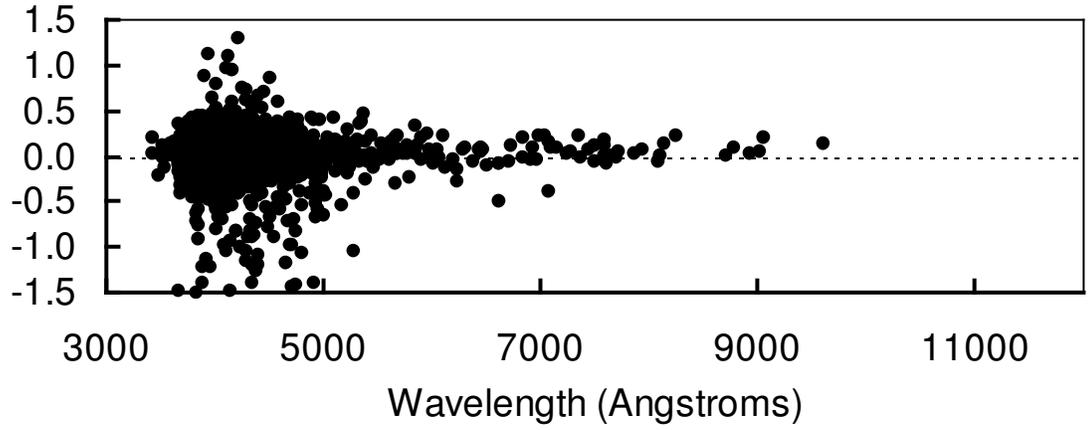
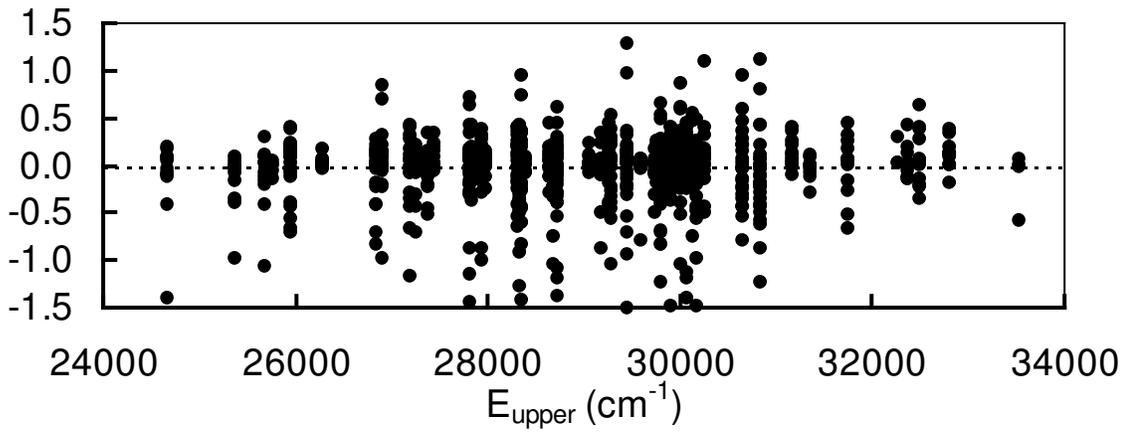

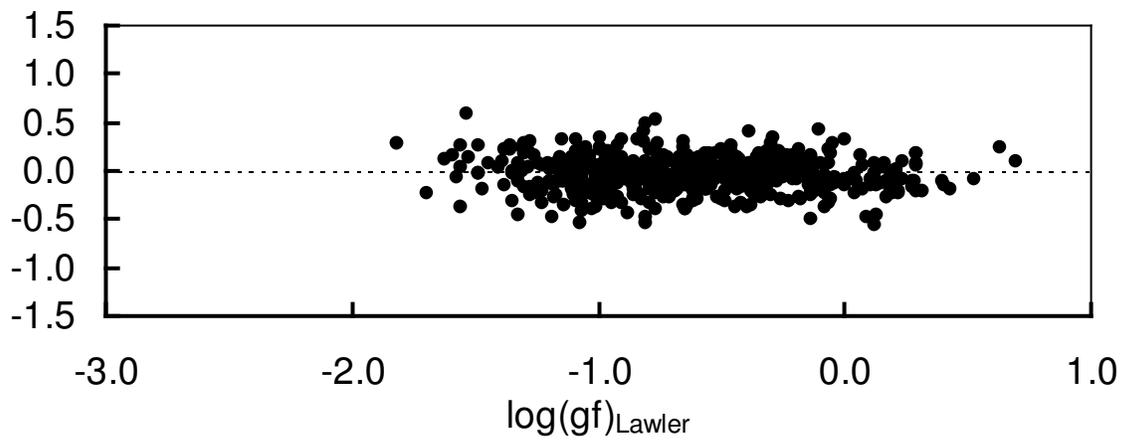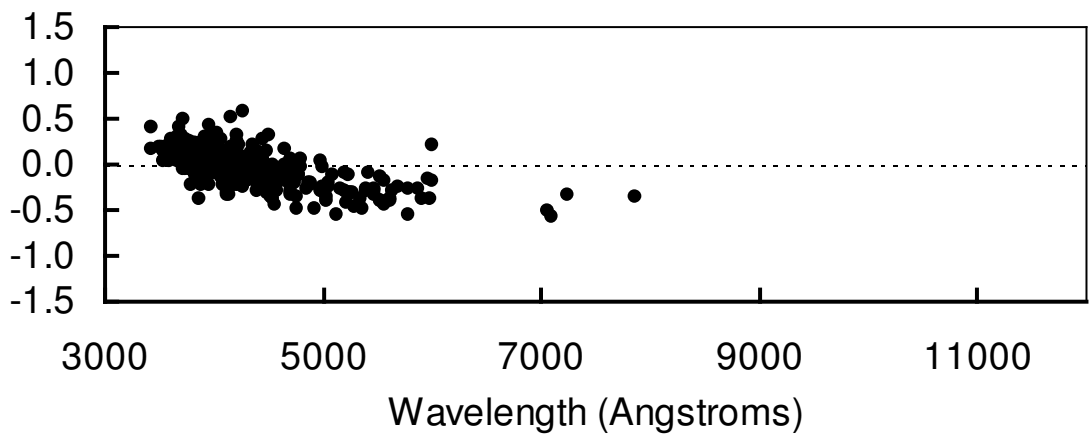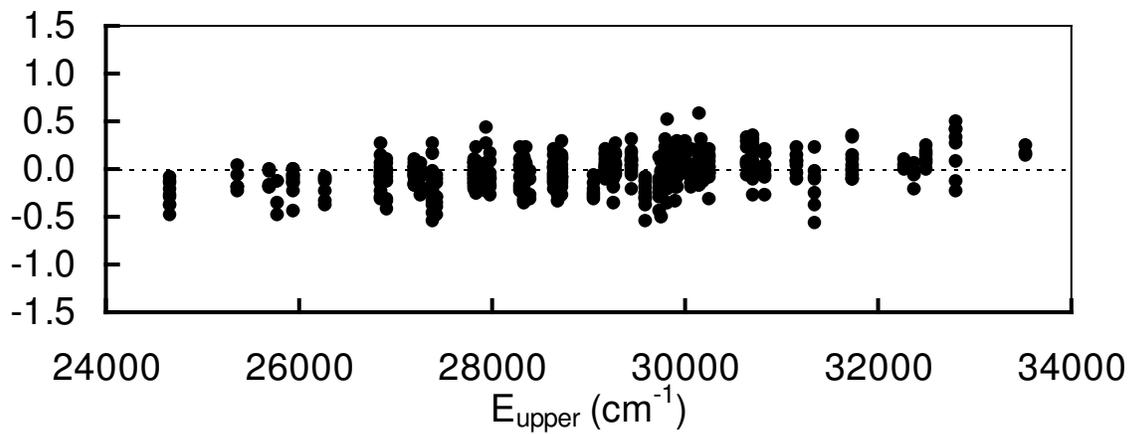

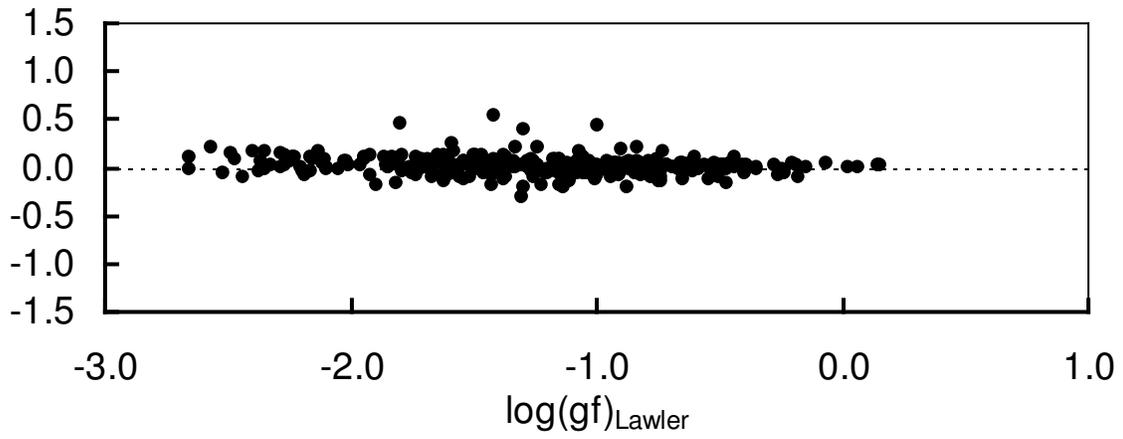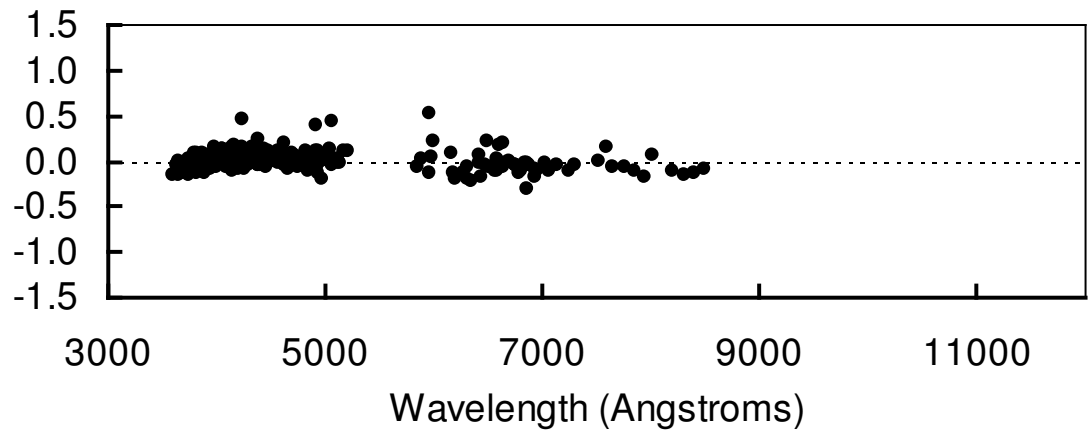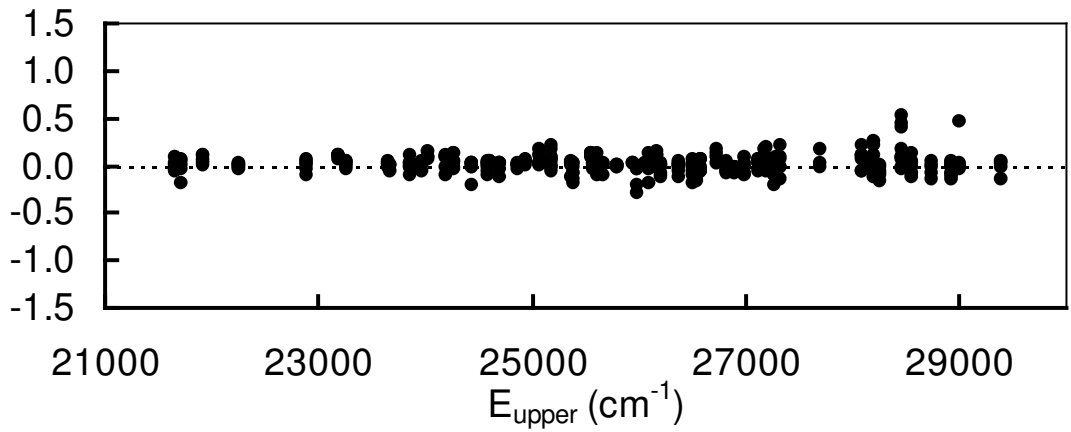

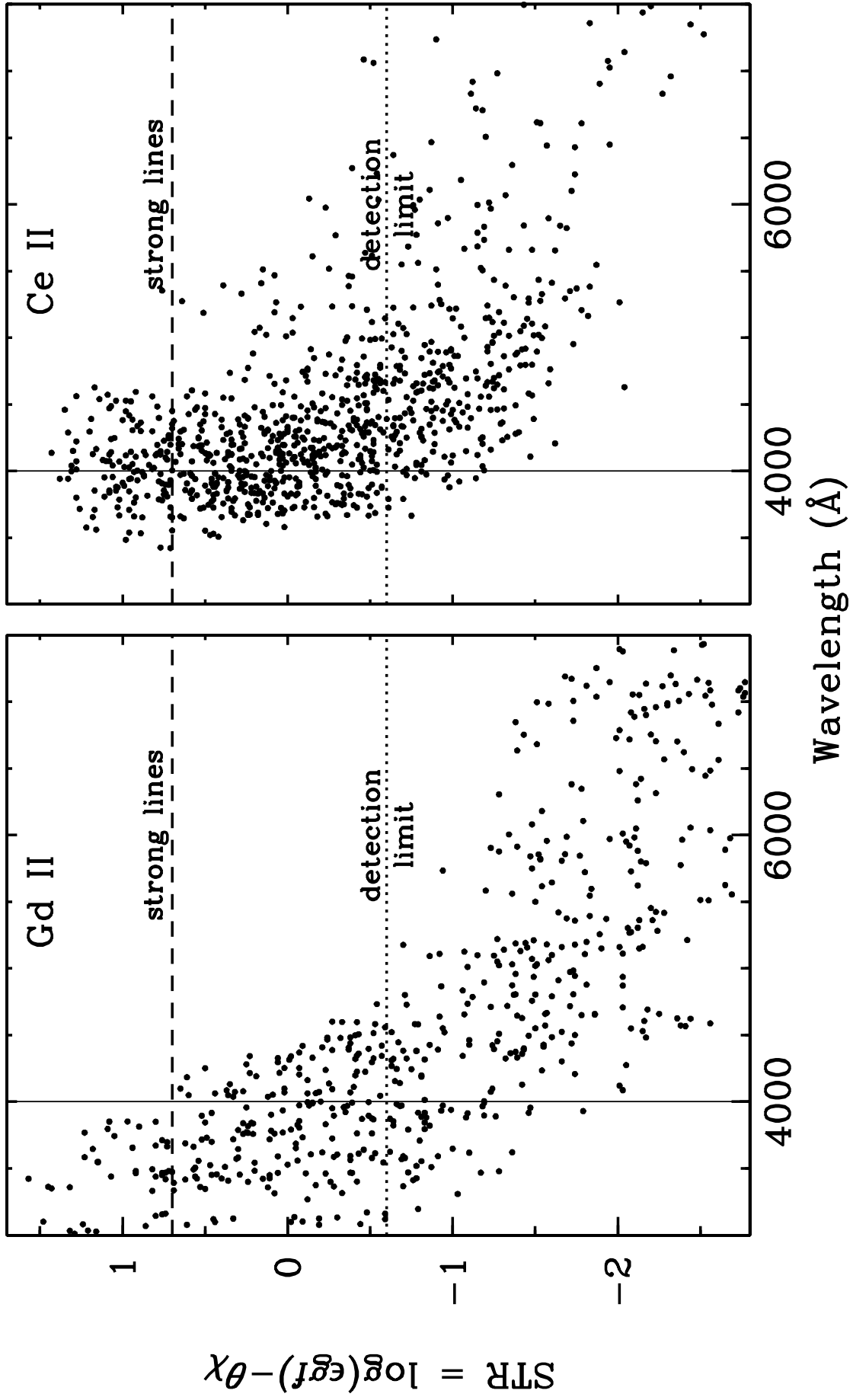

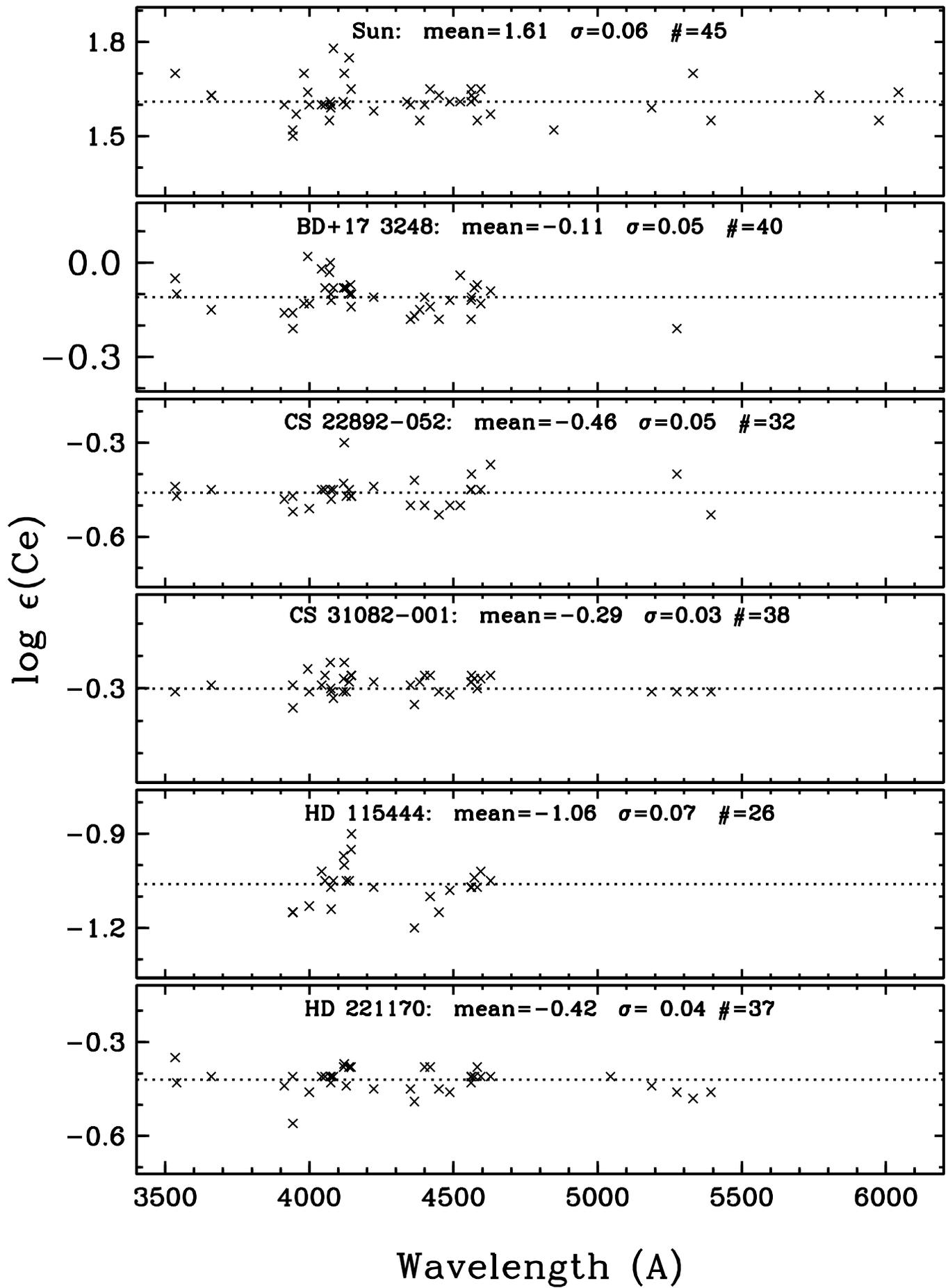

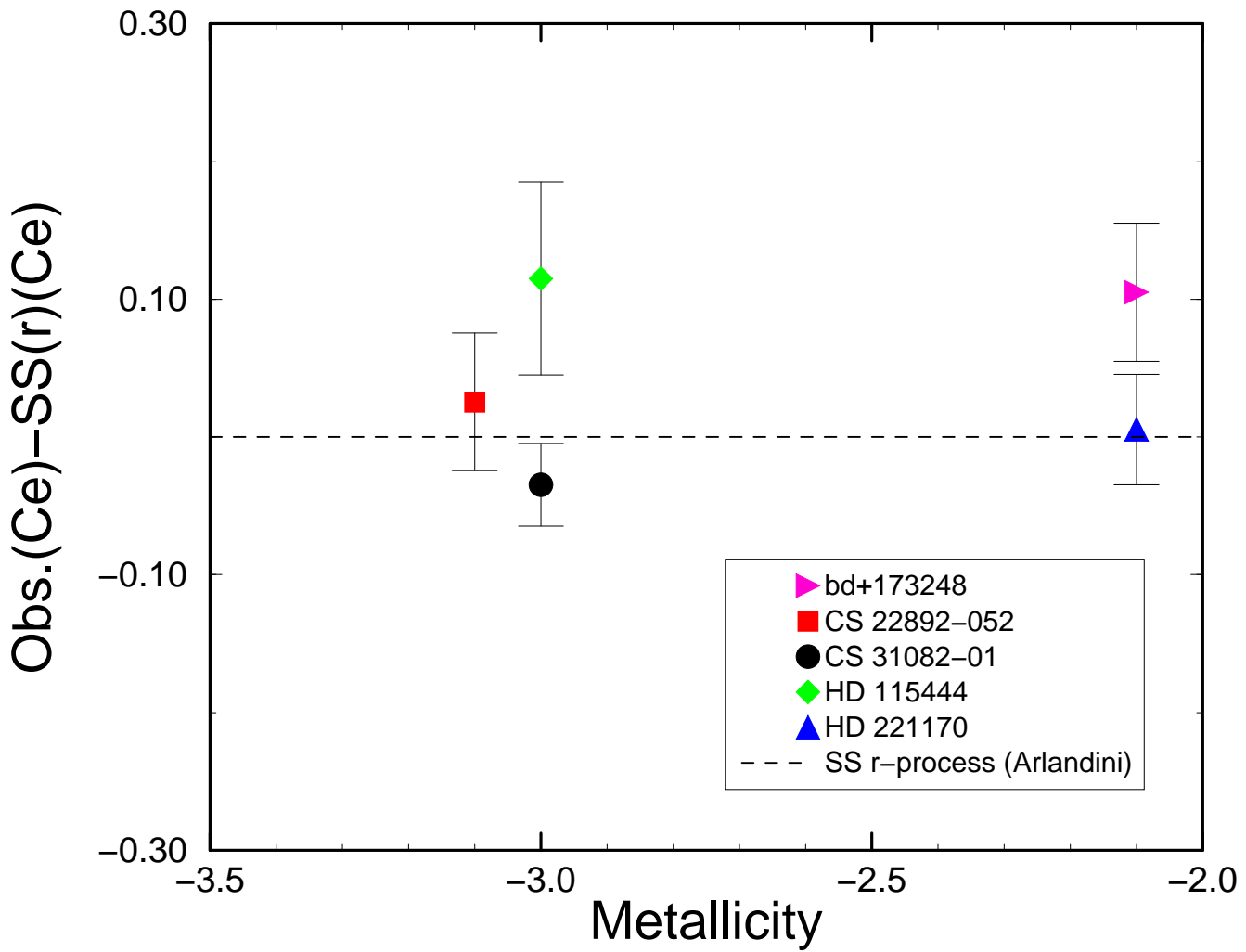

Table 1. Fourier transform spectra of Ce lamps used in this study. All were recorded using the 1.0 meter FTS on the McMath telescope at the National Solar Observatory, Kitt Peak, AZ.

| Index | Date | Serial Number | Lamp Type[1] | Buffer Gas | Lamp Current (mA) | Wavenumber Range (cm$^{-1}$) | Limit of Resolution (cm$^{-1}$) | Coadds | Beam Splitter | Filter | Detector[2] | Calibration[3] |
|---|---|---|---|---|---|---|---|---|---|---|---|---|
| 1 | 2002 Feb. 27 | 30 | Commercial HCD | Ar | 27 | 7929 – 34998 | 0.050 | 50 | UV | | S. B. Si Diode | Ar I & II WQH Lamp |
| 2 | 2002 Feb. 28 | 32 | Commercial HCD | Ar | 27 | 7929 – 34998 | 0.050 | 50 | UV | | S. B. Si Diode | Ar I & II WQH Lamp |
| 3 | 2002 Feb. 28 | 33 | Commercial HCD | Ar | 27 | 7929 – 34998 | 0.050 | 112 | UV | | S. B. Si Diode | Ar I & II WQH Lamp |
| 4 | 2002 Feb. 27 | 29 | Commercial HCD | Ar | 27 | 7929 – 34998 | 0.050 | 8 | UV | | S. B. Si Diode | Ar I & II WQH Lamp |
| 5 | 2002 Feb. 26 | 14 | Commercial HCD | Ar | 22 | 7929 – 34998 | 0.050 | 10 | UV | | S. B. Si Diode | Ar I & II WQH Lamp |
| 6 | 2002 Feb. 26 | 15 | Commercial HCD | Ar | 17 | 7929 – 34998 | 0.050 | 10 | UV | | S. B. Si Diode | Ar I & II WQH Lamp |

| # | Date | N | Lamp¹ | Fill Gas | Current (mA) | Range (Å) | Resolution (Å) | Exposure (min) | Filter | | Detector² | Radiometric Std³ |
|---|------|---|-------|----------|--------------|-----------|----------------|----------------|--------|---|-----------|------------------|
| 7 | 2002 Feb. 27 | 28 | HCD Commercial | Ne | 27 | 7929 – 34998 | 0.050 | 9 | UV | Diode | S. B. Si | WQH Lamp |
| 8 | 2002 Feb. 26 | 16 | HCD Commercial | Ne | 25 | 7929 – 34998 | 0.050 | 10 | UV | Diode | S. B. Si | WQH Lamp |
| 9 | 2002 Feb. 27 | 20 | HCD Commercial | Ne | 20 | 7929 – 34998 | 0.050 | 20 | UV | Diode | S. B. Si | WQH Lamp |
| 10 | 1985 Feb. 5 | 23 | EDL CeI$_3$ | Ar | | 7456 – 28808 | 0.035 | 8 | UV | GG375 | Diode S. B. Si | W Strip Lamp |
| 11 | 1985 Feb. 5 | 67 | EDL CeI$_3$ | Ar | | 7456 – 28808 | 0.035 | 8 | UV | GG375 | Diode S. B. Si | W Strip Lamp |
| 12 | 1985 Feb. 5 | 4 | EDL CeI$_3$ | Ar | | 3285-15050 | 0.018 | 3 | UV | RG715 | Diode InSb | W Strip Lamp |
| 13 | 1985 Feb. 5 | 5 | EDL CeI$_3$ | Ar | | 3285-15050 | 0.018 | 5 | UV | RG715 | InSb | W Strip Lamp |
| 14 | 1985 Feb. 6 | 35 | EDL CeI$_3$ | Ar | | 3285-15050 | 0.018 | 11 | UV | RG715 | InSb | W Strip Lamp |

[1]Lamp types include commercially available small sealed Hollow Cathode Discharge (HCD) lamps typically used in atomic absorption spectrophotometers and an Electrodeless Discharge Lamp (EDL) run with isotopically pure $^{140}$Ce introduced as an iodide salt.

[2]Detectors types include the Super Blue (S. B.) Si photodiode and Indium Antimonide (InSb) detector for the near infrared.

[3]Relative radiometric calibrations were based on selected sets of Ar I and Ar II lines, on a Tungsten-Quartz-Halogen (WQH) lamp calibrated as a secondary irradiance standard, and on a Tungsten (W) Strip Lamp calibrated as a secondary radiance standard.

Table 2. Atomic transition probabilities for Ce II organized by increasing wavelength in air, $\lambda_{air}$.

| $\lambda_{air}$ (Å) | $E_{upper}$ (cm$^{-1}$) | Parity | $J_{upp}$ | $E_{lower}$ (cm$^{-1}$) | Parity | $J_{low}$ | A-value (10$^6$ s$^{-1}$) | log(gf) |
|---|---|---|---|---|---|---|---|---|
| 3422.705 | 32802.165 | ev | 5.5 | 3593.882 | od | 4.5 | 19.4 ± 2.4 | -0.39 |
| 3426.205 | 30166.057 | ev | 3.5 | 987.611 | od | 4.5 | 15.5 ± 1.5 | -0.66 |
| 3485.053 | 28685.758 | ev | 2.5 | 0.000 | od | 3.5 | 24.9 ± 1.5 | -0.57 |
| 3507.941 | 29908.904 | ev | 4.5 | 1410.304 | od | 4.5 | 6.0 ± 1.0 | -0.96 |
| 3520.520 | 29807.078 | ev | 3.5 | 1410.304 | od | 4.5 | 8.3 ± 0.8 | -0.91 |
| 3530.019 | 30702.610 | ev | 4.5 | 2382.246 | od | 4.5 | 8.6 ± 0.9 | -0.80 |
| 3534.045 | 32492.038 | ev | 5.5 | 4203.934 | od | 6.5 | 32.4 ± 2.7 | -0.14 |
| 3539.079 | 30829.124 | ev | 3.5 | 2581.257 | od | 4.5 | 35.7 ± 2.7 | -0.27 |
| 3552.724 | 30702.610 | ev | 4.5 | 2563.233 | od | 5.5 | 9.8 ± 0.9 | -0.73 |
| 3555.001 | 30702.610 | ev | 4.5 | 2581.257 | od | 4.5 | 15.5 ± 1.5 | -0.53 |
| 3560.802 | 33531.388 | ev | 6.5 | 5455.845 | od | 7.5 | 73 ± 4 | 0.29 |
| 3577.456 | 31738.484 | ev | 5.5 | 3793.634 | od | 6.5 | 60 ± 4 | 0.14 |
| 3580.565 | 29794.517 | ev | 3.5 | 1873.934 | od | 3.5 | 3.3 ± 0.4 | -1.30 |
| 3604.195 | 28725.148 | ev | 4.5 | 987.611 | od | 4.5 | 2.6 ± 0.4 | -1.30 |
| 3632.092 | 30166.057 | ev | 3.5 | 2641.559 | od | 3.5 | 7.1 ± 0.8 | -0.95 |
| 3646.962 | 29794.517 | ev | 3.5 | 2382.246 | od | 4.5 | 15.0 ± 1.3 | -0.62 |
| 3653.104 | 30245.878 | ev | 4.5 | 2879.695 | od | 5.5 | 16.2 ± 1.8 | -0.49 |
| 3653.664 | 31155.623 | ev | 6.5 | 3793.634 | od | 6.5 | 22.8 ± 1.6 | -0.20 |
| 3654.934 | 29994.041 | ev | 2.5 | 2641.559 | od | 3.5 | 12.5 ± 1.4 | -0.82 |
| 3655.844 | 29908.904 | ev | 4.5 | 2563.233 | od | 5.5 | 44.9 ± 2.6 | -0.05 |
| 3658.256 | 29908.904 | ev | 4.5 | 2581.257 | od | 4.5 | 1.95 ± 0.21 | -1.41 |
| 3659.225 | 28730.712 | ev | 3.5 | 1410.304 | od | 4.5 | 13.2 ± 1.0 | -0.67 |
| 3659.970 | 28725.148 | ev | 4.5 | 1410.304 | od | 4.5 | 9.2 ± 1.0 | -0.73 |
| 3660.638 | 28297.473 | ev | 3.5 | 987.611 | od | 4.5 | 19.8 ± 1.4 | -0.50 |
| 3661.907 | 30180.096 | ev | 6.5 | 2879.695 | od | 5.5 | 1.32 ± 0.12 | -1.43 |
| 3665.488 | 30637.157 | ev | 2.5 | 3363.427 | od | 2.5 | 1.10 ± 0.18 | -1.88 |
| 3666.346 | 29908.904 | ev | 4.5 | 2641.559 | od | 3.5 | 0.97 ± 0.11 | -1.71 |
| 3667.978 | 30134.910 | ev | 5.5 | 2879.695 | od | 5.5 | 28.2 ± 1.9 | -0.17 |
| 3668.724 | 27249.669 | ev | 2.5 | 0.000 | od | 3.5 | 4.8 ± 0.4 | -1.23 |
| 3670.668 | 30829.124 | ev | 3.5 | 3593.882 | od | 4.5 | 2.15 ± 0.19 | -1.46 |
| 3671.938 | 29807.078 | ev | 3.5 | 2581.257 | od | 4.5 | 5.5 ± 0.6 | -1.05 |
| 3672.155 | 28634.516 | ev | 5.5 | 1410.304 | od | 4.5 | 3.0 ± 0.5 | -1.14 |
| 3673.633 | 29794.517 | ev | 3.5 | 2581.257 | od | 4.5 | 6.2 ± 0.6 | -1.00 |
| 3679.156 | 29807.078 | ev | 3.5 | 2634.666 | od | 2.5 | 4.24 ± 0.28 | -1.16 |
| 3680.089 | 29807.078 | ev | 3.5 | 2641.559 | od | 3.5 | 6.4 ± 0.5 | -0.98 |
| 3680.857 | 29794.517 | ev | 3.5 | 2634.666 | od | 2.5 | 2.17 ± 0.14 | -1.45 |
| 3682.083 | 32802.165 | ev | 5.5 | 5651.357 | od | 5.5 | 6.2 ± 0.7 | -0.82 |
| 3685.515 | 30829.124 | ev | 3.5 | 3703.594 | od | 3.5 | 1.48 ± 0.17 | -1.62 |
| 3687.799 | 30702.610 | ev | 4.5 | 3593.882 | od | 4.5 | 5.9 ± 0.4 | -0.92 |
| 3694.908 | 29438.817 | ev | 5.5 | 2382.246 | od | 4.5 | 5.0 ± 0.5 | -0.91 |
| 3702.785 | 30702.610 | ev | 4.5 | 3703.594 | od | 3.5 | 4.2 ± 0.4 | -1.06 |
| 3704.976 | 32802.165 | ev | 5.5 | 5819.113 | od | 4.5 | 6.2 ± 0.8 | -0.81 |
| 3709.287 | 31155.623 | ev | 6.5 | 4203.934 | od | 6.5 | 29.4 ± 1.8 | -0.07 |
| 3709.929 | 27934.638 | ev | 4.5 | 987.611 | od | 4.5 | 26.3 ± 1.5 | -0.26 |
| 3716.366 | 26900.354 | ev | 3.5 | 0.000 | od | 3.5 | 31.0 ± 1.9 | -0.29 |
| 3719.791 | 29438.817 | ev | 5.5 | 2563.233 | od | 5.5 | 8.0 ± 0.8 | -0.70 |
| 3722.100 | 32372.621 | od | 4.5 | 5513.709 | ev | 5.5 | 6.0 ± 0.6 | -0.90 |
| 3722.288 | 29438.817 | ev | 5.5 | 2581.257 | od | 4.5 | 5.2 ± 0.3 | -0.89 |
| 3722.763 | 29449.778 | ev | 1.5 | 2595.644 | od | 1.5 | 9.3 ± 0.6 | -1.11 |
| 3724.629 | 32492.038 | ev | 5.5 | 5651.357 | od | 5.5 | 6.2 ± 0.5 | -0.81 |
| 3725.603 | 30829.124 | ev | 3.5 | 3995.460 | od | 3.5 | 1.28 ± 0.15 | -1.67 |
| 3725.673 | 32802.165 | ev | 5.5 | 5969.007 | od | 5.5 | 20.0 ± 2.2 | -0.30 |
| 3726.456 | 31738.484 | ev | 5.5 | 4910.963 | od | 5.5 | 1.75 ± 0.18 | -1.36 |
| 3726.961 | 27811.496 | ev | 3.5 | 987.611 | od | 4.5 | 6.3 ± 0.5 | -0.98 |
| 3728.018 | 32492.038 | ev | 5.5 | 5675.763 | od | 4.5 | 29.0 ± 2.3 | -0.14 |
| 3728.180 | 29449.778 | ev | 1.5 | 2634.666 | od | 2.5 | 12.1 ± 1.1 | -0.99 |
| 3728.417 | 32269.252 | ev | 7.5 | 5455.845 | od | 7.5 | 43.1 ± 2.9 | 0.16 |
| 3728.637 | 28685.758 | ev | 2.5 | 1873.934 | od | 3.5 | 3.05 ± 0.28 | -1.42 |
| 3729.916 | 30166.057 | ev | 3.5 | 3363.427 | od | 2.5 | 3.1 ± 0.3 | -1.29 |
| 3732.462 | 29166.597 | ev | 4.5 | 2382.246 | od | 4.5 | 2.43 ± 0.21 | -1.29 |
| 3736.434 | 32372.621 | od | 4.5 | 5616.739 | ev | 4.5 | 2.13 ± 0.24 | -1.35 |
| 3744.010 | 30065.164 | ev | 3.5 | 3363.427 | od | 2.5 | 1.75 ± 0.17 | -1.53 |
| 3746.256 | 29281.374 | ev | 2.5 | 2595.644 | od | 1.5 | 2.19 ± 0.17 | -1.56 |
| 3748.055 | 32492.038 | ev | 5.5 | 5819.113 | od | 4.5 | 17.8 ± 1.1 | -0.35 |
| 3750.998 | 30245.878 | ev | 4.5 | 3593.882 | od | 4.5 | 6.5 ± 0.5 | -0.86 |
| 3751.742 | 29281.374 | ev | 2.5 | 2634.666 | od | 2.5 | 2.19 ± 0.16 | -1.56 |
| 3752.448 | 30637.157 | ev | 2.5 | 3995.460 | od | 3.5 | 3.5 ± 0.3 | -1.35 |
| 3755.789 | 33531.388 | ev | 6.5 | 6913.392 | od | 6.5 | 3.3 ± 0.5 | -1.01 |
| 3757.855 | 29166.597 | ev | 4.5 | 2563.233 | od | 5.5 | 6.8 ± 0.8 | -0.84 |
| 3760.403 | 29166.597 | ev | 4.5 | 2581.257 | od | 4.5 | 1.39 ± 0.13 | -1.53 |
| 3763.604 | 30829.124 | ev | 3.5 | 4266.397 | od | 3.5 | 6.9 ± 0.5 | -0.93 |
| 3764.115 | 29438.817 | ev | 5.5 | 2879.695 | od | 5.5 | 21.4 ± 1.3 | -0.26 |
| 3766.503 | 30245.878 | ev | 4.5 | 3703.594 | od | 3.5 | 5.5 ± 0.4 | -0.94 |
| 3766.681 | 30134.910 | ev | 5.5 | 3593.882 | od | 4.5 | 1.66 ± 0.17 | -1.37 |
| 3769.052 | 27934.638 | ev | 4.5 | 1410.304 | od | 4.5 | 1.51 ± 0.14 | -1.49 |
| 3771.600 | 30829.124 | ev | 3.5 | 4322.708 | od | 2.5 | 16.7 ± 1.0 | -0.55 |
| 3776.606 | 30065.164 | ev | 3.5 | 3593.882 | od | 4.5 | 10.0 ± 0.7 | -0.77 |
| 3777.662 | 28337.814 | ev | 2.5 | 1873.934 | od | 3.5 | 5.8 ± 0.6 | -1.13 |
| 3781.616 | 30702.610 | ev | 4.5 | 4266.397 | od | 3.5 | 25.6 ± 1.4 | -0.26 |
| 3786.632 | 27811.496 | ev | 3.5 | 1410.304 | od | 4.5 | 25.1 ± 1.3 | -0.36 |
| 3787.903 | 27379.949 | ev | 5.5 | 987.611 | od | 4.5 | 4.03 ± 0.28 | -0.98 |
| 3788.746 | 30180.096 | ev | 6.5 | 3793.634 | od | 6.5 | 16.5 ± 1.1 | -0.30 |
| 3791.002 | 30637.157 | ev | 2.5 | 4266.397 | od | 3.5 | 1.48 ± 0.15 | -1.72 |
| 3791.219 | 30829.124 | ev | 3.5 | 4459.872 | od | 3.5 | 1.43 ± 0.15 | -1.61 |

| | | | | | | | | | |
|---|---|---|---|---|---|---|---|---|---|
| 3792.324 | 30065.164 | ev | 3.5 | 3703.594 | od | 3.5 | 18.1 | ± 1.2 | -0.51 |
| 3794.210 | 28730.712 | ev | 3.5 | 2382.246 | od | 4.5 | 1.48 | ± 0.12 | -1.59 |
| 3795.011 | 28725.148 | ev | 4.5 | 2382.246 | od | 4.5 | 5.9 | ± 0.4 | -0.89 |
| 3795.246 | 30134.910 | ev | 5.5 | 3793.634 | od | 6.5 | 6.7 | ± 0.5 | -0.76 |
| 3799.032 | 29908.904 | ev | 4.5 | 3593.882 | od | 4.5 | 2.51 | ± 0.16 | -1.26 |
| 3800.322 | 30829.124 | ev | 3.5 | 4523.033 | od | 4.5 | 8.2 | ± 0.6 | -0.85 |
| 3801.526 | 33531.388 | ev | 6.5 | 7233.627 | od | 5.5 | 139 | ± 7 | 0.63 |
| 3803.096 | 29166.597 | ev | 4.5 | 2879.695 | od | 5.5 | 24.9 | ± 1.7 | -0.27 |
| 3808.113 | 28634.516 | ev | 5.5 | 2382.246 | od | 4.5 | 27.7 | ± 1.6 | -0.14 |
| 3808.382 | 30245.878 | ev | 4.5 | 3995.460 | od | 3.5 | 1.33 | ± 0.12 | -1.54 |
| 3809.218 | 31155.623 | ev | 6.5 | 4910.963 | od | 5.5 | 20.3 | ± 1.3 | -0.21 |
| 3809.497 | 30702.610 | ev | 4.5 | 4459.872 | od | 3.5 | 4.6 | ± 0.6 | -1.00 |
| 3814.938 | 29908.904 | ev | 4.5 | 3703.594 | od | 3.5 | 2.15 | ± 0.16 | -1.33 |
| 3815.793 | 27187.047 | ev | 3.5 | 987.611 | od | 4.5 | 15.1 | ± 1.1 | -0.58 |
| 3818.688 | 30702.610 | ev | 4.5 | 4523.033 | od | 4.5 | 4.19 | ± 0.30 | -1.04 |
| 3819.022 | 30637.157 | ev | 2.5 | 4459.872 | od | 3.5 | 27.2 | ± 2.6 | -0.45 |
| 3819.998 | 30166.057 | ev | 3.5 | 3995.460 | od | 3.5 | 2.08 | ± 0.17 | -1.44 |
| 3821.266 | 28725.148 | ev | 4.5 | 2563.233 | od | 5.5 | 7.6 | ± 0.4 | -0.78 |
| 3823.900 | 28725.148 | ev | 4.5 | 2581.257 | od | 4.5 | 15.1 | ± 1.1 | -0.48 |
| 3826.534 | 30637.157 | ev | 2.5 | 4511.257 | od | 2.5 | 2.7 | ± 0.5 | -1.44 |
| 3829.820 | 29807.078 | ev | 3.5 | 3703.594 | od | 3.5 | 1.73 | ± 0.21 | -1.52 |
| 3830.023 | 32492.038 | ev | 5.5 | 6389.942 | od | 4.5 | 8.3 | ± 0.5 | -0.66 |
| 3830.911 | 28730.712 | ev | 3.5 | 2634.666 | od | 2.5 | 9.3 | ± 0.6 | -0.79 |
| 3831.542 | 30829.124 | ev | 3.5 | 4737.373 | od | 2.5 | 6.0 | ± 1.0 | -0.98 |
| 3831.663 | 29794.517 | ev | 3.5 | 3703.594 | od | 3.5 | 4.1 | ± 0.4 | -1.15 |
| 3831.782 | 28685.758 | ev | 2.5 | 2595.644 | od | 1.5 | 4.7 | ± 0.4 | -1.21 |
| 3832.221 | 31738.484 | ev | 5.5 | 5651.357 | od | 5.5 | 8.1 | ± 0.7 | -0.67 |
| 3832.335 | 29449.778 | ev | 1.5 | 3363.427 | od | 2.5 | 3.88 | ± 0.27 | -1.47 |
| 3832.741 | 28725.148 | ev | 4.5 | 2641.559 | od | 3.5 | 3.9 | ± 0.4 | -1.07 |
| 3834.550 | 28634.516 | ev | 5.5 | 2563.233 | od | 5.5 | 12.4 | ± 0.8 | -0.49 |
| 3834.782 | 30065.164 | ev | 3.5 | 3995.460 | od | 3.5 | 7.9 | ± 0.5 | -0.86 |
| 3836.107 | 27934.638 | ev | 4.5 | 1873.934 | od | 3.5 | 8.5 | ± 0.5 | -0.73 |
| 3837.203 | 28634.516 | ev | 5.5 | 2581.257 | od | 4.5 | 3.70 | ± 0.28 | -1.01 |
| 3837.522 | 28685.758 | ev | 2.5 | 2634.666 | od | 2.5 | 1.30 | ± 0.10 | -1.76 |
| 3838.538 | 28685.758 | ev | 2.5 | 2641.559 | od | 3.5 | 55.0 | ± 3.0 | -0.14 |
| 3843.767 | 33531.388 | ev | 6.5 | 7522.622 | od | 5.5 | 14.0 | ± 1.3 | -0.36 |
| 3845.273 | 29994.041 | ev | 2.5 | 3995.460 | od | 3.5 | 4.2 | ± 0.5 | -1.25 |
| 3848.100 | 30245.878 | ev | 4.5 | 4266.397 | od | 3.5 | 13.1 | ± 0.9 | -0.54 |
| 3848.592 | 30180.096 | ev | 6.5 | 4203.934 | od | 6.5 | 27.4 | ± 1.6 | -0.07 |
| 3849.558 | 27379.949 | ev | 5.5 | 1410.304 | od | 4.5 | 3.7 | ± 0.3 | -1.01 |
| 3852.099 | 28334.756 | ev | 4.5 | 2382.246 | od | 4.5 | 3.7 | ± 0.3 | -1.08 |
| 3853.156 | 25945.396 | ev | 3.5 | 0.000 | od | 3.5 | 17.5 | ± 1.0 | -0.51 |
| 3854.185 | 27812.398 | ev | 2.5 | 1873.934 | od | 3.5 | 43.1 | ± 2.3 | -0.24 |
| 3854.319 | 27811.496 | ev | 3.5 | 1873.934 | od | 3.5 | 39.7 | ± 2.1 | -0.15 |
| 3855.298 | 30134.910 | ev | 5.5 | 4203.934 | od | 6.5 | 19.5 | ± 1.2 | -0.28 |
| 3857.025 | 31738.484 | ev | 5.5 | 5819.113 | od | 4.5 | 21.0 | ± 1.9 | -0.25 |

| | | | | | | | | |
|---|---|---|---|---|---|---|---|---|
| 3857.237 | 29281.374 | ev | 2.5 | 3363.427 | od | 2.5 | 13.0 ± 0.9 | -0.76 |
| 3857.641 | 28297.473 | ev | 3.5 | 2382.246 | od | 4.5 | 13.0 ± 0.8 | -0.63 |
| 3868.133 | 29438.817 | ev | 5.5 | 3593.882 | od | 4.5 | 5.7 ± 0.3 | -0.81 |
| 3870.867 | 31340.393 | od | 6.5 | 5513.709 | ev | 5.5 | 5.2 ± 0.4 | -0.79 |
| 3872.131 | 30829.124 | ev | 3.5 | 5010.870 | od | 2.5 | 2.70 ± 0.19 | -1.31 |
| 3873.126 | 29807.078 | ev | 3.5 | 3995.460 | od | 3.5 | 1.55 ± 0.11 | -1.55 |
| 3875.012 | 29794.517 | ev | 3.5 | 3995.460 | od | 3.5 | 8.0 ± 0.7 | -0.84 |
| 3875.056 | 30065.164 | ev | 3.5 | 4266.397 | od | 3.5 | 10.2 ± 0.7 | -0.74 |
| 3875.995 | 30637.157 | ev | 2.5 | 4844.644 | od | 1.5 | 2.24 ± 0.18 | -1.52 |
| 3876.126 | 30702.610 | ev | 4.5 | 4910.963 | od | 5.5 | 11.7 ± 0.8 | -0.58 |
| 3878.367 | 27187.047 | ev | 3.5 | 1410.304 | od | 4.5 | 24.9 ± 1.5 | -0.35 |
| 3879.460 | 31738.484 | ev | 5.5 | 5969.007 | od | 5.5 | 0.60 ± 0.08 | -1.79 |
| 3881.668 | 28634.516 | ev | 5.5 | 2879.695 | od | 5.5 | 6.7 ± 0.5 | -0.74 |
| 3881.867 | 28334.756 | ev | 4.5 | 2581.257 | od | 4.5 | 7.1 ± 1.4 | -0.79 |
| 3882.445 | 28345.313 | ev | 0.5 | 2595.644 | od | 1.5 | 144 ± 7 | -0.19 |
| 3883.437 | 32802.165 | ev | 5.5 | 7059.072 | od | 4.5 | 2.08 ± 0.23 | -1.25 |
| 3883.533 | 30065.164 | ev | 3.5 | 4322.708 | od | 2.5 | 4.7 ± 0.4 | -1.07 |
| 3883.576 | 28337.814 | ev | 2.5 | 2595.644 | od | 1.5 | 6.7 ± 0.5 | -1.04 |
| 3885.769 | 29994.041 | ev | 2.5 | 4266.397 | od | 3.5 | 3.37 ± 0.25 | -1.34 |
| 3886.493 | 30245.878 | ev | 4.5 | 4523.033 | od | 4.5 | 4.0 ± 0.4 | -1.05 |
| 3888.387 | 30829.124 | ev | 3.5 | 5118.806 | od | 2.5 | 8.6 ± 0.6 | -0.80 |
| 3889.297 | 29449.778 | ev | 1.5 | 3745.475 | od | 1.5 | 7.6 ± 0.6 | -1.16 |
| 3889.472 | 28337.814 | ev | 2.5 | 2634.666 | od | 2.5 | 3.7 ± 0.4 | -1.29 |
| 3889.982 | 31155.623 | ev | 6.5 | 5455.845 | od | 7.5 | 41.9 ± 2.3 | 0.12 |
| 3890.515 | 28337.814 | ev | 2.5 | 2641.559 | od | 3.5 | 3.01 ± 0.30 | -1.39 |
| 3890.744 | 27835.233 | ev | 1.5 | 2140.492 | od | 0.5 | 18.4 ± 1.3 | -0.78 |
| 3890.978 | 28334.756 | ev | 4.5 | 2641.559 | od | 3.5 | 7.3 ± 0.5 | -0.78 |
| 3894.292 | 29994.041 | ev | 2.5 | 4322.708 | od | 2.5 | 3.63 ± 0.21 | -1.31 |
| 3896.633 | 28297.473 | ev | 3.5 | 2641.559 | od | 3.5 | 1.41 ± 0.10 | -1.59 |
| 3896.802 | 30166.057 | ev | 3.5 | 4511.257 | od | 2.5 | 29.5 ± 1.6 | -0.27 |
| 3898.263 | 29438.817 | ev | 5.5 | 3793.634 | od | 6.5 | 16.7 ± 1.0 | -0.34 |
| 3898.670 | 29908.904 | ev | 4.5 | 4266.397 | od | 3.5 | 1.28 ± 0.08 | -1.53 |
| 3903.333 | 30134.910 | ev | 5.5 | 4523.033 | od | 4.5 | 6.7 ± 0.4 | -0.73 |
| 3904.337 | 30065.164 | ev | 3.5 | 4459.872 | od | 3.5 | 14.7 ± 1.1 | -0.57 |
| 3907.434 | 29750.547 | od | 5.5 | 4165.550 | ev | 4.5 | 6.5 ± 0.4 | -0.75 |
| 3908.404 | 32492.038 | ev | 5.5 | 6913.392 | od | 6.5 | 42.4 ± 2.5 | 0.07 |
| 3908.536 | 29281.374 | ev | 2.5 | 3703.594 | od | 3.5 | 23.9 ± 1.3 | -0.48 |
| 3909.310 | 29166.597 | ev | 4.5 | 3593.882 | od | 4.5 | 10.0 ± 0.6 | -0.64 |
| 3909.747 | 29735.413 | od | 4.5 | 4165.550 | ev | 4.5 | 4.78 ± 0.29 | -0.96 |
| 3912.188 | 30065.164 | ev | 3.5 | 4511.257 | od | 2.5 | 15.1 ± 0.9 | -0.56 |
| 3912.420 | 27934.638 | ev | 4.5 | 2382.246 | od | 4.5 | 24.7 ± 1.3 | -0.25 |
| 3913.992 | 30065.164 | ev | 3.5 | 4523.033 | od | 4.5 | 6.6 ± 0.4 | -0.92 |
| 3914.947 | 29281.374 | ev | 2.5 | 3745.475 | od | 1.5 | 5.1 ± 0.3 | -1.16 |
| 3916.140 | 29794.517 | ev | 3.5 | 4266.397 | od | 3.5 | 11.9 ± 0.7 | -0.66 |
| 3917.639 | 30637.157 | ev | 2.5 | 5118.806 | od | 2.5 | 19.9 ± 1.3 | -0.56 |
| 3919.803 | 31155.623 | ev | 6.5 | 5651.357 | od | 5.5 | 22.4 ± 1.5 | -0.14 |
| 3921.989 | 26900.354 | ev | 3.5 | 1410.304 | od | 4.5 | 0.21 ± 0.04 | -2.42 |
| 3922.863 | 29807.078 | ev | 3.5 | 4322.708 | od | 2.5 | 5.7 ± 0.6 | -0.98 |
| 3923.107 | 29994.041 | ev | 2.5 | 4511.257 | od | 2.5 | 44.2 ± 2.5 | -0.21 |
| 3927.380 | 28334.756 | ev | 4.5 | 2879.695 | od | 5.5 | 3.4 ± 0.3 | -1.11 |
| 3928.310 | 29908.904 | ev | 4.5 | 4459.872 | od | 3.5 | 5.9 ± 0.4 | -0.86 |
| 3930.792 | 32492.038 | ev | 5.5 | 7059.072 | od | 4.5 | 9.7 ± 0.7 | -0.57 |
| 3931.083 | 26841.384 | ev | 4.5 | 1410.304 | od | 4.5 | 14.6 ± 0.9 | -0.47 |
| 3931.366 | 27811.496 | ev | 3.5 | 2382.246 | od | 4.5 | 15.4 ± 0.9 | -0.54 |
| 3937.180 | 30829.124 | ev | 3.5 | 5437.422 | od | 3.5 | 0.80 ± 0.08 | -1.83 |
| 3939.657 | 27249.669 | ev | 2.5 | 1873.934 | od | 3.5 | 2.34 ± 0.15 | -1.49 |
| 3940.330 | 27934.638 | ev | 4.5 | 2563.233 | od | 5.5 | 23.3 ± 1.2 | -0.27 |
| 3940.970 | 28730.712 | ev | 3.5 | 3363.427 | od | 2.5 | 14.6 ± 0.9 | -0.57 |
| 3942.151 | 25359.686 | ev | 2.5 | 0.000 | od | 3.5 | 43.1 ± 2.4 | -0.22 |
| 3942.745 | 32269.252 | ev | 7.5 | 6913.392 | od | 6.5 | 132 ± 7 | 0.69 |
| 3943.131 | 27934.638 | ev | 4.5 | 2581.257 | od | 4.5 | 7.3 ± 0.4 | -0.77 |
| 3943.884 | 31738.484 | ev | 5.5 | 6389.942 | od | 4.5 | 55 ± 3 | 0.19 |
| 3944.092 | 29807.078 | ev | 3.5 | 4459.872 | od | 3.5 | 4.8 ± 0.3 | -1.05 |
| 3946.047 | 29794.517 | ev | 3.5 | 4459.872 | od | 3.5 | 0.39 ± 0.04 | -2.14 |
| 3947.115 | 30065.164 | ev | 3.5 | 4737.373 | od | 2.5 | 1.00 ± 0.10 | -1.73 |
| 3947.966 | 28685.758 | ev | 2.5 | 3363.427 | od | 2.5 | 12.9 ± 0.8 | -0.74 |
| 3949.404 | 27187.047 | ev | 3.5 | 1873.934 | od | 3.5 | 5.0 ± 0.3 | -1.03 |
| 3952.104 | 29807.078 | ev | 3.5 | 4511.257 | od | 2.5 | 5.7 ± 0.5 | -0.97 |
| 3952.532 | 27934.638 | ev | 4.5 | 2641.559 | od | 3.5 | 33.5 ± 1.7 | -0.11 |
| 3953.652 | 29281.374 | ev | 2.5 | 3995.460 | od | 3.5 | 16.4 ± 0.9 | -0.64 |
| 3953.944 | 29807.078 | ev | 3.5 | 4523.033 | od | 4.5 | 5.3 ± 0.4 | -1.00 |
| 3955.910 | 29794.517 | ev | 3.5 | 4523.033 | od | 4.5 | 8.7 ± 0.8 | -0.79 |
| 3956.278 | 30180.096 | ev | 6.5 | 4910.963 | od | 5.5 | 39.8 ± 2.1 | 0.12 |
| 3956.896 | 30702.610 | ev | 4.5 | 5437.422 | od | 3.5 | 10.4 ± 0.6 | -0.61 |
| 3957.957 | 32492.038 | ev | 5.5 | 7233.627 | od | 5.5 | 10.3 ± 1.2 | -0.54 |
| 3959.607 | 29449.778 | ev | 1.5 | 4201.893 | od | 1.5 | 16.6 ± 1.4 | -0.81 |
| 3960.909 | 27835.233 | ev | 1.5 | 2595.644 | od | 1.5 | 45.9 ± 2.4 | -0.36 |
| 3961.648 | 29438.817 | ev | 5.5 | 4203.934 | od | 6.5 | 1.45 ± 0.14 | -1.39 |
| 3963.365 | 30134.910 | ev | 5.5 | 4910.963 | od | 5.5 | 2.86 ± 0.23 | -1.09 |
| 3964.496 | 27812.398 | ev | 2.5 | 2595.644 | od | 1.5 | 15.7 ± 0.9 | -0.65 |
| 3967.042 | 27835.233 | ev | 1.5 | 2634.666 | od | 2.5 | 47.4 ± 2.4 | -0.35 |
| 3967.173 | 30637.157 | ev | 2.5 | 5437.422 | od | 3.5 | 18.5 ± 1.2 | -0.58 |
| 3967.431 | 32492.038 | ev | 5.5 | 7293.938 | od | 6.5 | 0.85 ± 0.08 | -1.62 |
| 3969.240 | 31155.623 | ev | 6.5 | 5969.007 | od | 5.5 | 2.2 ± 0.3 | -1.14 |
| 3970.640 | 27812.398 | ev | 2.5 | 2634.666 | od | 2.5 | 7.5 ± 0.5 | -0.97 |
| 3970.783 | 27811.496 | ev | 3.5 | 2634.666 | od | 2.5 | 0.61 ± 0.04 | -1.94 |
| 3971.681 | 29166.597 | ev | 4.5 | 3995.460 | od | 3.5 | 16.4 ± 0.9 | -0.41 |
| 3971.870 | 27811.496 | ev | 3.5 | 2641.559 | od | 3.5 | 4.3 ± 0.4 | -1.09 |
| 3974.199 | 30166.057 | ev | 3.5 | 5010.870 | od | 2.5 | 3.6 ± 0.3 | -1.16 |
| 3974.488 | 30829.124 | ev | 3.5 | 5675.763 | od | 4.5 | 1.07 ± 0.11 | -1.69 |
| 3978.646 | 29449.778 | ev | 1.5 | 4322.708 | od | 2.5 | 60 ± 3 | -0.24 |
| 3980.890 | 30829.124 | ev | 3.5 | 5716.216 | od | 3.5 | 32.2 ± 2.1 | -0.21 |

| | | | | | | | | | |
|---|---|---|---|---|---|---|---|---|---|
| 3982.901 | 31738.484 | ev | 5.5 | 6638.258 | od | 4.5 | 18.1 | ± 2.5 | -0.29 |
| 3983.288 | 29263.338 | od | 5.5 | 4165.550 | ev | 4.5 | 8.9 | ± 0.6 | -0.59 |
| 3984.671 | 32802.165 | ev | 5.5 | 7713.089 | od | 4.5 | 34.7 | ± 2.8 | 0.00 |
| 3990.100 | 27934.638 | ev | 4.5 | 2879.695 | od | 5.5 | 4.86 | ± 0.27 | -0.94 |
| 3990.688 | 30702.610 | ev | 4.5 | 5651.357 | od | 5.5 | 4.2 | ± 0.4 | -0.99 |
| 3991.325 | 30166.057 | ev | 3.5 | 5118.806 | od | 2.5 | 2.30 | ± 0.13 | -1.36 |
| 3992.380 | 28634.516 | ev | 5.5 | 3593.882 | od | 4.5 | 20.8 | ± 1.1 | -0.22 |
| 3993.819 | 32372.621 | od | 4.5 | 7341.007 | ev | 5.5 | 81 | ± 4 | 0.29 |
| 3994.580 | 30702.610 | ev | 4.5 | 5675.763 | od | 4.5 | 3.40 | ± 0.20 | -1.09 |
| 3994.648 | 26900.354 | ev | 3.5 | 1873.934 | od | 3.5 | 0.49 | ± 0.06 | -2.02 |
| 3995.425 | 28725.148 | ev | 4.5 | 3703.594 | od | 3.5 | 0.97 | ± 0.07 | -1.63 |
| 3996.475 | 29281.374 | ev | 2.5 | 4266.397 | od | 3.5 | 4.5 | ± 0.3 | -1.19 |
| 3997.269 | 30829.124 | ev | 3.5 | 5819.113 | od | 4.5 | 1.18 | ± 0.10 | -1.64 |
| 3999.237 | 27379.949 | ev | 5.5 | 2382.246 | od | 4.5 | 39.6 | ± 2.0 | 0.06 |
| 4001.047 | 30702.610 | ev | 4.5 | 5716.216 | od | 3.5 | 7.6 | ± 0.6 | -0.74 |
| 4001.563 | 29994.041 | ev | 2.5 | 5010.870 | od | 2.5 | 17.9 | ± 1.4 | -0.59 |
| 4001.724 | 28685.758 | ev | 2.5 | 3703.594 | od | 3.5 | 10.3 | ± 0.7 | -0.83 |
| 4002.822 | 32269.252 | ev | 7.5 | 7293.938 | od | 6.5 | 11.4 | ± 0.8 | -0.36 |
| 4002.971 | 28337.814 | ev | 2.5 | 3363.427 | od | 2.5 | 5.7 | ± 0.5 | -1.09 |
| 4003.767 | 32492.038 | ev | 5.5 | 7522.622 | od | 5.5 | 68 | ± 4 | 0.29 |
| 4004.083 | 26841.384 | ev | 4.5 | 1873.934 | od | 3.5 | 0.63 | ± 0.08 | -1.82 |
| 4005.492 | 29281.374 | ev | 2.5 | 4322.708 | od | 2.5 | 0.43 | ± 0.05 | -2.20 |
| 4005.633 | 25945.396 | ev | 3.5 | 987.611 | od | 4.5 | 9.8 | ± 0.6 | -0.73 |
| 4008.444 | 28685.758 | ev | 2.5 | 3745.475 | od | 1.5 | 1.70 | ± 0.14 | -1.61 |
| 4011.556 | 30637.157 | ev | 2.5 | 5716.216 | od | 3.5 | 6.1 | ± 0.7 | -1.05 |
| 4012.386 | 29438.817 | ev | 5.5 | 4523.033 | od | 4.5 | 67 | ± 3 | 0.29 |
| 4014.897 | 29166.597 | ev | 4.5 | 4266.397 | od | 3.5 | 25.9 | ± 1.4 | -0.20 |
| 4017.135 | 30829.124 | ev | 3.5 | 5942.798 | od | 3.5 | 1.54 | ± 0.14 | -1.53 |
| 4017.592 | 30702.610 | ev | 4.5 | 5819.113 | od | 4.5 | 4.12 | ± 0.24 | -1.00 |
| 4024.485 | 28634.516 | ev | 5.5 | 3793.634 | od | 6.5 | 18.6 | ± 1.0 | -0.27 |
| 4025.139 | 28345.313 | ev | 0.5 | 3508.470 | od | 0.5 | 34.5 | ± 2.0 | -0.78 |
| 4027.044 | 31738.484 | ev | 5.5 | 6913.392 | od | 6.5 | 2.43 | ± 0.23 | -1.15 |
| 4027.627 | 29281.374 | ev | 2.5 | 4459.872 | od | 3.5 | 4.6 | ± 0.6 | -1.17 |
| 4028.404 | 27379.949 | ev | 5.5 | 2563.233 | od | 5.5 | 18.5 | ± 1.0 | -0.27 |
| 4031.332 | 27379.949 | ev | 5.5 | 2581.257 | od | 4.5 | 19.7 | ± 1.1 | -0.24 |
| 4033.779 | 29794.517 | ev | 3.5 | 5010.870 | od | 2.5 | 1.34 | ± 0.09 | -1.58 |
| 4035.982 | 29281.374 | ev | 2.5 | 4511.257 | od | 2.5 | 0.47 | ± 0.06 | -2.17 |
| 4037.662 | 30702.610 | ev | 4.5 | 5942.798 | od | 3.5 | 14.3 | ± 1.0 | -0.46 |
| 4040.753 | 28334.756 | ev | 4.5 | 3593.882 | od | 4.5 | 66 | ± 3 | 0.21 |
| 4041.269 | 30702.610 | ev | 4.5 | 5964.896 | od | 3.5 | 6.8 | ± 0.5 | -0.78 |
| 4042.581 | 28725.148 | ev | 4.5 | 3995.460 | od | 3.5 | 41.1 | ± 2.1 | 0.00 |
| 4045.318 | 30637.157 | ev | 2.5 | 5924.204 | od | 1.5 | 7.6 | ± 0.5 | -0.95 |
| 4045.408 | 29449.778 | ev | 1.5 | 4737.373 | od | 2.5 | 1.73 | ± 0.15 | -1.77 |
| 4046.338 | 29166.597 | ev | 4.5 | 4459.872 | od | 3.5 | 28.5 | ± 1.6 | -0.16 |
| 4046.851 | 28297.473 | ev | 3.5 | 3593.882 | od | 4.5 | 0.67 | ± 0.08 | -1.88 |
| 4048.364 | 30637.157 | ev | 2.5 | 5942.798 | od | 3.5 | 3.03 | ± 0.23 | -1.35 |
| 4049.362 | 29807.078 | ev | 3.5 | 5118.806 | od | 2.5 | 2.5 | ± 0.3 | -1.31 |
| 4051.424 | 29794.517 | ev | 3.5 | 5118.806 | od | 2.5 | 9.5 | ± 0.8 | -0.73 |
| 4051.990 | 30637.157 | ev | 2.5 | 5964.896 | od | 3.5 | 11.9 | ± 1.1 | -0.76 |
| 4053.503 | 24663.053 | ev | 4.5 | 0.000 | od | 3.5 | 10.0 | ± 0.5 | -0.61 |
| 4058.248 | 28337.814 | ev | 2.5 | 3703.594 | od | 3.5 | 4.86 | ± 0.30 | -1.14 |
| 4058.751 | 28334.756 | ev | 4.5 | 3703.594 | od | 3.5 | 2.29 | ± 0.22 | -1.25 |
| 4061.416 | 27249.669 | ev | 2.5 | 2634.666 | od | 2.5 | 1.43 | ± 0.23 | -1.67 |
| 4062.937 | 27187.047 | ev | 3.5 | 2581.257 | od | 4.5 | 8.4 | ± 0.5 | -0.78 |
| 4063.045 | 29449.778 | ev | 1.5 | 4844.644 | od | 1.5 | 1.89 | ± 0.21 | -1.73 |
| 4063.920 | 28345.313 | ev | 0.5 | 3745.475 | od | 1.5 | 16.8 | ± 1.0 | -1.08 |
| 4064.798 | 30245.878 | ev | 4.5 | 5651.357 | od | 5.5 | 0.58 | ± 0.04 | -1.84 |
| 4064.904 | 28297.473 | ev | 3.5 | 3703.594 | od | 3.5 | 3.23 | ± 0.21 | -1.19 |
| 4068.836 | 30245.878 | ev | 4.5 | 5675.763 | od | 4.5 | 27.3 | ± 1.5 | -0.17 |
| 4071.072 | 29994.041 | ev | 2.5 | 5437.422 | od | 3.5 | 6.6 | ± 0.7 | -1.00 |
| 4071.775 | 27187.047 | ev | 3.5 | 2634.666 | od | 2.5 | 31.5 | ± 1.6 | -0.20 |
| 4072.918 | 27187.047 | ev | 3.5 | 2641.559 | od | 3.5 | 11.5 | ± 0.6 | -0.64 |
| 4073.474 | 28396.150 | od | 2.5 | 3854.012 | ev | 3.5 | 109 | ± 6 | 0.21 |
| 4074.644 | 25945.396 | ev | 3.5 | 1410.304 | od | 4.5 | 1.00 | ± 0.09 | -1.70 |
| 4075.546 | 30245.878 | ev | 4.5 | 5716.216 | od | 3.5 | 1.80 | ± 0.20 | -1.35 |
| 4075.700 | 30180.096 | ev | 6.5 | 5651.357 | od | 5.5 | 48.7 | ± 2.5 | 0.23 |
| 4075.847 | 29438.817 | ev | 5.5 | 4910.963 | od | 5.5 | 48.6 | ± 2.6 | 0.16 |
| 4079.672 | 31738.484 | ev | 5.5 | 7233.627 | od | 5.5 | 13.2 | ± 1.1 | -0.40 |
| 4080.438 | 27379.949 | ev | 5.5 | 2879.695 | od | 5.5 | 6.7 | ± 0.4 | -0.70 |
| 4082.098 | 30166.057 | ev | 3.5 | 5675.763 | od | 4.5 | 1.04 | ± 0.11 | -1.68 |
| 4083.222 | 30134.910 | ev | 5.5 | 5651.357 | od | 5.5 | 62 | ± 3 | 0.27 |
| 4085.236 | 29908.904 | ev | 4.5 | 5437.422 | od | 3.5 | 28.1 | ± 1.8 | -0.15 |
| 4086.433 | 28730.712 | ev | 3.5 | 4266.397 | od | 3.5 | 5.3 | ± 0.6 | -0.97 |
| 4087.362 | 28725.148 | ev | 4.5 | 4266.397 | od | 3.5 | 7.5 | ± 0.4 | -0.73 |
| 4088.852 | 30166.057 | ev | 3.5 | 5716.216 | od | 3.5 | 17.5 | ± 1.1 | -0.46 |
| 4088.997 | 27812.398 | ev | 2.5 | 3363.427 | od | 2.5 | 4.62 | ± 0.27 | -1.16 |
| 4089.148 | 27811.496 | ev | 3.5 | 3363.427 | od | 2.5 | 1.38 | ± 0.11 | -1.56 |
| 4089.738 | 31738.484 | ev | 5.5 | 7293.938 | od | 6.5 | 6.1 | ± 0.6 | -0.74 |
| 4091.046 | 29281.374 | ev | 2.5 | 4844.644 | od | 1.5 | 2.70 | ± 0.23 | -1.39 |
| 4092.075 | 28634.516 | ev | 5.5 | 4203.934 | od | 6.5 | 2.17 | ± 0.25 | -1.18 |
| 4092.715 | 30245.878 | ev | 4.5 | 5819.113 | od | 4.5 | 6.3 | ± 0.4 | -0.80 |
| 4093.956 | 28685.758 | ev | 2.5 | 4266.397 | od | 3.5 | 11.7 | ± 0.8 | -0.75 |
| 4098.985 | 30065.164 | ev | 3.5 | 5675.763 | od | 4.5 | 9.0 | ± 0.7 | -0.74 |
| 4101.769 | 31340.393 | od | 6.5 | 6967.547 | ev | 6.5 | 31.4 | ± 1.7 | 0.05 |
| 4106.133 | 30166.057 | ev | 3.5 | 5819.113 | od | 4.5 | 6.6 | ± 0.4 | -0.87 |
| 4106.907 | 28337.814 | ev | 2.5 | 3995.460 | od | 3.5 | 5.3 | ± 0.3 | -1.10 |
| 4107.177 | 27934.638 | ev | 4.5 | 3593.882 | od | 4.5 | 0.35 | ± 0.06 | -2.05 |
| 4107.423 | 28334.756 | ev | 4.5 | 3995.460 | od | 3.5 | 28.9 | ± 1.5 | -0.14 |
| 4108.828 | 29449.778 | ev | 1.5 | 5118.806 | od | 2.5 | 0.40 | ± 0.06 | -2.39 |
| 4109.539 | 27835.233 | ev | 1.5 | 3508.470 | od | 0.5 | 5.1 | ± 0.4 | -1.28 |
| 4110.835 | 26900.354 | ev | 3.5 | 2581.257 | od | 4.5 | 4.07 | ± 0.28 | -1.08 |
| 4111.393 | 30134.910 | ev | 5.5 | 5819.113 | od | 4.5 | 14.6 | ± 1.0 | -0.35 |

| | | | | | | | | |
|---|---|---|---|---|---|---|---|---|
| 4111.922 | 30702.610 | ev | 4.5 | 6389.942 | od | 4.5 | 4.35 ± 0.27 | -0.96 |
| 4113.544 | 30245.878 | ev | 4.5 | 5942.798 | od | 3.5 | 1.91 ± 0.20 | -1.32 |
| 4113.725 | 28297.473 | ev | 3.5 | 3995.460 | od | 3.5 | 7.1 ± 0.5 | -0.84 |
| 4117.288 | 30245.878 | ev | 4.5 | 5964.896 | od | 3.5 | 13.9 ± 1.0 | -0.45 |
| 4117.768 | 26841.384 | ev | 4.5 | 2563.233 | od | 5.5 | 0.23 ± 0.04 | -2.23 |
| 4117.823 | 29994.041 | ev | 2.5 | 5716.216 | od | 3.5 | 3.4 ± 0.4 | -1.28 |
| 4117.985 | 30245.878 | ev | 4.5 | 5969.007 | od | 5.5 | 4.3 ± 0.4 | -0.96 |
| 4118.143 | 29892.677 | od | 3.5 | 5616.739 | ev | 4.5 | 66 ± 4 | 0.13 |
| 4119.008 | 28730.712 | ev | 3.5 | 4459.872 | od | 3.5 | 14.7 ± 0.8 | -0.53 |
| 4119.883 | 26900.354 | ev | 3.5 | 2634.666 | od | 2.5 | 18.7 ± 1.2 | -0.42 |
| 4119.953 | 28725.148 | ev | 4.5 | 4459.872 | od | 3.5 | 2.30 ± 0.19 | -1.23 |
| 4120.827 | 26841.384 | ev | 4.5 | 2581.257 | od | 4.5 | 16.6 ± 0.9 | -0.37 |
| 4121.266 | 29908.904 | ev | 4.5 | 5651.357 | od | 5.5 | 0.39 ± 0.04 | -2.00 |
| 4121.591 | 29166.597 | ev | 4.5 | 4910.963 | od | 5.5 | 3.08 ± 0.20 | -1.11 |
| 4123.220 | 30065.164 | ev | 3.5 | 5819.113 | od | 4.5 | 15.8 ± 1.3 | -0.49 |
| 4123.869 | 31155.623 | ev | 6.5 | 6913.392 | od | 6.5 | 68 ± 4 | 0.39 |
| 4124.787 | 29750.547 | od | 5.5 | 5513.709 | ev | 5.5 | 25.7 ± 1.4 | -0.10 |
| 4125.416 | 29908.904 | ev | 4.5 | 5675.763 | od | 4.5 | 1.20 ± 0.11 | -1.51 |
| 4125.773 | 27934.638 | ev | 4.5 | 3703.594 | od | 3.5 | 2.67 ± 0.18 | -1.17 |
| 4126.652 | 28685.758 | ev | 2.5 | 4459.872 | od | 3.5 | 7.4 ± 0.6 | -0.95 |
| 4127.099 | 30166.057 | ev | 3.5 | 5942.798 | od | 3.5 | 3.55 ± 0.29 | -1.14 |
| 4127.364 | 29735.413 | od | 4.5 | 5513.709 | ev | 5.5 | 79 ± 4 | 0.31 |
| 4127.748 | 28730.712 | ev | 3.5 | 4511.257 | od | 2.5 | 22.4 ± 1.2 | -0.34 |
| 4128.061 | 27811.496 | ev | 3.5 | 3593.882 | od | 4.5 | 11.8 ± 1.1 | -0.62 |
| 4128.360 | 31738.484 | ev | 5.5 | 7522.622 | od | 5.5 | 16.9 ± 1.7 | -0.28 |
| 4129.174 | 30180.096 | ev | 6.5 | 5969.007 | od | 5.5 | 4.6 ± 0.4 | -0.78 |
| 4130.705 | 28725.148 | ev | 4.5 | 4523.033 | od | 4.5 | 24.9 ± 1.3 | -0.20 |
| 4132.315 | 29908.904 | ev | 4.5 | 5716.216 | od | 3.5 | 8.1 ± 0.7 | -0.68 |
| 4132.626 | 30829.124 | ev | 3.5 | 6638.258 | od | 4.5 | 8.9 ± 1.0 | -0.74 |
| 4135.424 | 28685.758 | ev | 2.5 | 4511.257 | od | 2.5 | 21.2 ± 1.4 | -0.49 |
| 4136.750 | 29449.778 | ev | 1.5 | 5283.029 | od | 0.5 | 5.4 ± 0.5 | -1.25 |
| 4136.895 | 30134.910 | ev | 5.5 | 5969.007 | od | 5.5 | 5.9 ± 0.4 | -0.74 |
| 4137.466 | 29281.374 | ev | 2.5 | 5118.806 | od | 2.5 | 21.1 ± 1.3 | -0.49 |
| 4137.645 | 28327.071 | od | 5.5 | 4165.550 | ev | 4.5 | 82 ± 4 | 0.40 |
| 4140.747 | 28345.313 | ev | 0.5 | 4201.893 | od | 1.5 | 12.5 ± 0.7 | -1.19 |
| 4142.034 | 28337.814 | ev | 2.5 | 4201.893 | od | 1.5 | 1.82 ± 0.17 | -1.55 |
| 4142.397 | 29750.547 | od | 5.5 | 5616.739 | ev | 4.5 | 54.3 ± 2.8 | 0.22 |
| 4142.825 | 29807.078 | ev | 3.5 | 5675.763 | od | 4.5 | 15.6 ± 0.9 | -0.49 |
| 4144.362 | 30065.164 | ev | 3.5 | 5942.798 | od | 3.5 | 3.40 ± 0.23 | -1.15 |
| 4144.492 | 27975.619 | od | 4.5 | 3854.012 | ev | 3.5 | 17.9 ± 1.0 | -0.34 |
| 4144.847 | 30637.157 | ev | 2.5 | 6517.619 | od | 2.5 | 7.6 ± 0.8 | -0.93 |
| 4144.996 | 29735.413 | od | 4.5 | 5616.739 | ev | 4.5 | 49.1 ± 2.6 | 0.10 |
| 4145.486 | 30637.157 | ev | 2.5 | 6521.332 | od | 1.5 | 1.56 ± 0.14 | -1.62 |
| 4146.232 | 28634.516 | ev | 5.5 | 4523.033 | od | 4.5 | 24.6 ± 1.4 | -0.12 |
| 4148.162 | 30065.164 | ev | 3.5 | 5964.896 | od | 3.5 | 8.4 ± 0.6 | -0.76 |
| 4149.143 | 32372.621 | od | 4.5 | 8278.054 | ev | 5.5 | 3.72 ± 0.27 | -1.02 |
| 4149.781 | 29807.078 | ev | 3.5 | 5716.216 | od | 3.5 | 8.2 ± 0.6 | -0.77 |
| 4149.966 | 29908.904 | ev | 4.5 | 5819.113 | od | 4.5 | 34.0 ± 2.1 | -0.06 |
| 4150.404 | 30637.157 | ev | 2.5 | 6549.908 | od | 2.5 | 0.97 ± 0.10 | -1.82 |
| 4151.970 | 29591.873 | od | 6.5 | 5513.709 | ev | 5.5 | 75 ± 4 | 0.43 |
| 4153.126 | 25945.396 | ev | 3.5 | 1873.934 | od | 3.5 | 7.6 ± 0.4 | -0.80 |
| 4153.406 | 29994.041 | ev | 2.5 | 5924.204 | od | 1.5 | 2.03 ± 0.21 | -1.50 |
| 4160.106 | 28297.473 | ev | 3.5 | 4266.397 | od | 3.5 | 3.6 ± 0.3 | -1.12 |
| 4160.440 | 29994.041 | ev | 2.5 | 5964.896 | od | 3.5 | 0.59 ± 0.06 | -2.04 |
| 4162.872 | 28337.814 | ev | 2.5 | 4322.708 | od | 2.5 | 1.54 ± 0.15 | -1.62 |
| 4165.599 | 31340.393 | od | 6.5 | 7341.007 | ev | 5.5 | 90 ± 5 | 0.52 |
| 4165.850 | 32802.165 | ev | 5.5 | 8804.224 | od | 4.5 | 6.6 ± 0.7 | -0.68 |
| 4166.649 | 28730.712 | ev | 3.5 | 4737.373 | od | 2.5 | 10.3 ± 0.6 | -0.67 |
| 4167.582 | 29807.078 | ev | 3.5 | 5819.113 | od | 4.5 | 2.39 ± 0.14 | -1.30 |
| 4169.766 | 29794.517 | ev | 3.5 | 5819.113 | od | 4.5 | 13.6 ± 1.0 | -0.55 |
| 4169.877 | 28297.473 | ev | 3.5 | 4322.708 | od | 2.5 | 21.2 ± 1.1 | -0.35 |
| 4171.384 | 29908.904 | ev | 4.5 | 5942.798 | od | 3.5 | 3.2 ± 0.4 | -1.08 |
| 4172.152 | 26841.384 | ev | 4.5 | 2879.695 | od | 5.5 | 3.26 ± 0.19 | -1.07 |
| 4174.470 | 28685.758 | ev | 2.5 | 4737.373 | od | 2.5 | 4.9 ± 0.3 | -1.11 |
| 4175.233 | 29908.904 | ev | 4.5 | 5964.896 | od | 3.5 | 4.2 ± 0.3 | -0.95 |
| 4176.076 | 27934.638 | ev | 4.5 | 3995.460 | od | 3.5 | 3.84 ± 0.26 | -1.00 |
| 4179.075 | 31155.623 | ev | 6.5 | 7233.627 | od | 5.5 | 3.11 ± 0.22 | -0.94 |
| 4185.331 | 27249.669 | ev | 2.5 | 3363.427 | od | 2.5 | 16.4 ± 0.9 | -0.59 |
| 4187.322 | 28334.756 | ev | 4.5 | 4459.872 | od | 3.5 | 14.2 ± 0.8 | -0.43 |
| 4189.183 | 29807.078 | ev | 3.5 | 5942.798 | od | 3.5 | 3.45 ± 0.22 | -1.14 |
| 4189.638 | 31155.623 | ev | 6.5 | 7293.938 | od | 6.5 | 3.41 ± 0.18 | -0.90 |
| 4192.754 | 29281.374 | ev | 2.5 | 5437.422 | od | 3.5 | 4.2 ± 0.3 | -1.18 |
| 4193.065 | 29807.078 | ev | 3.5 | 5964.896 | od | 3.5 | 6.6 ± 0.6 | -0.85 |
| 4193.283 | 32372.621 | od | 4.5 | 8531.678 | ev | 3.5 | 41.7 ± 2.4 | 0.04 |
| 4193.871 | 28297.473 | ev | 3.5 | 4459.872 | od | 3.5 | 15.6 ± 0.8 | -0.48 |
| 4195.276 | 29794.517 | ev | 3.5 | 5964.896 | od | 3.5 | 2.80 ± 0.23 | -1.23 |
| 4195.815 | 28337.814 | ev | 2.5 | 4511.257 | od | 2.5 | 6.8 ± 0.4 | -0.97 |
| 4196.332 | 27187.047 | ev | 3.5 | 3363.427 | od | 2.5 | 21.6 ± 1.2 | -0.34 |
| 4197.510 | 27812.398 | ev | 2.5 | 3995.460 | od | 3.5 | 5.8 ± 0.4 | -1.03 |
| 4197.669 | 27811.496 | ev | 3.5 | 3995.460 | od | 3.5 | 9.1 ± 0.6 | -0.72 |
| 4198.429 | 28334.756 | ev | 4.5 | 4523.033 | od | 4.5 | 8.2 ± 0.5 | -0.66 |
| 4198.721 | 27975.619 | od | 4.5 | 4165.550 | ev | 4.5 | 18.0 ± 1.0 | -0.32 |
| 4202.712 | 29438.817 | ev | 5.5 | 5651.357 | od | 5.5 | 0.57 ± 0.09 | -1.74 |
| 4202.931 | 28297.473 | ev | 3.5 | 4511.257 | od | 2.5 | 22.0 ± 1.2 | -0.33 |
| 4202.958 | 27379.949 | ev | 5.5 | 3593.882 | od | 4.5 | 10.8 ± 0.6 | -0.46 |
| 4204.717 | 30166.057 | ev | 3.5 | 6389.942 | od | 4.5 | 2.47 ± 0.20 | -1.28 |
| 4205.013 | 28297.473 | ev | 3.5 | 4523.033 | od | 4.5 | 0.32 ± 0.05 | -2.17 |
| 4207.028 | 29438.817 | ev | 5.5 | 5675.763 | od | 4.5 | 0.107 ± 0.016 | -2.47 |
| 4209.406 | 29263.338 | od | 5.5 | 5513.709 | ev | 5.5 | 7.8 ± 0.5 | -0.61 |
| 4210.233 | 30134.910 | ev | 5.5 | 6389.942 | od | 4.5 | 1.15 ± 0.06 | -1.44 |
| 4213.035 | 29166.597 | ev | 4.5 | 5437.422 | od | 3.5 | 4.30 ± 0.24 | -0.94 |
| 4214.033 | 28634.516 | ev | 5.5 | 4910.963 | od | 5.5 | 11.1 ± 0.8 | -0.45 |
| 4222.597 | 24663.053 | ev | 4.5 | 987.611 | od | 4.5 | 26.7 ± 1.4 | -0.15 |
| 4223.881 | 27934.638 | ev | 4.5 | 4266.397 | od | 3.5 | 4.5 ± 0.3 | -0.92 |
| 4227.747 | 29263.338 | od | 5.5 | 5616.739 | ev | 4.5 | 25.0 ± 1.3 | -0.10 |
| 4228.295 | 30702.610 | ev | 4.5 | 7059.072 | od | 4.5 | 5.0 ± 0.3 | -0.87 |
| 4230.119 | 27835.233 | ev | 1.5 | 4201.893 | od | 1.5 | 3.90 ± 0.24 | -1.38 |
| 4230.180 | 31155.623 | ev | 6.5 | 7522.622 | od | 5.5 | 0.96 ± 0.15 | -1.44 |
| 4231.327 | 30829.124 | ev | 3.5 | 7202.529 | od | 2.5 | 0.51 ± 0.04 | -1.96 |
| 4232.561 | 29438.817 | ev | 5.5 | 5819.113 | od | 4.5 | 5.3 ± 0.3 | -0.77 |
| 4233.959 | 28730.712 | ev | 3.5 | 5118.806 | od | 2.5 | 1.68 ± 0.15 | -1.44 |
| 4234.210 | 27812.398 | ev | 2.5 | 4201.893 | od | 1.5 | 14.3 ± 0.9 | -0.64 |
| 4234.728 | 30245.878 | ev | 4.5 | 6638.258 | od | 4.5 | 4.8 ± 0.3 | -0.89 |
| 4236.016 | 28337.814 | ev | 2.5 | 4737.373 | od | 2.5 | 15.0 ± 0.8 | -0.62 |
| 4236.354 | 32372.621 | od | 4.5 | 8774.064 | ev | 4.5 | 10.3 ± 0.9 | -0.56 |
| 4238.553 | 27379.949 | ev | 5.5 | 3793.634 | od | 6.5 | 0.59 ± 0.07 | -1.72 |
| 4239.909 | 27432.782 | od | 4.5 | 3854.012 | ev | 3.5 | 36.8 ± 1.9 | 0.00 |
| 4242.359 | 29281.374 | ev | 2.5 | 5716.216 | od | 3.5 | 0.60 ± 0.07 | -2.01 |
| 4242.720 | 25945.396 | ev | 3.5 | 2382.246 | od | 4.5 | 11.9 ± 0.6 | -0.59 |
| 4245.973 | 27811.496 | ev | 3.5 | 4266.397 | od | 3.5 | 15.4 ± 0.9 | -0.48 |
| 4246.713 | 25681.488 | ev | 1.5 | 2140.492 | od | 0.5 | 25.6 ± 1.3 | -0.56 |
| 4247.447 | 26900.354 | ev | 3.5 | 3363.427 | od | 2.5 | 2.07 ± 0.18 | -1.35 |
| 4248.671 | 29043.854 | od | 6.5 | 5513.709 | ev | 5.5 | 40.3 ± 2.1 | 0.18 |
| 4250.692 | 29908.904 | ev | 4.5 | 6389.942 | od | 4.5 | 1.23 ± 0.09 | -1.48 |
| 4251.855 | 27835.233 | ev | 1.5 | 4322.708 | od | 2.5 | 7.3 ± 0.6 | -1.10 |
| 4253.362 | 27249.669 | ev | 2.5 | 3745.475 | od | 1.5 | 17.4 ± 0.9 | -0.55 |
| 4254.000 | 28345.313 | ev | 0.5 | 4844.644 | od | 1.5 | 1.38 ± 0.16 | -2.13 |
| 4254.728 | 30134.910 | ev | 5.5 | 6638.258 | od | 4.5 | 0.88 ± 0.08 | -1.54 |
| 4255.358 | 28337.814 | ev | 2.5 | 4844.644 | od | 1.5 | 3.31 ± 0.23 | -1.27 |
| 4255.782 | 29166.597 | ev | 4.5 | 5675.763 | od | 4.5 | 27.0 ± 1.4 | -0.13 |
| 4255.989 | 27812.398 | ev | 2.5 | 4322.708 | od | 2.5 | 6.6 ± 0.6 | -0.97 |
| 4256.152 | 27811.496 | ev | 3.5 | 4322.708 | od | 2.5 | 8.3 ± 0.6 | -0.74 |
| 4256.702 | 25359.686 | ev | 2.5 | 1873.934 | od | 3.5 | 0.138 ± 0.013 | -2.65 |
| 4257.119 | 27187.047 | ev | 3.5 | 3703.594 | od | 3.5 | 3.33 ± 0.24 | -1.14 |
| 4258.394 | 29994.041 | ev | 2.5 | 6517.619 | od | 2.5 | 3.8 ± 0.3 | -1.20 |
| 4259.068 | 29994.041 | ev | 2.5 | 6521.332 | od | 1.5 | 2.60 ± 0.20 | -1.37 |
| 4259.594 | 29438.817 | ev | 5.5 | 5969.007 | od | 5.5 | 0.73 ± 0.08 | -1.62 |
| 4259.744 | 30702.610 | ev | 4.5 | 7233.627 | od | 5.5 | 7.5 ± 0.5 | -0.69 |
| 4264.368 | 30702.610 | ev | 4.5 | 7259.075 | od | 3.5 | 10.2 ± 0.7 | -0.56 |
| 4269.176 | 29807.078 | ev | 3.5 | 6389.942 | od | 4.5 | 1.78 ± 0.17 | -1.41 |
| 4270.184 | 27934.638 | ev | 4.5 | 4523.033 | od | 4.5 | 19.6 ± 1.2 | -0.27 |
| 4278.859 | 25945.396 | ev | 3.5 | 2581.257 | od | 4.5 | 6.3 ± 0.3 | -0.86 |
| 4279.941 | 30637.157 | ev | 2.5 | 7278.922 | od | 1.5 | 0.57 ± 0.07 | -2.03 |
| 4280.136 | 29281.374 | ev | 2.5 | 5924.204 | od | 1.5 | 12.6 ± 0.7 | -0.68 |
| 4280.987 | 27812.398 | ev | 2.5 | 4459.872 | od | 3.5 | 6.9 ± 0.4 | -0.95 |
| 4281.153 | 27811.496 | ev | 3.5 | 4459.872 | od | 3.5 | 4.8 ± 0.3 | -0.98 |
| 4283.546 | 29281.374 | ev | 2.5 | 5942.798 | od | 3.5 | 3.30 ± 0.23 | -1.26 |
| 4287.144 | 32372.621 | od | 4.5 | 9053.629 | ev | 3.5 | 1.72 ± 0.15 | -1.32 |
| 4287.606 | 29281.374 | ev | 2.5 | 5964.896 | od | 3.5 | 1.49 ± 0.13 | -1.61 |
| 4288.663 | 25945.396 | ev | 3.5 | 2634.666 | od | 2.5 | 5.25 ± 0.28 | -0.94 |
| 4289.447 | 26900.354 | ev | 3.5 | 3593.882 | od | 4.5 | 8.0 ± 0.6 | -0.75 |
| 4289.932 | 25945.396 | ev | 3.5 | 2641.559 | od | 3.5 | 58.4 ± 3.0 | 0.11 |
| 4290.428 | 27812.398 | ev | 2.5 | 4511.257 | od | 2.5 | 1.63 ± 0.15 | -1.57 |
| 4290.594 | 27811.496 | ev | 3.5 | 4511.257 | od | 2.5 | 1.51 ± 0.10 | -1.48 |
| 4291.874 | 28730.712 | ev | 3.5 | 5437.422 | od | 3.5 | 0.238 ± 0.028 | -2.28 |
| 4292.580 | 29807.078 | ev | 3.5 | 6517.619 | od | 2.5 | 8.3 ± 0.7 | -0.74 |
| 4292.764 | 27811.496 | ev | 3.5 | 4523.033 | od | 4.5 | 6.3 ± 0.4 | -0.85 |
| 4292.900 | 28725.148 | ev | 4.5 | 5437.422 | od | 3.5 | 1.61 ± 0.13 | -1.35 |
| 4294.897 | 29794.517 | ev | 3.5 | 6517.619 | od | 2.5 | 1.80 ± 0.13 | -1.40 |
| 4296.051 | 29908.904 | ev | 4.5 | 6638.258 | od | 4.5 | 5.1 ± 0.5 | -0.85 |
| 4296.681 | 27432.782 | od | 4.5 | 4165.550 | ev | 4.5 | 53.2 ± 2.7 | 0.17 |
| 4296.779 | 30180.096 | ev | 6.5 | 6913.392 | od | 6.5 | 23.5 ± 1.3 | -0.04 |
| 4299.087 | 27249.669 | ev | 2.5 | 3995.460 | od | 3.5 | 4.71 ± 0.28 | -1.11 |
| 4299.357 | 24663.053 | ev | 4.5 | 1410.304 | od | 4.5 | 9.9 ± 0.5 | -0.56 |
| 4300.328 | 26841.384 | ev | 4.5 | 3593.882 | od | 4.5 | 22.1 ± 1.1 | -0.21 |
| 4300.863 | 29794.517 | ev | 3.5 | 6549.908 | od | 2.5 | 3.80 ± 0.26 | -1.07 |
| 4304.717 | 29166.597 | ev | 4.5 | 5942.798 | od | 3.5 | 7.4 ± 0.4 | -0.69 |
| 4305.140 | 30134.910 | ev | 5.5 | 6913.392 | od | 6.5 | 20.3 ± 1.2 | -0.17 |
| 4305.605 | 28337.814 | ev | 2.5 | 5118.806 | od | 2.5 | 2.93 ± 0.26 | -1.31 |
| 4306.726 | 27378.515 | od | 5.5 | 4165.550 | ev | 4.5 | 24.2 ± 1.3 | -0.09 |
| 4309.580 | 29166.597 | ev | 4.5 | 5969.007 | od | 5.5 | 4.7 ± 0.3 | -0.88 |
| 4309.735 | 26900.354 | ev | 3.5 | 3703.594 | od | 3.5 | 15.5 ± 1.0 | -0.46 |
| 4310.696 | 27187.047 | ev | 3.5 | 3995.460 | od | 3.5 | 5.1 ± 0.4 | -0.94 |
| 4311.585 | 30245.878 | ev | 4.5 | 7059.072 | od | 4.5 | 7.5 ± 0.5 | -0.68 |
| 4312.853 | 30702.610 | ev | 4.5 | 7522.622 | od | 5.5 | 0.89 ± 0.09 | -1.60 |
| 4313.099 | 28297.473 | ev | 3.5 | 5118.806 | od | 2.5 | 2.26 ± 0.14 | -1.30 |
| 4313.592 | 27379.949 | ev | 5.5 | 4203.934 | od | 6.5 | 0.77 ± 0.06 | -1.59 |
| 4314.932 | 29807.078 | ev | 3.5 | 6638.258 | od | 4.5 | 3.04 ± 0.22 | -1.17 |
| 4317.273 | 29794.517 | ev | 3.5 | 6638.258 | od | 4.5 | 1.29 ± 0.23 | -1.54 |
| 4320.719 | 26841.384 | ev | 4.5 | 3703.594 | od | 3.5 | 14.5 ± 0.8 | -0.39 |
| 4324.785 | 30829.124 | ev | 3.5 | 7713.089 | od | 4.5 | 16.7 ± 1.3 | -0.43 |
| 4326.479 | 30166.057 | ev | 3.5 | 7059.072 | od | 4.5 | 0.76 ± 0.08 | -1.77 |
| 4330.441 | 25681.488 | ev | 1.5 | 2595.644 | od | 1.5 | 13.0 ± 0.7 | -0.83 |
| 4331.757 | 32802.165 | ev | 5.5 | 9723.335 | od | 4.5 | 13.1 ± 0.9 | -0.35 |
| 4332.319 | 30134.910 | ev | 5.5 | 7059.072 | od | 4.5 | 0.56 ± 0.06 | -1.72 |
| 4332.472 | 27812.398 | ev | 2.5 | 4737.373 | od | 2.5 | 1.29 ± 0.11 | -1.66 |
| 4332.641 | 27811.496 | ev | 3.5 | 4737.373 | od | 2.5 | 2.54 ± 0.21 | -1.24 |
| 4332.703 | 28725.148 | ev | 4.5 | 5651.357 | od | 5.5 | 9.2 ± 0.5 | -0.59 |
| 4337.291 | 28725.148 | ev | 4.5 | 5675.763 | od | 4.5 | 1.03 ± 0.07 | -1.54 |
| 4337.387 | 29438.817 | ev | 5.5 | 6389.942 | od | 4.5 | 1.19 ± 0.11 | -1.40 |
| 4337.594 | 27249.669 | ev | 2.5 | 4201.893 | od | 1.5 | 1.87 ± 0.20 | -1.50 |
| 4337.773 | 25681.488 | ev | 1.5 | 2634.666 | od | 2.5 | 43.5 ± 2.2 | -0.31 |
| 4342.135 | 27934.638 | ev | 4.5 | 4910.963 | od | 5.5 | 1.74 ± 0.19 | -1.31 |
| 4343.866 | 28730.712 | ev | 3.5 | 5716.216 | od | 3.5 | 2.42 ± 0.19 | -1.26 |
| 4344.917 | 28725.148 | ev | 4.5 | 5716.216 | od | 3.5 | 1.18 ± 0.13 | -1.48 |
| 4345.453 | 30065.164 | ev | 3.5 | 7059.072 | od | 4.5 | 2.70 ± 0.22 | -1.21 |
| 4348.585 | 30702.610 | ev | 4.5 | 7713.089 | od | 4.5 | 1.25 ± 0.10 | -1.45 |
| 4349.768 | 27249.669 | ev | 2.5 | 4266.397 | od | 3.5 | 11.1 ± 0.6 | -0.73 |
| 4349.789 | 28634.516 | ev | 5.5 | 5651.357 | od | 5.5 | 14.0 ± 0.9 | -0.32 |
| 4352.707 | 27812.398 | ev | 2.5 | 4844.644 | od | 1.5 | 32.9 ± 1.9 | -0.25 |
| 4355.923 | 30829.124 | ev | 3.5 | 7878.328 | od | 3.5 | 0.60 ± 0.07 | -1.86 |
| 4356.744 | 30180.096 | ev | 6.5 | 7233.627 | od | 5.5 | 1.77 ± 0.18 | -1.15 |
| 4359.064 | 31738.484 | ev | 5.5 | 8804.224 | od | 4.5 | 5.5 ± 0.9 | -0.72 |
| 4360.169 | 29449.778 | ev | 1.5 | 6521.332 | od | 1.5 | 9.5 ± 0.7 | -0.97 |
| 4361.652 | 27187.047 | ev | 3.5 | 4266.397 | od | 3.5 | 4.9 ± 0.3 | -0.95 |
| 4364.653 | 26900.354 | ev | 3.5 | 3995.460 | od | 3.5 | 29.5 ± 1.9 | -0.17 |
| 4365.341 | 30134.910 | ev | 5.5 | 7233.627 | od | 5.5 | 0.62 ± 0.10 | -1.67 |
| 4365.511 | 28337.814 | ev | 2.5 | 5437.422 | od | 3.5 | 2.56 ± 0.19 | -1.36 |
| 4366.094 | 28334.756 | ev | 4.5 | 5437.422 | od | 3.5 | 0.229 ± 0.028 | -2.18 |
| 4368.226 | 30180.096 | ev | 6.5 | 7293.938 | od | 6.5 | 3.40 ± 0.23 | -0.87 |
| 4369.235 | 29892.677 | od | 3.5 | 7011.804 | ev | 4.5 | 5.6 ± 0.4 | -0.89 |
| 4372.394 | 27187.047 | ev | 3.5 | 4322.708 | od | 2.5 | 2.82 ± 0.16 | -1.19 |
| 4373.215 | 28297.473 | ev | 3.5 | 5437.422 | od | 3.5 | 1.78 ± 0.14 | -1.39 |
| 4373.814 | 27379.949 | ev | 5.5 | 4523.033 | od | 4.5 | 9.7 ± 0.6 | -0.48 |
| 4375.170 | 29908.904 | ev | 4.5 | 7059.072 | od | 4.5 | 2.89 ± 0.21 | -1.08 |
| 4375.919 | 26841.384 | ev | 4.5 | 3995.460 | od | 3.5 | 14.9 ± 0.8 | -0.37 |
| 4376.868 | 30134.910 | ev | 5.5 | 7293.938 | od | 6.5 | 0.74 ± 0.06 | -1.59 |
| 4380.053 | 27835.233 | ev | 1.5 | 5010.870 | od | 2.5 | 8.6 ± 0.8 | -1.01 |
| 4381.080 | 30637.157 | ev | 2.5 | 7818.147 | od | 1.5 | 2.63 ± 0.23 | -1.34 |
| 4381.773 | 28634.516 | ev | 5.5 | 5819.113 | od | 4.5 | 4.6 ± 0.3 | -0.80 |
| 4382.165 | 28327.071 | od | 5.5 | 5513.709 | ev | 5.5 | 39.2 ± 2.2 | 0.13 |
| 4384.439 | 27812.398 | ev | 2.5 | 5010.870 | od | 2.5 | 0.66 ± 0.05 | -1.94 |
| 4386.366 | 29994.041 | ev | 2.5 | 7202.529 | od | 2.5 | 4.8 ± 0.3 | -1.08 |
| 4386.696 | 27249.669 | ev | 2.5 | 4459.872 | od | 3.5 | 14.6 ± 0.8 | -0.60 |
| 4386.827 | 24663.053 | ev | 4.5 | 1873.934 | od | 3.5 | 14.3 ± 0.7 | -0.38 |
| 4387.059 | 28730.712 | ev | 3.5 | 5942.798 | od | 3.5 | 4.4 ± 0.4 | -0.99 |
| 4388.005 | 29750.547 | od | 5.5 | 6967.547 | ev | 6.5 | 11.8 ± 0.6 | -0.39 |
| 4390.760 | 32492.038 | ev | 5.5 | 9723.335 | od | 4.5 | 0.42 ± 0.05 | -1.84 |
| 4391.317 | 28730.712 | ev | 3.5 | 5964.896 | od | 3.5 | 1.70 ± 0.14 | -1.41 |
| 4391.659 | 25359.686 | ev | 2.5 | 2595.644 | od | 1.5 | 52.9 ± 2.7 | -0.04 |
| 4393.184 | 28725.148 | ev | 4.5 | 5969.007 | od | 5.5 | 9.3 ± 0.6 | -0.57 |
| 4396.609 | 27249.669 | ev | 2.5 | 4511.257 | od | 2.5 | 2.33 ± 0.16 | -1.39 |
| 4397.183 | 29794.517 | ev | 3.5 | 7059.072 | od | 4.5 | 0.70 ± 0.07 | -1.79 |
| 4397.276 | 29994.041 | ev | 2.5 | 7259.075 | od | 3.5 | 2.87 ± 0.21 | -1.30 |
| 4398.783 | 27187.047 | ev | 3.5 | 4459.872 | od | 3.5 | 7.5 ± 0.5 | -0.76 |
| 4399.200 | 25359.686 | ev | 2.5 | 2634.666 | od | 2.5 | 21.1 ± 1.1 | -0.44 |
| 4399.474 | 29735.413 | od | 4.5 | 7011.804 | ev | 4.5 | 0.94 ± 0.10 | -1.56 |
| 4399.542 | 30245.878 | ev | 4.5 | 7522.622 | od | 5.5 | 1.72 ± 0.13 | -1.30 |
| 4400.535 | 25359.686 | ev | 2.5 | 2641.559 | od | 3.5 | 3.18 ± 0.22 | -1.26 |
| 4400.865 | 27835.233 | ev | 1.5 | 5118.806 | od | 2.5 | 9.6 ± 0.7 | -0.95 |
| 4405.293 | 27812.398 | ev | 2.5 | 5118.806 | od | 2.5 | 1.96 ± 0.19 | -1.47 |
| 4405.468 | 27811.496 | ev | 3.5 | 5118.806 | od | 2.5 | 3.9 ± 0.4 | -1.04 |
| 4407.272 | 28334.756 | ev | 4.5 | 5651.357 | od | 5.5 | 6.6 ± 0.4 | -0.71 |
| 4412.020 | 28334.756 | ev | 4.5 | 5675.763 | od | 4.5 | 4.9 ± 0.3 | -0.84 |
| 4416.900 | 26900.354 | ev | 3.5 | 4266.397 | od | 3.5 | 7.4 ± 0.5 | -0.76 |
| 4418.780 | 29591.873 | od | 6.5 | 6967.547 | ev | 6.5 | 45.1 ± 2.5 | 0.27 |
| 4427.069 | 25945.396 | ev | 3.5 | 3363.427 | od | 2.5 | 9.9 ± 0.6 | -0.63 |
| 4427.208 | 28297.473 | ev | 3.5 | 5716.216 | od | 3.5 | 0.63 ± 0.10 | -1.83 |
| 4427.916 | 26900.354 | ev | 3.5 | 4322.708 | od | 2.5 | 16.7 ± 1.1 | -0.41 |
| 4428.438 | 26841.384 | ev | 4.5 | 4266.397 | od | 3.5 | 9.0 ± 0.5 | -0.58 |
| 4432.912 | 27835.233 | ev | 1.5 | 5283.029 | od | 0.5 | 5.3 ± 0.3 | -1.20 |
| 4433.738 | 29807.078 | ev | 3.5 | 7259.075 | od | 3.5 | 1.84 ± 0.13 | -1.36 |
| 4436.209 | 29794.517 | ev | 3.5 | 7259.075 | od | 3.5 | 1.70 ± 0.13 | -1.40 |
| 4440.110 | 28334.756 | ev | 4.5 | 5819.113 | od | 4.5 | 0.44 ± 0.05 | -1.89 |
| 4443.747 | 27934.638 | ev | 4.5 | 5437.422 | od | 3.5 | 6.3 ± 0.4 | -0.73 |
| 4449.330 | 27379.949 | ev | 5.5 | 4910.963 | od | 5.5 | 31.0 ± 1.8 | 0.04 |
| 4449.633 | 30637.157 | ev | 2.5 | 8169.698 | od | 1.5 | 10.2 ± 1.0 | -0.74 |
| 4450.732 | 27975.619 | od | 4.5 | 5513.709 | ev | 5.5 | 23.0 ± 1.2 | -0.17 |
| 4450.854 | 30637.157 | ev | 2.5 | 8175.863 | od | 2.5 | 2.17 ± 0.23 | -1.41 |
| 4452.504 | 30166.057 | ev | 3.5 | 7713.089 | od | 4.5 | 1.23 ± 0.18 | -1.53 |
| 4453.158 | 27187.047 | ev | 3.5 | 4737.373 | od | 2.5 | 1.94 ± 0.19 | -1.34 |
| 4454.982 | 26900.354 | ev | 3.5 | 4459.872 | od | 3.5 | 3.16 ± 0.24 | -1.12 |
| 4457.768 | 30829.124 | ev | 3.5 | 8402.668 | od | 3.5 | 4.3 ± 0.4 | -0.99 |
| 4458.831 | 28345.313 | ev | 0.5 | 5924.204 | od | 1.5 | 1.49 ± 0.17 | -2.05 |
| 4459.077 | 30166.057 | ev | 3.5 | 7746.185 | od | 2.5 | 1.08 ± 0.10 | -1.59 |
| 4460.207 | 26268.203 | od | 3.5 | 3854.012 | ev | 3.5 | 80 ± 4 | 0.28 |
| 4461.133 | 29750.547 | od | 5.5 | 7341.007 | ev | 5.5 | 23.1 ± 1.3 | -0.08 |
| 4462.032 | 27249.669 | ev | 2.5 | 4844.644 | od | 1.5 | 0.338 ± 0.028 | -2.22 |
| 4465.437 | 29449.778 | ev | 1.5 | 7061.838 | od | 0.5 | 5.8 ± 0.4 | -1.16 |
| 4466.720 | 26841.384 | ev | 4.5 | 4459.872 | od | 3.5 | 0.48 ± 0.05 | -1.84 |
| 4467.073 | 29438.817 | ev | 5.5 | 7059.072 | od | 4.5 | 1.05 ± 0.15 | -1.42 |
| 4468.025 | 27812.398 | ev | 2.5 | 5437.422 | od | 3.5 | 0.50 ± 0.06 | -2.04 |
| 4469.508 | 30245.878 | ev | 4.5 | 7878.328 | od | 3.5 | 0.44 ± 0.04 | -1.88 |
| 4471.241 | 27975.619 | od | 4.5 | 5616.739 | ev | 4.5 | 56.0 ± 2.9 | 0.23 |
| 4472.082 | 28297.473 | ev | 3.5 | 5942.798 | od | 3.5 | 0.81 ± 0.08 | -1.71 |
| 4472.602 | 30065.164 | ev | 3.5 | 7713.089 | od | 4.5 | 6.3 ± 0.6 | -0.82 |
| 4472.715 | 25945.396 | ev | 3.5 | 3593.882 | od | 4.5 | 10.6 ± 0.6 | -0.60 |
| 4479.235 | 30065.164 | ev | 3.5 | 7746.185 | od | 2.5 | 7.3 ± 0.5 | -0.75 |
| 4479.361 | 26841.384 | ev | 4.5 | 4523.033 | od | 4.5 | 12.0 ± 0.7 | -0.44 |
| 4479.419 | 25681.488 | ev | 1.5 | 3363.427 | od | 2.5 | 9.8 ± 0.7 | -0.93 |
| 4483.059 | 30702.610 | ev | 4.5 | 8402.668 | od | 3.5 | 0.84 ± 0.11 | -1.60 |
| 4483.893 | 29263.338 | od | 5.5 | 6967.547 | ev | 6.5 | 34.9 ± 1.8 | 0.10 |
| 4485.515 | 30166.057 | ev | 3.5 | 7878.328 | od | 3.5 | 9.3 ± 0.8 | -0.65 |
| 4486.909 | 24663.053 | ev | 4.5 | 2382.246 | od | 4.5 | 22.0 ± 1.2 | -0.18 |
| 4492.947 | 29591.873 | od | 6.5 | 7341.007 | ev | 5.5 | 1.99 ± 0.13 | -1.07 |
| 4493.554 | 29994.041 | ev | 2.5 | 7746.185 | od | 2.5 | 0.59 ± 0.07 | -1.97 |
| 4494.217 | 28634.516 | ev | 5.5 | 6389.942 | od | 4.5 | 5.9 ± 0.5 | -0.67 |
| 4495.385 | 27249.669 | ev | 2.5 | 5010.870 | od | 2.5 | 7.1 ± 0.5 | -0.89 |
| 4496.256 | 30637.157 | ev | 2.5 | 8402.668 | od | 3.5 | 8.0 ± 0.7 | -0.84 |
| 4499.507 | 27934.638 | ev | 4.5 | 5716.216 | od | 3.5 | 0.52 ± 0.06 | -1.80 |
| 4508.079 | 27187.047 | ev | 3.5 | 5010.870 | od | 2.5 | 3.39 ± 0.27 | -1.08 |
| 4508.721 | 25681.488 | ev | 1.5 | 3508.470 | od | 0.5 | 2.10 ± 0.16 | -1.59 |
| 4509.255 | 29892.677 | od | 3.5 | 7722.285 | ev | 2.5 | 3.48 ± 0.24 | -1.07 |
| 4510.763 | 26900.354 | ev | 3.5 | 4737.373 | od | 2.5 | 0.41 ± 0.05 | -2.00 |
| 4520.404 | 29994.041 | ev | 2.5 | 7878.328 | od | 3.5 | 1.96 ± 0.26 | -1.44 |
| 4523.075 | 26268.203 | od | 3.5 | 4165.550 | ev | 4.5 | 34.2 ± 1.8 | -0.08 |
| 4524.849 | 29807.078 | ev | 3.5 | 7713.089 | od | 4.5 | 1.11 ± 0.09 | -1.56 |
| 4527.348 | 24663.053 | ev | 4.5 | 2581.257 | od | 4.5 | 21.0 ± 1.2 | -0.19 |
| 4527.423 | 29794.517 | ev | 3.5 | 7713.089 | od | 4.5 | 4.6 ± 0.5 | -0.95 |
| 4527.953 | 29281.374 | ev | 2.5 | 7202.529 | od | 2.5 | 2.07 ± 0.16 | -1.42 |
| 4528.473 | 29043.854 | od | 6.5 | 6967.547 | ev | 6.5 | 45.2 ± 2.4 | 0.29 |
| 4534.219 | 29794.517 | ev | 3.5 | 7746.185 | od | 2.5 | 2.58 ± 0.30 | -1.20 |
| 4537.874 | 29908.904 | ev | 4.5 | 7878.328 | od | 3.5 | 4.0 ± 0.4 | -0.90 |
| 4539.745 | 24663.053 | ev | 4.5 | 2641.559 | od | 3.5 | 26.7 ± 1.4 | -0.08 |
| 4544.953 | 25359.686 | ev | 2.5 | 3363.427 | od | 2.5 | 8.8 ± 0.5 | -0.79 |
| 4545.867 | 27934.638 | ev | 4.5 | 5942.798 | od | 3.5 | 1.89 ± 0.17 | -1.23 |
| 4551.291 | 27934.638 | ev | 4.5 | 5969.007 | od | 5.5 | 12.2 ± 0.8 | -0.42 |
| 4554.545 | 25945.396 | ev | 3.5 | 3995.460 | od | 3.5 | 1.86 ± 0.11 | -1.33 |
| 4555.608 | 28334.756 | ev | 4.5 | 6389.942 | od | 4.5 | 1.07 ± 0.09 | -1.48 |
| 4558.068 | 29166.597 | ev | 4.5 | 7233.627 | od | 5.5 | 0.49 ± 0.06 | -1.82 |
| 4558.598 | 26841.384 | ev | 4.5 | 4910.963 | od | 5.5 | 3.38 ± 0.21 | -0.98 |
| 4559.243 | 29449.778 | ev | 1.5 | 7522.458 | od | 0.5 | 1.15 ± 0.19 | -1.84 |
| 4560.280 | 29263.338 | od | 5.5 | 7341.007 | ev | 5.5 | 40.7 ± 2.1 | 0.18 |
| 4560.958 | 27432.782 | od | 4.5 | 5513.709 | ev | 5.5 | 17.8 ± 1.0 | -0.26 |
| 4562.359 | 25766.355 | od | 4.5 | 3854.012 | ev | 3.5 | 51.8 ± 2.6 | 0.21 |
| 4568.036 | 30166.057 | ev | 3.5 | 8280.946 | od | 2.5 | 0.40 ± 0.04 | -2.00 |
| 4569.670 | 32802.165 | ev | 5.5 | 10924.876 | od | 4.5 | 5.8 ± 0.5 | -0.66 |
| 4572.278 | 27378.515 | od | 5.5 | 5513.709 | ev | 5.5 | 43.7 ± 2.2 | 0.22 |
| 4576.799 | 30245.878 | ev | 4.5 | 8402.668 | od | 3.5 | 0.80 ± 0.08 | -1.60 |
| 4582.050 | 29994.041 | ev | 2.5 | 8175.863 | od | 2.5 | 0.52 ± 0.06 | -2.00 |
| 4582.406 | 28337.814 | ev | 2.5 | 6521.332 | od | 1.5 | 6.0 ± 0.4 | -0.95 |
| 4582.499 | 27432.782 | od | 4.5 | 5616.739 | ev | 4.5 | 14.1 ± 0.8 | -0.35 |
| 4588.416 | 28337.814 | ev | 2.5 | 6549.908 | od | 2.5 | 2.93 ± 0.21 | -1.26 |
| 4589.374 | 24663.053 | ev | 4.5 | 2879.695 | od | 5.5 | 0.336 ± 0.027 | -1.97 |
| 4591.116 | 30702.610 | ev | 4.5 | 8927.514 | od | 5.5 | 12.2 ± 1.0 | -0.41 |
| 4593.926 | 27378.515 | od | 5.5 | 5616.739 | ev | 4.5 | 31.1 ± 1.6 | 0.07 |
| 4596.928 | 28297.473 | ev | 3.5 | 6549.908 | od | 2.5 | 1.16 ± 0.12 | -1.53 |
| 4601.374 | 32372.621 | od | 4.5 | 10646.070 | ev | 5.5 | 4.9 ± 0.4 | -0.81 |
| 4604.226 | 29994.041 | ev | 2.5 | 8280.946 | od | 2.5 | 2.45 ± 0.30 | -1.33 |
| 4606.400 | 29043.854 | od | 6.5 | 7341.007 | ev | 5.5 | 19.5 ± 1.0 | -0.06 |
| 4613.033 | 28730.712 | ev | 3.5 | 7059.072 | od | 4.5 | 2.34 ± 0.22 | -1.22 |
| 4613.528 | 32372.621 | od | 4.5 | 10703.305 | ev | 4.5 | 0.92 ± 0.13 | -1.53 |
| 4618.930 | 29166.597 | ev | 4.5 | 7522.622 | od | 5.5 | 0.99 ± 0.09 | -1.50 |
| 4621.743 | 30829.124 | ev | 3.5 | 9198.326 | od | 3.5 | 0.86 ± 0.11 | -1.66 |
| 4623.477 | 25945.396 | ev | 3.5 | 4322.708 | od | 2.5 | 0.53 ± 0.03 | -1.87 |
| 4625.290 | 25359.686 | ev | 2.5 | 3745.475 | od | 1.5 | 0.41 ± 0.03 | -2.10 |
| 4628.161 | 25766.355 | od | 4.5 | 4165.550 | ev | 4.5 | 42.5 ± 2.2 | 0.14 |
| 4630.183 | 29994.041 | ev | 2.5 | 8402.668 | od | 3.5 | 0.14 ± 0.03 | -2.55 |
| 4633.601 | 28634.516 | ev | 5.5 | 7059.072 | od | 4.5 | 1.58 ± 0.23 | -1.21 |
| 4636.741 | 27379.949 | ev | 5.5 | 5819.113 | od | 4.5 | 0.66 ± 0.07 | -1.59 |
| 4640.214 | 27934.638 | ev | 4.5 | 6389.942 | od | 4.5 | 0.341 ± 0.027 | -1.96 |
| 4643.774 | 28730.712 | ev | 3.5 | 7202.529 | od | 2.5 | 1.03 ± 0.10 | -1.57 |
| 4644.216 | 29807.078 | ev | 3.5 | 8280.946 | od | 2.5 | 5.1 ± 0.4 | -0.88 |
| 4647.422 | 27187.047 | ev | 3.5 | 5675.763 | od | 4.5 | 2.04 ± 0.18 | -1.28 |
| 4648.513 | 29908.904 | ev | 4.5 | 8402.668 | od | 3.5 | 0.51 ± 0.05 | -1.78 |
| 4654.278 | 25681.488 | ev | 1.5 | 4201.893 | od | 1.5 | 8.7 ± 0.6 | -0.95 |
| 4656.003 | 28730.712 | ev | 3.5 | 7259.075 | od | 3.5 | 2.9 ± 0.4 | -1.12 |
| 4656.178 | 27187.047 | ev | 3.5 | 5716.216 | od | 3.5 | 0.85 ± 0.10 | -1.66 |
| 4657.210 | 28725.148 | ev | 4.5 | 7259.075 | od | 3.5 | 1.59 ± 0.18 | -1.29 |
| 4657.828 | 29281.374 | ev | 2.5 | 7818.147 | od | 1.5 | 2.22 ± 0.17 | -1.36 |
| 4659.101 | 29735.413 | od | 4.5 | 8278.054 | ev | 5.5 | 0.25 ± 0.03 | -2.10 |
| 4659.938 | 29166.597 | ev | 4.5 | 7713.089 | od | 4.5 | 2.16 ± 0.22 | -1.15 |
| 4664.149 | 25945.396 | ev | 3.5 | 4511.257 | od | 2.5 | 0.41 ± 0.07 | -1.97 |
| 4666.713 | 25945.396 | ev | 3.5 | 4523.033 | od | 4.5 | 1.00 ± 0.15 | -1.58 |
| 4666.889 | 27811.496 | ev | 3.5 | 6389.942 | od | 4.5 | 0.41 ± 0.05 | -1.97 |
| 4670.098 | 28685.758 | ev | 2.5 | 7278.922 | od | 1.5 | 3.4 ± 0.3 | -1.18 |
| 4670.627 | 29807.078 | ev | 3.5 | 8402.668 | od | 3.5 | 0.67 ± 0.09 | -1.75 |
| 4670.725 | 26841.384 | ev | 4.5 | 5437.422 | od | 3.5 | 3.08 ± 0.19 | -1.00 |
| 4671.395 | 28634.516 | ev | 5.5 | 7233.627 | od | 5.5 | 0.76 ± 0.13 | -1.53 |
| 4678.600 | 27187.047 | ev | 3.5 | 5819.113 | od | 4.5 | 1.17 ± 0.11 | -1.51 |
| 4679.412 | 25359.686 | ev | 2.5 | 3995.460 | od | 3.5 | 0.57 ± 0.08 | -1.95 |
| 4680.119 | 29892.677 | od | 3.5 | 8531.678 | ev | 3.5 | 12.6 ± 1.0 | -0.48 |
| 4680.442 | 28327.071 | od | 5.5 | 6967.547 | ev | 6.5 | 1.57 ± 0.10 | -1.21 |
| 4680.986 | 32372.621 | od | 4.5 | 11015.579 | ev | 3.5 | 5.4 ± 0.4 | -0.75 |
| 4684.598 | 28634.516 | ev | 5.5 | 7293.938 | od | 6.5 | 10.3 ± 0.8 | -0.39 |
| 4686.770 | 30134.910 | ev | 5.5 | 8804.224 | od | 4.5 | 3.38 ± 0.28 | -0.87 |
| 4687.608 | 29750.547 | od | 5.5 | 8423.672 | ev | 6.5 | 0.85 ± 0.08 | -1.48 |
| 4689.479 | 30245.878 | ev | 4.5 | 8927.514 | od | 5.5 | 2.47 ± 0.17 | -1.09 |
| 4690.160 | 28327.071 | od | 5.5 | 7011.804 | ev | 4.5 | 1.48 ± 0.11 | -1.23 |
| 4690.479 | 29591.873 | od | 6.5 | 8278.054 | ev | 5.5 | 2.15 ± 0.16 | -1.00 |
| 4692.009 | 27249.669 | ev | 2.5 | 5942.798 | od | 3.5 | 2.27 ± 0.21 | -1.35 |
| 4694.320 | 27934.638 | ev | 4.5 | 6638.258 | od | 4.5 | 0.47 ± 0.05 | -1.81 |
| 4694.872 | 27811.496 | ev | 3.5 | 6517.619 | od | 2.5 | 2.91 ± 0.22 | -1.11 |
| 4695.375 | 29994.041 | ev | 2.5 | 8702.444 | od | 1.5 | 1.20 ± 0.09 | -1.62 |
| 4698.886 | 28334.756 | ev | 4.5 | 7059.072 | od | 4.5 | 0.36 ± 0.05 | -1.93 |
| 4702.002 | 27811.496 | ev | 3.5 | 6549.908 | od | 2.5 | 2.34 ± 0.18 | -1.21 |
| 4707.932 | 28327.071 | od | 5.5 | 7092.265 | ev | 5.5 | 1.81 ± 0.17 | -1.14 |
| 4710.198 | 26900.354 | ev | 3.5 | 5675.763 | od | 4.5 | 0.61 ± 0.08 | -1.79 |
| 4714.017 | 30134.910 | ev | 5.5 | 8927.514 | od | 5.5 | 11.7 ± 0.7 | -0.33 |
| 4714.831 | 29735.413 | od | 4.5 | 8531.678 | ev | 3.5 | 7.5 ± 0.5 | -0.60 |
| 4717.881 | 26841.384 | ev | 4.5 | 5651.357 | od | 5.5 | 2.15 ± 0.25 | -1.14 |
| 4719.193 | 26900.354 | ev | 3.5 | 5716.216 | od | 3.5 | 0.29 ± 0.04 | -2.12 |
| 4722.293 | 25681.488 | ev | 1.5 | 4511.257 | od | 2.5 | 2.55 ± 0.17 | -1.47 |
| 4722.746 | 29591.873 | od | 6.5 | 8423.672 | ev | 6.5 | 0.98 ± 0.07 | -1.34 |
| 4723.321 | 26841.384 | ev | 4.5 | 5675.763 | od | 4.5 | 1.25 ± 0.08 | -1.38 |
| 4730.101 | 28337.814 | ev | 2.5 | 7202.529 | od | 2.5 | 6.4 ± 0.4 | -0.89 |
| 4732.366 | 26841.384 | ev | 4.5 | 5716.216 | od | 3.5 | 0.83 ± 0.07 | -1.55 |
| 4733.835 | 29892.677 | od | 3.5 | 8774.064 | ev | 4.5 | 1.77 ± 0.17 | -1.32 |
| 4737.271 | 29892.677 | od | 3.5 | 8789.380 | ev | 2.5 | 29.0 ± 1.8 | -0.11 |
| 4739.147 | 28297.473 | ev | 3.5 | 7202.529 | od | 2.5 | 3.11 ± 0.24 | -1.08 |
| 4742.227 | 26900.354 | ev | 3.5 | 5819.113 | od | 4.5 | 0.68 ± 0.07 | -1.74 |
| 4744.944 | 24663.053 | ev | 4.5 | 3593.882 | od | 4.5 | 1.90 ± 0.12 | -1.19 |
| 4747.260 | 28337.814 | ev | 2.5 | 7278.922 | od | 1.5 | 0.80 ± 0.10 | -1.79 |
| 4749.234 | 30829.124 | ev | 3.5 | 9778.986 | od | 2.5 | 2.05 ± 0.20 | -1.26 |
| 4757.841 | 28725.148 | ev | 4.5 | 7713.089 | od | 4.5 | 5.7 ± 0.4 | -0.71 |
| 4759.927 | 29807.078 | ev | 3.5 | 8804.224 | od | 4.5 | 2.63 ± 0.17 | -1.14 |
| 4762.840 | 27379.949 | ev | 5.5 | 6389.942 | od | 4.5 | 0.194 ± 0.021 | -2.10 |
| 4763.735 | 28327.071 | od | 5.5 | 7341.007 | ev | 5.5 | 0.145 ± 0.027 | -2.23 |
| 4763.912 | 29263.338 | od | 5.5 | 8278.054 | ev | 5.5 | 3.53 ± 0.23 | -0.84 |
| 4768.791 | 27975.619 | od | 4.5 | 7011.804 | ev | 4.5 | 4.7 ± 0.4 | -0.80 |
| 4769.782 | 24663.053 | ev | 4.5 | 3703.594 | od | 3.5 | 0.245 ± 0.025 | -2.08 |
| 4773.941 | 28396.150 | od | 2.5 | 7454.951 | ev | 1.5 | 20.0 ± 1.2 | -0.39 |
| 4775.463 | 25945.396 | ev | 3.5 | 5010.870 | od | 2.5 | 1.20 ± 0.19 | -1.49 |
| 4787.165 | 27975.619 | od | 4.5 | 7092.265 | ev | 5.5 | 1.37 ± 0.15 | -1.33 |
| 4788.230 | 29281.374 | ev | 2.5 | 8402.668 | od | 3.5 | 1.73 ± 0.19 | -1.45 |
| 4789.682 | 26841.384 | ev | 4.5 | 5969.007 | od | 5.5 | 0.93 ± 0.13 | -1.49 |
| 4792.944 | 30637.157 | ev | 2.5 | 9778.986 | od | 2.5 | 0.87 ± 0.13 | -1.75 |
| 4795.184 | 25359.686 | ev | 2.5 | 4511.257 | od | 2.5 | 1.89 ± 0.22 | -1.41 |
| 4795.554 | 28725.148 | ev | 4.5 | 7878.328 | od | 3.5 | 2.13 ± 0.25 | -1.13 |
| 4797.343 | 29892.677 | od | 3.5 | 9053.629 | ev | 3.5 | 0.49 ± 0.05 | -1.87 |
| 4797.426 | 29735.413 | od | 4.5 | 8896.729 | ev | 5.5 | 1.08 ± 0.07 | -1.43 |
| 4797.850 | 25681.488 | ev | 1.5 | 4844.644 | od | 1.5 | 0.27 ± 0.03 | -2.43 |
| 4804.633 | 28685.758 | ev | 2.5 | 7878.328 | od | 3.5 | 0.76 ± 0.08 | -1.80 |
| 4812.505 | 27835.233 | ev | 1.5 | 7061.838 | od | 0.5 | 1.97 ± 0.19 | -1.56 |
| 4835.674 | 28396.150 | od | 2.5 | 7722.285 | ev | 2.5 | 5.8 ± 0.4 | -0.92 |
| 4847.914 | 28334.756 | ev | 4.5 | 7713.089 | od | 4.5 | 5.0 ± 0.5 | -0.75 |
| 4848.263 | 29043.854 | od | 6.5 | 8423.672 | ev | 6.5 | 0.97 ± 0.07 | -1.32 |
| 4850.902 | 27811.496 | ev | 3.5 | 7202.529 | od | 2.5 | 1.52 ± 0.15 | -1.37 |
| 4858.729 | 29892.677 | od | 3.5 | 9316.912 | ev | 3.5 | 4.9 ± 0.4 | -0.86 |
| 4861.821 | 25681.488 | ev | 1.5 | 5118.806 | od | 2.5 | 0.42 ± 0.06 | -2.23 |
| 4863.673 | 28730.712 | ev | 3.5 | 8175.863 | od | 2.5 | 0.95 ± 0.11 | -1.57 |
| 4863.838 | 30829.124 | ev | 3.5 | 10274.971 | od | 3.5 | 1.80 ± 0.24 | -1.29 |
| 4873.999 | 29438.817 | ev | 5.5 | 8927.514 | od | 5.5 | 4.7 ± 0.5 | -0.69 |
| 4874.211 | 26900.354 | ev | 3.5 | 6389.942 | od | 4.5 | 0.25 ± 0.05 | -2.14 |
| 4882.463 | 32802.165 | ev | 5.5 | 12326.417 | od | 6.5 | 35.8 ± 2.4 | 0.19 |
| 4887.074 | 28334.756 | ev | 4.5 | 7878.328 | od | 3.5 | 0.31 ± 0.05 | -1.96 |
| 4890.351 | 30166.057 | ev | 3.5 | 9723.335 | od | 4.5 | 1.71 ± 0.16 | -1.31 |
| 4895.558 | 27432.782 | od | 4.5 | 7011.804 | ev | 4.5 | 0.72 ± 0.04 | -1.59 |
| 4895.998 | 28297.473 | ev | 3.5 | 7878.328 | od | 3.5 | 1.30 ± 0.10 | -1.43 |
| 4896.152 | 29735.413 | od | 4.5 | 9316.912 | ev | 3.5 | 0.34 ± 0.05 | -1.91 |
| 4897.959 | 27378.515 | od | 5.5 | 6967.547 | ev | 6.5 | 0.292 ± 0.028 | -1.90 |
| 4901.396 | 24663.053 | ev | 4.5 | 4266.397 | od | 3.5 | 0.54 ± 0.08 | -1.71 |
| 4903.700 | 30166.057 | ev | 3.5 | 9778.986 | od | 2.5 | 1.62 ± 0.19 | -1.33 |
| 4912.919 | 25359.686 | ev | 2.5 | 5010.870 | od | 2.5 | 0.20 ± 0.03 | -2.35 |
| 4914.607 | 30065.164 | ev | 3.5 | 9723.335 | od | 4.5 | 1.59 ± 0.18 | -1.34 |
| 4914.924 | 27432.782 | od | 4.5 | 7092.265 | ev | 5.5 | 4.3 ± 0.3 | -0.81 |
| 4917.939 | 28730.712 | ev | 3.5 | 8402.668 | od | 3.5 | 0.38 ± 0.04 | -1.95 |
| 4928.072 | 27378.515 | od | 5.5 | 7092.265 | ev | 5.5 | 1.15 ± 0.08 | -1.30 |
| 4932.112 | 25945.396 | ev | 3.5 | 5675.763 | od | 4.5 | 0.30 ± 0.04 | -2.06 |
| 4946.624 | 30245.878 | ev | 4.5 | 10035.711 | od | 5.5 | 2.00 ± 0.19 | -1.14 |
| 4952.134 | 30829.124 | ev | 3.5 | 10641.442 | od | 2.5 | 0.37 ± 0.06 | -1.96 |
| 4962.103 | 29043.854 | od | 6.5 | 8896.729 | ev | 5.5 | 0.39 ± 0.04 | -1.69 |
| 4967.241 | 25945.396 | ev | 3.5 | 5819.113 | od | 4.5 | 0.29 ± 0.04 | -2.07 |
| 4968.395 | 28297.473 | ev | 3.5 | 8175.863 | od | 2.5 | 1.30 ± 0.12 | -1.41 |
| 4977.201 | 27379.949 | ev | 5.5 | 7293.938 | od | 6.5 | 2.26 ± 0.27 | -1.00 |
| 4984.433 | 28337.814 | ev | 2.5 | 8280.946 | od | 2.5 | 4.2 ± 0.4 | -1.02 |
| 4991.014 | 31340.393 | od | 6.5 | 11309.972 | ev | 7.5 | 4.0 ± 0.3 | -0.68 |
| 4994.478 | 28297.473 | ev | 3.5 | 8280.946 | od | 2.5 | 2.7 ± 0.3 | -1.09 |
| 4994.727 | 29794.517 | ev | 3.5 | 9778.986 | od | 2.5 | 5.4 ± 0.5 | -0.79 |
| 4997.956 | 25945.396 | ev | 3.5 | 5942.798 | od | 3.5 | 0.148 ± 0.018 | -2.35 |
| 5000.958 | 27249.669 | ev | 2.5 | 7259.075 | od | 3.5 | 0.85 ± 0.08 | -1.72 |
| 5002.779 | 28685.758 | ev | 2.5 | 8702.444 | od | 1.5 | 6.6 ± 0.6 | -0.83 |
| 5011.759 | 28396.150 | od | 2.5 | 8448.641 | ev | 2.5 | 14.3 ± 1.0 | -0.49 |
| 5015.138 | 27812.398 | ev | 2.5 | 7878.328 | od | 3.5 | 0.67 ± 0.06 | -1.82 |
| 5015.365 | 27811.496 | ev | 3.5 | 7878.328 | od | 3.5 | 0.41 ± 0.05 | -1.91 |
| 5018.448 | 28725.148 | ev | 4.5 | 8804.224 | od | 4.5 | 0.30 ± 0.05 | -1.95 |
| 5022.652 | 30829.124 | ev | 3.5 | 10924.876 | od | 4.5 | 0.63 ± 0.10 | -1.72 |
| 5022.867 | 28327.071 | od | 5.5 | 8423.672 | ev | 6.5 | 9.2 ± 0.7 | -0.38 |
| 5027.339 | 31340.393 | od | 6.5 | 11454.701 | ev | 6.5 | 3.2 ± 0.5 | -0.77 |
| 5028.263 | 31340.393 | od | 6.5 | 11458.353 | ev | 5.5 | 2.68 ± 0.26 | -0.85 |
| 5037.800 | 27975.619 | od | 4.5 | 8131.217 | ev | 4.5 | 7.0 ± 0.6 | -0.57 |
| 5039.273 | 30637.157 | ev | 2.5 | 10798.555 | od | 2.5 | 1.43 ± 0.14 | -1.49 |
| 5044.023 | 29591.873 | od | 6.5 | 9771.956 | ev | 7.5 | 13.5 ± 1.0 | -0.14 |
| 5049.701 | 28725.148 | ev | 4.5 | 8927.514 | od | 5.5 | 0.37 ± 0.04 | -1.85 |
| 5072.924 | 28634.516 | ev | 5.5 | 8927.514 | od | 5.5 | 1.56 ± 0.20 | -1.14 |
| 5075.355 | 27975.619 | od | 4.5 | 8278.054 | ev | 5.5 | 11.5 ± 0.8 | -0.35 |
| 5083.279 | 27379.949 | ev | 5.5 | 7713.089 | od | 4.5 | 0.50 ± 0.06 | -1.64 |
| 5085.216 | 27835.233 | ev | 1.5 | 8175.863 | od | 2.5 | 0.52 ± 0.06 | -2.10 |
| 5089.487 | 28345.313 | ev | 0.5 | 8702.444 | od | 1.5 | 1.62 ± 0.21 | -1.90 |
| 5105.227 | 26841.384 | ev | 4.5 | 7259.075 | od | 3.5 | 1.22 ± 0.11 | -1.32 |
| 5106.217 | 29892.677 | od | 3.5 | 10314.162 | ev | 4.5 | 0.57 ± 0.08 | -1.75 |
| 5112.543 | 27835.233 | ev | 1.5 | 8280.946 | od | 2.5 | 0.94 ± 0.12 | -1.83 |
| 5117.946 | 29591.873 | od | 6.5 | 10058.226 | ev | 6.5 | 2.8 ± 0.3 | -0.81 |
| 5118.349 | 29807.078 | ev | 3.5 | 10274.971 | od | 3.5 | 0.96 ± 0.11 | -1.52 |
| 5141.737 | 29166.597 | ev | 4.5 | 9723.335 | od | 4.5 | 0.40 ± 0.03 | -1.80 |
| 5144.844 | 27249.669 | ev | 2.5 | 7818.147 | od | 1.5 | 1.12 ± 0.16 | -1.58 |
| 5145.156 | 28327.071 | od | 5.5 | 8896.729 | ev | 5.5 | 1.93 ± 0.22 | -1.04 |
| 5146.910 | 30065.164 | ev | 3.5 | 10641.442 | od | 2.5 | 1.17 ± 0.13 | -1.43 |
| 5147.565 | 29735.413 | od | 4.5 | 10314.162 | ev | 4.5 | 12.5 ± 0.9 | -0.30 |
| 5150.860 | 27811.496 | ev | 3.5 | 8402.668 | od | 3.5 | 0.58 ± 0.07 | -1.73 |
| 5163.547 | 29449.778 | ev | 1.5 | 10088.640 | od | 1.5 | 0.47 ± 0.09 | -2.12 |
| 5187.458 | 29043.854 | od | 6.5 | 9771.956 | ev | 7.5 | 26.2 ± 1.8 | 0.17 |
| 5191.633 | 26268.203 | od | 3.5 | 7011.804 | ev | 4.5 | 7.8 ± 0.6 | -0.60 |
| 5192.092 | 32372.621 | od | 4.5 | 13117.922 | ev | 4.5 | 1.55 ± 0.24 | -1.20 |
| 5208.108 | 29994.041 | ev | 2.5 | 10798.555 | od | 2.5 | 0.42 ± 0.08 | -1.99 |
| 5210.339 | 26900.354 | ev | 3.5 | 7713.089 | od | 4.5 | 2.64 ± 0.30 | -1.07 |
| 5216.701 | 25681.488 | ev | 1.5 | 6517.619 | od | 2.5 | 1.17 ± 0.13 | -1.72 |
| 5217.711 | 25681.488 | ev | 1.5 | 6521.332 | od | 1.5 | 1.07 ± 0.15 | -1.76 |
| 5225.505 | 25681.488 | ev | 1.5 | 6549.908 | od | 2.5 | 2.6 ± 0.3 | -1.38 |
| 5226.402 | 26841.384 | ev | 4.5 | 7713.089 | od | 4.5 | 1.08 ± 0.16 | -1.35 |
| 5232.918 | 29750.547 | od | 5.5 | 10646.070 | ev | 5.5 | 10.0 ± 0.9 | -0.31 |
| 5234.019 | 27378.515 | od | 5.5 | 8278.054 | ev | 5.5 | 3.3 ± 0.3 | -0.79 |
| 5237.067 | 29735.413 | od | 4.5 | 10646.070 | ev | 5.5 | 5.8 ± 0.5 | -0.62 |
| 5241.333 | 27249.669 | ev | 2.5 | 8175.863 | od | 2.5 | 0.73 ± 0.08 | -1.75 |
| 5241.777 | 29892.677 | od | 3.5 | 10820.486 | ev | 2.5 | 2.52 ± 0.26 | -1.08 |
| 5252.817 | 29735.413 | od | 4.5 | 10703.305 | ev | 4.5 | 1.72 ± 0.15 | -1.15 |
| 5262.938 | 29449.778 | ev | 1.5 | 10454.272 | od | 1.5 | 1.18 ± 0.25 | -1.71 |
| 5265.506 | 29994.041 | ev | 2.5 | 11007.799 | od | 1.5 | 0.26 ± 0.05 | -2.20 |
| 5265.677 | 29043.854 | od | 6.5 | 10058.226 | ev | 6.5 | 10.0 ± 0.7 | -0.23 |
| 5274.229 | 27378.515 | od | 5.5 | 8423.672 | ev | 6.5 | 26.8 ± 1.6 | 0.13 |
| 5276.457 | 24663.053 | ev | 4.5 | 5716.216 | od | 3.5 | 0.102 ± 0.010 | -2.37 |
| 5283.386 | 27975.619 | od | 4.5 | 9053.629 | ev | 3.5 | 1.06 ± 0.09 | -1.35 |
| 5294.855 | 29750.547 | od | 5.5 | 10869.541 | ev | 4.5 | 0.43 ± 0.05 | -1.66 |
| 5295.155 | 30829.124 | ev | 3.5 | 11949.189 | od | 3.5 | 0.53 ± 0.09 | -1.75 |
| 5304.408 | 27249.669 | ev | 2.5 | 8402.668 | od | 3.5 | 0.52 ± 0.07 | -1.88 |
| 5327.492 | 29449.778 | ev | 1.5 | 10684.441 | od | 0.5 | 1.00 ± 0.15 | -1.77 |
| 5330.556 | 25766.355 | od | 4.5 | 7011.804 | ev | 4.5 | 9.3 ± 0.7 | -0.40 |
| 5350.020 | 25945.396 | ev | 3.5 | 7259.075 | od | 3.5 | 0.127 ± 0.017 | -2.36 |
| 5353.524 | 25766.355 | od | 4.5 | 7092.265 | ev | 5.5 | 28.8 ± 1.7 | 0.09 |
| 5369.864 | 29263.338 | od | 5.5 | 10646.070 | ev | 5.5 | 0.20 ± 0.04 | -1.98 |
| 5383.812 | 30829.124 | ev | 3.5 | 12260.088 | od | 3.5 | 0.40 ± 0.06 | -1.86 |
| 5386.773 | 28337.814 | ev | 2.5 | 9778.986 | od | 2.5 | 7.5 ± 0.6 | -0.71 |
| 5390.523 | 26268.203 | od | 3.5 | 7722.285 | ev | 2.5 | 0.35 ± 0.04 | -1.91 |
| 5393.392 | 27432.782 | od | 4.5 | 8896.729 | ev | 5.5 | 19.9 ± 1.3 | -0.06 |
| 5398.508 | 28297.473 | ev | 3.5 | 9778.986 | od | 2.5 | 1.60 ± 0.13 | -1.25 |
| 5409.229 | 27378.515 | od | 5.5 | 8896.729 | ev | 5.5 | 9.8 ± 0.7 | -0.29 |
| 5413.051 | 29794.517 | ev | 3.5 | 11325.781 | od | 2.5 | 0.51 ± 0.06 | -1.75 |
| 5433.927 | 29043.854 | od | 6.5 | 10646.070 | ev | 5.5 | 0.53 ± 0.04 | -1.48 |
| 5435.105 | 29263.338 | od | 5.5 | 10869.541 | ev | 4.5 | 0.361 ± 0.029 | -1.72 |
| 5459.193 | 31340.393 | od | 6.5 | 13027.758 | ev | 6.5 | 7.5 ± 0.9 | -0.33 |
| 5464.203 | 29750.547 | od | 5.5 | 11454.701 | ev | 6.5 | 5.8 ± 0.6 | -0.50 |
| 5468.371 | 29591.873 | od | 6.5 | 11309.972 | ev | 7.5 | 13.4 ± 1.1 | -0.07 |
| 5505.941 | 25359.686 | ev | 2.5 | 7202.529 | od | 2.5 | 0.54 ± 0.05 | -1.84 |
| 5512.064 | 26268.203 | od | 3.5 | 8131.217 | ev | 4.5 | 11.3 ± 1.0 | -0.39 |
| 5513.118 | 29591.873 | od | 6.5 | 11458.353 | ev | 5.5 | 1.47 ± 0.19 | -1.03 |
| 5518.489 | 27432.782 | od | 4.5 | 9316.912 | ev | 3.5 | 4.9 ± 0.4 | -0.65 |
| 5524.450 | 32372.621 | od | 4.5 | 14276.298 | ev | 5.5 | 2.44 ± 0.27 | -0.95 |
| 5547.079 | 28297.473 | ev | 3.5 | 10274.971 | od | 3.5 | 0.194 ± 0.027 | -2.15 |
| 5550.033 | 28327.071 | od | 5.5 | 10314.162 | ev | 4.5 | 2.00 ± 0.20 | -0.96 |
| 5561.445 | 29735.413 | od | 4.5 | 11759.467 | ev | 5.5 | 2.83 ± 0.26 | -0.88 |
| 5610.253 | 26268.203 | od | 3.5 | 8448.641 | ev | 2.5 | 5.9 ± 0.5 | -0.65 |
| 5613.694 | 29263.338 | od | 5.5 | 11454.701 | ev | 6.5 | 4.1 ± 0.4 | -0.64 |
| 5637.359 | 29043.854 | od | 6.5 | 11309.972 | ev | 7.5 | 3.6 ± 0.3 | -0.62 |
| 5654.219 | 28327.071 | od | 5.5 | 10646.070 | ev | 5.5 | 0.248 ± 0.028 | -1.85 |
| 5659.517 | 25945.396 | ev | 3.5 | 8280.946 | od | 2.5 | 0.262 ± 0.027 | -2.00 |
| 5660.476 | 27975.619 | od | 4.5 | 10314.162 | ev | 4.5 | 0.51 ± 0.04 | -1.61 |
| 5667.960 | 29735.413 | od | 4.5 | 12097.276 | ev | 3.5 | 1.56 ± 0.16 | -1.12 |
| 5683.745 | 29043.854 | od | 6.5 | 11454.701 | ev | 6.5 | 2.03 ± 0.16 | -0.86 |
| 5684.925 | 29043.854 | od | 6.5 | 11458.353 | ev | 5.5 | 0.77 ± 0.08 | -1.28 |
| 5733.693 | 29892.677 | od | 3.5 | 12456.746 | ev | 3.5 | 1.59 ± 0.17 | -1.20 |
| 5768.891 | 27975.619 | od | 4.5 | 10646.070 | ev | 5.5 | 6.0 ± 0.5 | -0.52 |
| 5771.975 | 27378.515 | od | 5.5 | 10058.226 | ev | 6.5 | 1.38 ± 0.12 | -1.08 |
| 5788.007 | 27975.619 | od | 4.5 | 10703.305 | ev | 4.5 | 0.85 ± 0.12 | -1.37 |
| 5822.479 | 27811.496 | ev | 3.5 | 10641.442 | od | 2.5 | 0.29 ± 0.05 | -1.92 |
| 5832.290 | 25945.396 | ev | 3.5 | 8804.224 | od | 4.5 | 0.54 ± 0.06 | -1.65 |
| 5839.973 | 27432.782 | od | 4.5 | 10314.162 | ev | 4.5 | 0.23 ± 0.05 | -1.92 |
| 5842.096 | 29438.817 | ev | 5.5 | 12326.417 | od | 6.5 | 0.58 ± 0.06 | -1.45 |
| 5858.546 | 27378.515 | od | 5.5 | 10314.162 | ev | 4.5 | 1.08 ± 0.19 | -1.18 |
| 5895.679 | 28297.473 | ev | 3.5 | 11340.598 | od | 3.5 | 0.44 ± 0.06 | -1.73 |
| 5898.083 | 24663.053 | ev | 4.5 | 7713.089 | od | 4.5 | 0.53 ± 0.06 | -1.56 |
| 5959.688 | 29892.677 | od | 3.5 | 13117.922 | ev | 4.5 | 4.8 ± 0.5 | -0.69 |
| 5968.060 | 31155.623 | ev | 6.5 | 14404.400 | od | 7.5 | 1.33 ± 0.13 | -1.00 |
| 5975.818 | 27432.782 | od | 4.5 | 10703.305 | ev | 4.5 | 6.6 ± 0.6 | -0.45 |
| 5995.265 | 27378.515 | od | 5.5 | 10703.305 | ev | 4.5 | 1.62 ± 0.17 | -0.98 |
| 5995.448 | 29892.677 | od | 3.5 | 13217.976 | ev | 3.5 | 2.01 ± 0.19 | -1.06 |
| 6013.568 | 29892.677 | od | 3.5 | 13268.218 | ev | 2.5 | 1.72 ± 0.23 | -1.13 |
| 6034.205 | 28327.071 | od | 5.5 | 11759.467 | ev | 5.5 | 3.5 ± 0.4 | -0.64 |
| 6035.476 | 29591.873 | od | 6.5 | 13027.758 | ev | 6.5 | 2.37 ± 0.25 | -0.74 |
| 6043.373 | 26268.203 | od | 3.5 | 9725.733 | ev | 3.5 | 7.5 ± 0.8 | -0.48 |
| 6069.634 | 27811.496 | ev | 3.5 | 11340.598 | od | 3.5 | 0.76 ± 0.11 | -1.47 |
| 6101.243 | 28334.756 | ev | 4.5 | 11949.189 | od | 3.5 | 0.29 ± 0.04 | -1.79 |
| 6108.747 | 29892.677 | od | 3.5 | 13527.239 | ev | 4.5 | 4.2 ± 0.5 | -0.73 |
| 6183.954 | 25945.396 | ev | 3.5 | 9778.986  | od | 2.5 | 0.89 ± 0.10   | -1.39 |
| 6225.397 | 31340.393 | od | 6.5 | 15281.602 | ev | 6.5 | 0.49 ± 0.05   | -1.40 |
| 6232.448 | 25766.355 | od | 4.5 | 9725.733  | ev | 3.5 | 2.2  ± 0.3    | -0.89 |
| 6272.026 | 28396.150 | od | 2.5 | 12456.746 | ev | 3.5 | 11.1 ± 1.1    | -0.40 |
| 6296.145 | 27975.619 | od | 4.5 | 12097.276 | ev | 3.5 | 0.66 ± 0.08   | -1.41 |
| 6371.109 | 28396.150 | od | 2.5 | 12704.634 | ev | 1.5 | 6.6  ± 0.7    | -0.62 |
| 6428.509 | 27811.496 | ev | 3.5 | 12260.088 | od | 3.5 | 0.35 ± 0.05   | -1.77 |
| 6441.986 | 27975.619 | od | 4.5 | 12456.746 | ev | 3.5 | 0.42 ± 0.05   | -1.58 |
| 6449.503 | 26841.384 | ev | 4.5 | 11340.598 | od | 3.5 | 0.126 ± 0.016 | -2.10 |
| 6466.888 | 29735.413 | od | 4.5 | 14276.298 | ev | 5.5 | 3.6  ± 0.5    | -0.65 |
| 6507.163 | 29750.547 | od | 5.5 | 14387.112 | ev | 4.5 | 1.39 ± 0.15   | -0.97 |
| 6608.464 | 28396.150 | od | 2.5 | 13268.218 | ev | 2.5 | 0.52 ± 0.07   | -1.69 |
| 6609.726 | 29750.547 | od | 5.5 | 14625.503 | ev | 5.5 | 0.69 ± 0.08   | -1.27 |
| 6616.346 | 29735.413 | od | 4.5 | 14625.503 | ev | 5.5 | 0.85 ± 0.10   | -1.25 |
| 6706.051 | 29735.413 | od | 4.5 | 14827.623 | ev | 3.5 | 1.92 ± 0.24   | -0.89 |
| 6720.280 | 29263.338 | od | 5.5 | 14387.112 | ev | 4.5 | 1.52 ± 0.13   | -0.91 |
| 6829.583 | 28297.473 | ev | 3.5 | 13659.329 | od | 4.5 | 0.134 ± 0.022 | -2.13 |
| 6829.727 | 29263.338 | od | 5.5 | 14625.503 | ev | 5.5 | 1.68 ± 0.15   | -0.85 |
| 6905.307 | 30637.157 | ev | 2.5 | 16159.536 | od | 3.5 | 0.84 ± 0.11   | -1.44 |
| 6919.282 | 27975.619 | od | 4.5 | 13527.239 | ev | 4.5 | 1.41 ± 0.14   | -0.99 |
| 6960.365 | 29892.677 | od | 3.5 | 15529.576 | ev | 2.5 | 0.19 ± 0.04   | -1.95 |
| 6983.822 | 27432.782 | od | 4.5 | 13117.922 | ev | 4.5 | 0.88 ± 0.12   | -1.19 |
| 7025.915 | 29794.517 | ev | 3.5 | 15565.420 | od | 2.5 | 0.45 ± 0.07   | -1.57 |
| 7061.753 | 29750.547 | od | 5.5 | 15593.660 | ev | 6.5 | 8.1  ± 0.9    | -0.14 |
| 7075.546 | 29263.338 | od | 5.5 | 15134.048 | ev | 4.5 | 0.27 ± 0.04   | -1.62 |
| 7086.353 | 31340.393 | od | 6.5 | 17232.652 | ev | 7.5 | 12.6 ± 1.4    |  0.12 |
| 7142.824 | 25945.396 | ev | 3.5 | 11949.189 | od | 3.5 | 0.128 ± 0.018 | -2.11 |
| 7238.371 | 26268.203 | od | 3.5 | 12456.746 | ev | 3.5 | 1.96 ± 0.20   | -0.91 |
| 7275.857 | 29892.677 | od | 3.5 | 16152.377 | ev | 3.5 | 0.133 ± 0.022 | -2.07 |
| 7349.808 | 29794.517 | ev | 3.5 | 16192.466 | od | 4.5 | 0.16 ± 0.03   | -1.98 |
| 7360.097 | 29735.413 | od | 4.5 | 16152.377 | ev | 3.5 | 0.51 ± 0.07   | -1.38 |
| 7439.467 | 29892.677 | od | 3.5 | 16454.555 | ev | 2.5 | 0.33 ± 0.05   | -1.66 |
| 7486.722 | 27835.233 | ev | 1.5 | 14481.930 | od | 2.5 | 0.33 ± 0.06   | -1.96 |
| 7496.946 | 27432.782 | od | 4.5 | 14097.689 | ev | 3.5 | 0.70 ± 0.08   | -1.23 |
| 7577.678 | 28327.071 | od | 5.5 | 15134.048 | ev | 4.5 | 0.43 ± 0.05   | -1.35 |
| 7583.580 | 25945.396 | ev | 3.5 | 12762.641 | od | 4.5 | 0.065 ± 0.011 | -2.35 |
| 7584.818 | 27432.782 | od | 4.5 | 14252.178 | ev | 3.5 | 0.39 ± 0.05   | -1.47 |
| 7596.363 | 28396.150 | od | 2.5 | 15235.579 | ev | 1.5 | 0.66 ± 0.10   | -1.47 |
| 7660.593 | 26268.203 | od | 3.5 | 13217.976 | ev | 3.5 | 0.27 ± 0.03   | -1.72 |
| 7690.200 | 26268.203 | od | 3.5 | 13268.218 | ev | 2.5 | 0.63 ± 0.07   | -1.35 |
| 7711.746 | 30134.910 | ev | 5.5 | 17171.245 | od | 5.5 | 0.29 ± 0.04   | -1.51 |
| 7851.195 | 28327.071 | od | 5.5 | 15593.660 | ev | 6.5 | 1.99 ± 0.25   | -0.66 |
| 7939.063 | 26268.203 | od | 3.5 | 13675.722 | ev | 2.5 | 0.58 ± 0.07   | -1.36 |
| 8068.524 | 32372.621 | od | 4.5 | 19982.187 | ev | 4.5 | 1.45 ± 0.24   | -0.85 |
| 8088.901 | 29591.873 | od | 6.5 | 17232.652 | ev | 7.5 | 1.47 ± 0.19   | -0.69 |
| 8137.064 | 25945.396 | ev | 3.5 | 13659.329 | od | 4.5 | 0.139 ± 0.020 | -1.96 |
| 8233.415 | 28334.756 | ev | 4.5 | 16192.466 | od | 4.5 | 0.18 ± 0.03 | -1.73 |
| 8693.756 | 29892.677 | od | 3.5 | 18393.327 | ev | 3.5 | 0.61 ± 0.11 | -1.26 |
| 8777.180 | 31340.393 | od | 6.5 | 19950.340 | ev | 6.5 | 0.81 ± 0.12 | -0.88 |
| 8926.473 | 25681.488 | ev | 1.5 | 14481.930 | od | 2.5 | 0.34 ± 0.05 | -1.78 |
| 9014.550 | 27249.669 | ev | 2.5 | 16159.536 | od | 3.5 | 0.099 ± 0.017 | -2.14 |
| 9050.375 | 29750.547 | od | 5.5 | 18704.313 | ev | 5.5 | 0.37 ± 0.06 | -1.27 |
| 9602.004 | 29892.677 | od | 3.5 | 19481.040 | ev | 4.5 | 0.21 ± 0.03 | -1.64 |
| 10433.058 | 27975.619 | od | 4.5 | 18393.327 | ev | 3.5 | 0.102 ± 0.018 | -1.78 |
| 10783.013 | 27975.619 | od | 4.5 | 18704.313 | ev | 5.5 | 0.33 ± 0.05 | -1.24 |

Table 3. Cerium abundances from individual lines in the Sun and the r-process rich metal-poor giant stars.

| λ (Å) | E. P. (eV) | log(gf) | log ε Sun | log ε BD +17°3248 | log ε CS 22892 | log ε CS 31082 | log ε HD 115444 | log ε HD 221170 |
|---|---|---|---|---|---|---|---|---|
| 3534.045 | 0.521 | -0.140 | 1.70 | -0.05 | -0.44 | -0.31 | ... | -0.35 |
| 3539.079 | 0.320 | -0.270 | ... | -0.10 | -0.47 | ... | ... | -0.43 |
| 3659.225 | 0.175 | -0.670 | 1.63 | -0.15 | -0.45 | -0.29 | ... | -0.41 |
| 3659.970 | 0.175 | -0.730 | 1.63 | ... | ... | ... | ... | ... |
| 3912.420 | 0.295 | -0.250 | 1.60 | -0.16 | -0.48 | ... | ... | -0.44 |
| 3942.151 | 0.000 | -0.220 | 1.52 | -0.16 | -0.47 | -0.29 | -1.15 | -0.41 |
| 3942.745 | 0.857 | 0.690 | 1.50 | -0.21 | -0.52 | -0.36 | -1.15 | -0.56 |
| 3953.652 | 0.495 | -0.640 | 1.57 | ... | ... | ... | ... | ... |
| 3980.890 | 0.708 | -0.210 | 1.70 | -0.13 | ... | ... | ... | ... |
| 3993.819 | 0.909 | 0.290 | 1.64 | 0.02 | ... | -0.24 | ... | ... |
| 3999.237 | 0.295 | 0.060 | 1.60 | -0.13 | -0.51 | -0.31 | -1.13 | -0.46 |
| 4042.581 | 0.495 | 0.000 | 1.60 | -0.02 | -0.45 | -0.29 | -1.02 | -0.41 |
| 4053.503 | 0.000 | -0.610 | 1.60 | -0.08 | -0.45 | -0.26 | -1.05 | -0.41 |
| 4068.836 | 0.703 | -0.170 | 1.55 | -0.03 | ... | ... | ... | ... |
| 4072.918 | 0.327 | -0.640 | 1.61 | 0.00 | ... | -0.22 | ... | -0.41 |
| 4073.474 | 0.477 | 0.210 | 1.59 | -0.10 | -0.45 | -0.30 | -1.07 | -0.43 |
| 4075.700 | 0.700 | 0.230 | 1.60 | -0.12 | -0.48 | -0.31 | -1.14 | -0.41 |
| 4083.222 | 0.700 | 0.270 | 1.78 | -0.08 | -0.45 | -0.33 | -1.05 | -0.41 |
| 4117.288 | 0.739 | -0.450 | 1.61 | ... | ... | -0.31 | ... | ... |
| 4118.143 | 0.696 | 0.130 | ... | -0.08 | -0.43 | -0.27 | -0.97 | -0.38 |
| 4120.827 | 0.320 | -0.370 | 1.70 | -0.08 | -0.30 | -0.22 | -1.00 | -0.37 |
| 4127.364 | 0.683 | 0.310 | 1.60 | -0.08 | -0.47 | -0.31 | -1.05 | -0.44 |
| 4137.645 | 0.516 | 0.400 | 1.75 | -0.10 | -0.45 | -0.28 | -1.05 | -0.38 |
| 4142.397 | 0.696 | 0.220 | ... | -0.07 | ... | ... | ... | ... |
| 4144.996 | 0.696 | 0.100 | 1.65 | -0.14 | -0.47 | -0.26 | -0.95 | -0.38 |
| 4146.232 | 0.560 | -0.120 | ... | -0.10 | -0.47 | -0.26 | -0.90 | ... |
| 4222.597 | 0.122 | -0.150 | 1.58 | -0.11 | -0.44 | -0.28 | -1.07 | -0.45 |
| 4337.773 | 0.326 | -0.310 | 1.61 | ... | ... | ... | ... | ... |
| 4349.768 | 0.529 | -0.730 | ... | ... | ... | ... | ... | ... |
| 4349.789 | 0.700 | -0.320 | 1.60 | -0.18 | -0.50 | -0.29 | ... | -0.45 |
| 4364.653 | 0.495 | -0.170 | ... | -0.17 | -0.42 | -0.35 | -1.20 | -0.49 |
| 4382.165 | 0.683 | 0.130 | 1.55 | -0.15 | ... | -0.28 | ... | ... |
| 4399.200 | 0.326 | -0.440 | 1.60 | -0.11 | -0.50 | -0.26 | ... | -0.38 |
| 4418.780 | 0.863 | 0.270 | 1.65 | -0.14 | ... | -0.26 | -1.10 | -0.38 |
| 4449.330 | 0.608 | 0.040 | 1.63 | -0.18 | -0.53 | -0.31 | -1.15 | -0.45 |
| 4486.909 | 0.295 | -0.180 | 1.61 | -0.12 | -0.50 | -0.32 | -1.08 | -0.46 |
| 4523.075 | 0.516 | -0.080 | 1.61 | -0.04 | -0.50 | ... | ... | ... |
| 4560.280 | 0.909 | 0.180 | 1.65 | -0.18 | -0.45 | -0.28 | -1.07 | -0.43 |
| 4560.958 | 0.683 | -0.260 | 1.61 | -0.12 | -0.45 | -0.28 | -1.07 | -0.41 |
| 4562.359 | 0.477 | 0.210 | 1.63 | -0.11 | -0.40 | -0.26 | -1.07 | -0.41 |
| 4572.278 | 0.683 | 0.220 | 1.62 | -0.08 | ... | -0.27 | -1.04 | -0.41 |
| 4582.499 | 0.696 | -0.350 | 1.55 | -0.07 | ... | -0.30 | -1.07 | -0.38 |
| 4593.926 | 0.696 | 0.070 | 1.65 | -0.13 | -0.45 | -0.27 | -1.02 | -0.41 |
| 4628.161 | 0.516 | 0.140 | 1.57 | -0.09 | -0.37 | -0.26 | -1.05 | -0.41 |
| 4847.914 | 0.956 | -0.750 | 1.52 | ... | ... | ... | ... | ... |
| 5044.023 | 1.211 | -0.140 | ... | ... | ... | ... | ... | -0.41 |
| 5187.458 | 1.211 | 0.170 | 1.59 | ... | ... | -0.31 | ... | -0.44 |
| 5274.229 | 1.044 | 0.130 | ... | -0.21 | -0.40 | -0.31 | ... | -0.46 |
| 5330.556 | 0.869 | -0.400 | 1.70 | ... | ... | -0.31 | ... | -0.48 |
| 5393.392 | 1.102 | -0.060 | 1.55 | ... | -0.53 | -0.31 | ... | -0.46 |
| 5768.891 | 1.319 | -0.520 | 1.63 | ... | ... | ... | ... | ... |
| 5975.818 | 1.326 | -0.450 | 1.55 | ... | ... | ... | ... | ... |
| 6043.373 | 1.205 | -0.480 | 1.64 | ... | ... | ... | ... | ... |
| Mean | | | 1.61 | -0.11 | -0.46 | -0.29 | -1.06 | -0.42 |
| Unc. | | | 0.01 | 0.01 | 0.01 | 0.01 | 0.01 | 0.01 |
| σ | | | 0.06 | 0.05 | 0.05 | 0.03 | 0.07 | 0.04 |
| number of lines | | | 45 | 40 | 32 | 38 | 26 | 37 |

Table 4. Master table of laboratory data measured or compiled as part of this program on RE ions. Machine-readable (MR) tables of **log(*gf*)** values and Complete Line Component Patterns (**CLCP**) are now available for this data either in the original publication or in this Appendix.

| Spectrum | Z | log(*gf*) data | Isotope shifts and/or hfs constants | Location of Machine Readable (MR) Tables |
|---|---|---|---|---|
| La II | 57 | Lawler et al. 2001a | Lawler et al. 2001a, Ivans et al. 2006 | **log(gf)**: MR Table 5 below<br><br>**CLCP**: MR Table 4 in Ivans et al. 2006 |
| Ce II | 58 | This paper | ………… | **log(gf)**: MR Table 2 above<br><br>………… |
| Pr II | 59 | Companion paper: Sneden et al. 2009 | Companion paper: Sneden et al. 2009 | **log(gf)**: Companion paper<br><br>**CLCP:** MR Table 11 in Sneden et al. 2009 |
| Nd II | 60 | Den Hartog et al. 2003 | Roederer et al. 2008 | **log(gf)**: MR Table 3 in Den Hartog et al. 2003<br><br>**CLCP:** MR Table 5 in Roederer et al. 2008 |
| Pm II | 61 | no stable isotopes | no stable isotopes | no stable isotopes |
| Sm II | 62 | Lawler et al. 2006 | Roederer et al. 2008 | **log(*gf*)**: MR Table 2 of Lawler et al. 2006 |

| | | | | CLCP: MR Table 4 of Roederer et al. 2008 |
|---|---|---|---|---|
| Eu II | 63 | Lawler et al. 2001b | Lawler et al. 2001b, Ivans et al. 2006 | **log(gf)**: MR Table 6 below<br><br>**CLCP**: Mr Table 6 of Ivans et al. 2006 |
| Gd II | 64 | Den Hartog et al. 2006 | ………… | **log($gf$)**: MR Table 3 of Den Hartog et al. 2006<br><br>………… |
| Tb II | 65 | Lawler et al. 2001c | Lawler et al. 2001d | **log($gf$)**: MR Table 1 of Lawler et al. 2001c<br><br>**CLCP**: MR Table 7 below |
| Dy II | 66 | Wickliffe et al. 2000 | ………… | **log($gf$)**: MR Table 8 below<br><br>………… |
| Ho II | 67 | Lawler et al. 2004 | Lawler et al. 2004 | **log($gf$)**: MR Table 9 below<br><br>**CLCP**: MR Table 10 below |
| Er II | 68 | Lawler et al. 2008b | ………… | **log($gf$)**: MR Table 3 of Lawler et al. 2008b<br><br>………… |
| Tm II | 69 | Wickliffe & Lawler 1997 | ………… | **log($gf$)**: MR Table 11 below<br><br>………… |
| Yb II | 70 | Companion paper: Sneden et al. 2009 | Companion paper: Sneden et al. 2009 | **log(gf)**: Companion paper<br><br>**CLCP**: MR Table 12 in Sneden et al. 2009 |
| Lu II | 71 | Den Hartog et al. 1998, Quinet et al. 1999, Fedchak et al. 2000 | Sneden et al. 2003a | **log($gf$)**: MR Table 13 below<br><br>**CLCP**: MR Table 14 below |

Table 5. Experimental Atomic Transition Probabilities for La II arranged by wavenumber from Lawler et al. (2001a).

| Wavenumber (cm$^{-1}$) | $\lambda_{air}$ (Å) | Upper Level Energy (cm$^{-1}$) | J | Lower Level Energy (cm$^{-1}$) | J | Transition Probability (10$^6$ s$^{-1}$) | log$_{10}$(gf) |
|---|---|---|---|---|---|---|---|
| 27549.30 | 3628.82 | 28565.40 | 4 | 1016.10 | 3 | 4.0 ± 0.6 | -1.15 |
| 27423.91 | 3645.41 | 27423.91 | 1 | 0.00 | 2 | 43 ± 4 | -0.59 |
| 27299.15 | 3662.07 | 28315.25 | 3 | 1016.10 | 3 | 3.0 ± 0.6 | -1.37 |
| 26920.79 | 3713.54 | 28315.25 | 3 | 1394.46 | 2 | 10.9 ± 1.7 | -0.80 |
| 26594.70 | 3759.08 | 28565.40 | 4 | 1970.70 | 4 | 49 ± 4 | -0.03 |

Notes. – Table 5 available in its entirety via the link to the machine-readable version above.

Table 6. Experimental Atomic Transition Probabilities for Eu II arranged by wavenumber from Lawler et al. (2001b).

| Wavenumber (cm$^{-1}$) | $\lambda_{air}$ (Å) | Upper Level Energy (cm$^{-1}$) | J | Lower Level Energy (cm$^{-1}$) | J | Transition Probability (10$^6$ s$^{-1}$) | log$_{10}$(gf) |
|---|---|---|---|---|---|---|---|
| 27104.07 | 3688.43 | 27104.07 | 3 | 0.00 | 4 | 15.0 ± 1.1 | -0.67 |
| 26838.50 | 3724.93 | 26838.50 | 4 | 0.00 | 4 | 43.3 ± 2.6 | -0.09 |
| 26172.83 | 3819.67 | 26172.83 | 5 | 0.00 | 4 | 135 ± 7 | 0.51 |
| 25587.14 | 3907.11 | 27256.35 | 2 | 1669.21 | 3 | 129 ± 6 | 0.17 |
| 25434.86 | 3930.50 | 27104.07 | 3 | 1669.21 | 3 | 114 ± 6 | 0.27 |

Notes. – Table 6 available in its entirety via the link to the machine-readable version above.

Table 7. Hyperfine structure line component patterns for $^{159}$Tb II computed from hfs constants of Lawler et al. (2001b), energy levels of Martin et al. (1978), and the standard index of air (Edlén 1953).

| Wavenumber (cm$^{-1}$) | $\lambda_{air}$ (Å) | $F_{upp}$ | $F_{low}$ | Component Position (cm$^{-1}$) | Component Position (Å) | Strength |
|---|---|---|---|---|---|---|
| 32563.20 | 3070.059 | 8.5 | 8.5 | -0.14668 | 0.013830 | 0.29338 |
| 32563.20 | 3070.059 | 8.5 | 7.5 | 0.24576 | -0.023171 | 0.00662 |
| 32563.20 | 3070.059 | 7.5 | 8.5 | -0.42038 | 0.039636 | 0.00662 |
| 32563.20 | 3070.059 | 7.5 | 7.5 | -0.02794 | 0.002634 | 0.25128 |
| 32563.20 | 3070.059 | 7.5 | 6.5 | 0.31834 | -0.030014 | 0.00877 |

Notes. – Center-of-gravity wavenumbers and air wavelengths, $\lambda_{air}$, are given with component positions relative to those values. Strengths are normalized to sum to 1. Table 7 is available in its entirety via the link to the machine-readable version above.

Table 8. Experimental Atomic Transition Probabilities for Dy I (levels have integral J) and Dy II (levels have half integral J) from Wickliffe et al. (2000).

| Wavenumber (cm$^{-1}$) | $\lambda_{air}$ (Å) | Upper Level | | | Lower Level | | | Univ. of Wisconsin | | | NIST | | |
|---|---|---|---|---|---|---|---|---|---|---|---|---|---|
| | | Energy (cm$^{-1}$) | Parity | J | Energy (cm$^{-1}$) | Parity | J | Transition Probability (10$^6$ s$^{-1}$) | Uncert. (%) | $\log_{10}(gf)$ | Transition Probability | Uncert. (%) | $\log_{10}(gf)$ |
| 34921.87 | 2862.69 | 34921.87 | od | 7.0 | 0.00 | ev | 8.0 | 7.9 | 6 | -0.84 | ….. | ….. | ….. |
| 33311.52 | 3001.09 | 33311.52 | od | 7.0 | 0.00 | ev | 8.0 | 1.45 | 6 | -1.53 | ….. | ….. | ….. |
| 33165.77 | 3014.28 | 33165.77 | od | 8.0 | 0.00 | ev | 8.0 | 1.18 | 7 | -1.56 | ….. | ….. | ….. |
| 33035.57 | 3026.16 | 33035.57 | od | 9.5 | 0.00 | ev | 8.5 | 3.82 | 6 | -0.98 | ….. | ….. | ….. |
| 32490.90 | 3076.89 | 33319.21 | od | 6.5 | 828.31 | ev | 7.5 | 1.35 | 10 | -1.57 | ….. | ….. | ….. |

Notes. –Table 8 is available in its entirety via the link to the machine-readable version above.

Table 9. Experimental Atomic Transition Probabilities for Ho II arranged by wavenumber from Lawler et al. (2004).

| Wavenumber (cm$^{-1}$) | $\lambda_{air}$ (Å) | Upper Level Energy (cm$^{-1}$) | J | Lower Level Energy (cm$^{-1}$) | J | Transition Probability (10$^6$ s$^{-1}$) | log$_{10}$(gf) This Expt. | Other |
|---|---|---|---|---|---|---|---|---|
| 30918.86 | 3233.34 | 31556.61 | 7 | 637.75 | 7 | 4.1±0.8 | -1.02 | |
| 29899.58 | 3343.57 | 29899.58 | 7 | 0.00 | 8 | 14.1±1.1 | -0.45 | -0.54[a] |
| 29450.17 | 3394.59 | 35067.62 | 6 | 5617.46 | 7 | 16.1±1.1 | -0.44 | |
| 29412.51 | 3398.94 | 29412.51 | 8 | 0.00 | 8 | 87±5 | 0.41 | 0.27[b] |
| 29325.35 | 3409.04 | 29325.35 | 7 | 0.00 | 8 | 3.05±0.21 | -1.10 | |

[a]VALD database as described in Kupka et al. (1999), value determined using the method of Magazzù & Cowley (1986)

[b]VALD database as described in Kupka et al. (1999), value originally from Gorshkov & Komarovskii (1979)

Notes. – Table 9 available in its entirety via the link to the machine-readable version above.

Table 10. Hyperfine structure line component patterns for Ho II computed using hfs constants and improved energy levels from Lawler et al. (2004) and the standard index of air (Edlén 1953).

| Wavenumber (cm$^{-1}$) | $\lambda_{air}$ (Å) | $F_{upp}$ | $F_{low}$ | Component Position (cm$^{-1}$) | Component Position (Å) | Strength |
|---|---|---|---|---|---|---|
| 30918.859 | 3233.3387 | 10.5 | 10.5 | 0.81266 | -0.084984 | 0.17569 |
| 30918.859 | 3233.3387 | 10.5 | 9.5 | 0.80394 | -0.084073 | 0.00764 |
| 30918.859 | 3233.3387 | 9.5 | 10.5 | 0.46722 | -0.048860 | 0.00764 |
| 30918.859 | 3233.3387 | 9.5 | 9.5 | 0.45850 | -0.047949 | 0.14620 |
| 30918.859 | 3233.3387 | 9.5 | 8.5 | 0.45885 | -0.047985 | 0.01283 |

Notes. – Center-of-gravity wavenumbers and air wavelengths, $\lambda_{air}$, are given with component positions relative to those values. Strengths are normalized to sum to 1. Table 10 is available in its entirety via the link to the machine-readable version above.

Table 11. Experimental Atomic Transition Probabilities for Tm II from Wickliffe & Lawler (1997).

| Wavenumber (cm$^{-1}$) | $\lambda_{air}$ (Å) | Upper Level Energy (cm$^{-1}$) | Parity | J | Lower Level Energy (cm$^{-1}$) | Parity | J | Transition Probability (10$^6$ s$^{-1}$) | Uncert. (%) | log$_{10}$(gf) |
|---|---|---|---|---|---|---|---|---|---|---|
| 34913.84 | 2863.35 | 34913.84 | ev | 3 | 0.00 | od | 4 | 1.05 | 8 | -2.04 |
| 34842.39 | 2869.22 | 47299.68 | od | 7 | 12457.29 | ev | 6 | 225 | 6 | 0.62 |
| 34634.42 | 2886.45 | 34871.37 | ev | 4 | 236.95 | od | 3 | 1.16 | 11 | -1.88 |
| 34580.75 | 2890.93 | 34580.75 | ev | 5 | 0.00 | od | 4 | 5.23 | 5 | -1.14 |
| 34307.52 | 2913.96 | 34307.52 | ev | 4 | 0.00 | od | 4 | 1.11 | 5 | -1.89 |

Notes. –Table 11 is available in its entirety via the link to the machine-readable version above.

Table 12. Improved Lu II energy levels (± 0.008 cm$^{-1}$) and NIST (Martin et al. 1978) energy levels for comparison.

| Energy (cm$^{-1}$) | | J |
|---|---|---|
| This Expt. | NIST | |
| 0.000 | 0.00 | 0 |
| 11796.108 | 11796.24 | 1 |
| 12435.229 | 12435.32 | 2 |
| 14198.978 | 14199.08 | 3 |
| 17332.428 | 17332.58 | 2 |
| 27264.269 | 27264.40 | 0 |
| 28503.149 | 28503.16 | 1 |
| 32453.115 | 32453.26 | 2 |
| 41224.821 | 41224.96 | 2 |
| 44918.529 | 44918.68 | 3 |
| 48536.758 | 48536.83 | 4 |
| 45458.326 | 45458.56 | 2 |

Table 13. Experimental atomic transition probabilities for Lu II from odd-parity upper levels organized by increasing wavelength in air. Updated energy levels from Table 12 are used for transitions if available both for the upper and lower levels, otherwise NIST (Martin et al. 1978) energy levels are used. Wavelengths are computed using the standard index of air (Edlén 1953).

| $\lambda_{air}$ (Å) | Upper Level Energy (cm$^{-1}$) | J | Lower Level Energy (cm$^{-1}$) | J | Transition Probability (10$^6$ s$^{-1}$) | log$_{10}$(gf) |
|---|---|---|---|---|---|---|
| 2571.23 | 53079.33 | 3 | 14199.08 | 3 | 39.8 ± 2.7 | -0.56 |
| 2754.17 | 48733.19 | 4 | 12435.32 | 2 | 103 ± 7 | 0.02 |
| 2796.63 | 53079.33 | 3 | 17332.58 | 2 | 181 ± 9 | 0.17 |
| 2894.84 | 48733.19 | 4 | 14199.08 | 3 | 186 ± 12 | 0.32 |
| 2911.3915 | 48536.758 | 4 | 14198.978 | 3 | 245 ± 16 | 0.45 |
| 3077.6109 | 44918.529 | 3 | 12435.229 | 2 | 131 ± 7 | 0.12 |
| 3254.3173 | 44918.529 | 3 | 14198.978 | 3 | 64 ± 4 | -0.15 |
| 3397.0669 | 41224.821 | 2 | 11796.108 | 1 | 75 ± 5 | -0.19 |
| 3472.4832 | 41224.821 | 2 | 12435.229 | 2 | 71 ± 5 | -0.19 |
| 3507.3810 | 28503.149 | 1 | 0.000 | 0 | 12.5 ± 1.1 | -1.16 |
| 3623.9806 | 44918.529 | 3 | 17332.428 | 2 | 11.9 ± 1.4 | -0.79 |
| 4184.2533 | 41224.821 | 2 | 17332.428 | 2 | 21.0 ± 2.0 | -0.56 |
| 4839.6201 | 32453.115 | 2 | 11796.108 | 1 | 0.45 ± 0.05 | -2.10 |
| 4994.1393 | 32453.115 | 2 | 12435.229 | 2 | 4.73 ± 0.28 | -1.05 |
| 5476.6884 | 32453.115 | 2 | 14198.978 | 3 | 21.2 ± 1.1 | -0.32 |
| 5983.8429 | 28503.149 | 1 | 11796.108 | 1 | 4.3 ± 0.4 | -1.16 |
| 6221.8596 | 28503.149 | 1 | 12435.229 | 2 | 9.9 ± 0.9 | -0.76 |
| 6463.1065 | 27264.269 | 0 | 11796.108 | 1 | 15.4 ± 0.8 | -1.01 |
| 8459.16 | 41224.96 | 2 | 29406.70 | 2 | 2.5 ± 0.5 | -0.87 |

Notes. – A machine-readable version of Table 13 is available via the above link.

Table 14. Hyperfine structure line component patterns for $^{175}$Lu II computed from hfs constants of Sneden et al. (2003a), energy levels of Table 12, and the standard index of air (Edlén 1953).

| Wavenumber (cm$^{-1}$) | $\lambda_{air}$ (Å) | $F_{upp}$ | $F_{low}$ | Component Position (cm$^{-1}$) | Component Position (Å) | Strength |
|---|---|---|---|---|---|---|
| 34337.780 | 2911.3915 | 7.5 | 6.5 | -0.53033 | 0.044968 | 0.22222 |
| 34337.780 | 2911.3915 | 6.5 | 6.5 | -0.62922 | 0.053353 | 0.02618 |
| 34337.780 | 2911.3915 | 6.5 | 5.5 | -0.22156 | 0.018786 | 0.16827 |
| 34337.780 | 2911.3915 | 5.5 | 6.5 | -0.67036 | 0.056841 | 0.00160 |
| 34337.780 | 2911.3915 | 5.5 | 5.5 | -0.26270 | 0.022274 | 0.04196 |

Notes. – Center-of-gravity wavenumbers and air wavelengths, $\lambda_{air}$, are given with component positions relative to those values. Strengths are normalized to sum to 1. Table 14 is available in its entirety via the link to the machine-readable version above.